\documentclass[letterpaper,english,twoside]{IEEEtran}
\usepackage[latin9]{inputenc}
\usepackage{bm}
\usepackage{amsthm}
\usepackage{amsmath}
\usepackage{amssymb}
\usepackage{esint}

\makeatletter

\pdfpageheight\paperheight
\pdfpagewidth\paperwidth

\theoremstyle{plain}
\newtheorem{thm}{\protect\theoremname}
\theoremstyle{remark}
\newtheorem{rem}[thm]{\protect\remarkname}
\theoremstyle{definition}
\newtheorem{example}[thm]{\protect\examplename}
\theoremstyle{plain}
\newtheorem{prop}[thm]{\protect\propositionname}
\theoremstyle{plain}
\newtheorem{lem}[thm]{\protect\lemmaname}
\theoremstyle{definition}
\newtheorem{defn}[thm]{\protect\definitionname}
\theoremstyle{plain}
\newtheorem{cor}[thm]{\protect\corollaryname}

\IEEEoverridecommandlockouts
\usepackage{amsmath,amssymb,bbm,stmaryrd}
\usepackage{bm}
\usepackage{tikz}
\usepackage{stfloats}
\usepackage{pgfplots}
\usetikzlibrary{spy,shapes,decorations,matrix,positioning}
\usepackage{babel,flushend}
\usepackage{scalefnt}
\pgfplotsset{compat=newest}

\newif\ifextfig
\extfigtrue

\renewcommand{\epsilon}{\varepsilon}
\newcommand{\eps}{\epsilon}

\global\long\def\emap{\epsilon^{\mathrm{MAP}}}

\renewcommand{\t}{\top}

\newcommand{\R}{\mathbb{R}}
\newcommand{\Xca}{0}
\newcommand{\Xcb}{x_{\max}}
\newcommand{\Xc}{\mathcal{X}}
\newcommand{\Xm}{\mathcal{X}_{\circ}}
\newcommand{\Eca}{0}
\newcommand{\Ecb}{\epsilon_{\max}}
\newcommand{\Ec}{\mathcal{E}}
\newcommand{\Em}{\mathcal{E}_{\circ}}
\renewcommand{\SS}{\bm{S}}
\newcommand{\tx}{\tilde{\bm{x}}}
\newcommand{\ti}{i_0}
\newcommand{\x}{\bm{x}}
\newcommand{\y}{\bm{y}}
\newcommand{\z}{\bm{z}}
\newcommand{\A}{\bm{A}}
\newcommand{\T}{\bm{T}}
\newcommand{\f}{\bm{f}}
\newcommand{\g}{\bm{g}}

\renewcommand{\c}{\bm{c}}

\newcommand{\h}{\mathsf{h}}

\providecommand{\norm}[1]{\left\lVert#1\right\rVert}

\newcommand{\argmin}{\operatornamewithlimits{argmin}}
\newcommand{\X}{\bm{X}}

\renewcommand{\d}{\text{d}}

\pgfplotsset{tick label style={
font=\Large}}

\makeatother

\providecommand{\corollaryname}{Corollary}
\providecommand{\definitionname}{Definition}
\providecommand{\examplename}{Example}
\providecommand{\lemmaname}{Lemma}
\providecommand{\propositionname}{Proposition}
\providecommand{\remarkname}{Remark}
\providecommand{\theoremname}{Theorem}

\begin{document}
\global\long\def\snr{\textsf{snr}}

\global\long\def\mmse{\textsf{mmse}}

\global\long\def\xf{\tilde{x}}

\global\long\def\argmin{\operatornamewithlimits{arg\, min}}

\global\long\def\fdef{\colon}

\global\long\def\ir#1#2{[#1\mathbin:#2]}

\global\long\def\sXc{\tilde{\mathcal{X}}}

\global\long\def\Xm{\mathcal{X}_{\circ}}

\global\long\def\Xca{0}

\global\long\def\xmax{x_{\max}}

\global\long\def\Xcb{\xmax}

\global\long\def\Yc{\mathcal{Y}}

\global\long\def\sYc{\tilde{\mathcal{Y}}}

\global\long\def\ymax{y_{\max}}

\global\long\def\Ec{\mathcal{E}}

\global\long\def\Em{\mathcal{E}_{\circ}}

\global\long\def\Eca{0}

\global\long\def\emax{\epsilon_{\max}}

\global\long\def\Ecb{\emax}

\global\long\def\R{\mathbb{R}}

\global\long\def\d{\textup{\ensuremath{\mathrm{d}}}}

\global\long\def\c{\bm{c}}

\global\long\def\x{\bm{x}}

\global\long\def\y{\bm{y}}

\global\long\def\z{\bm{z}}

\global\long\def\f{\bm{f}}

\global\long\def\g{\bm{g}}

\global\long\def\h{\bm{h}}

\global\long\def\Vs{V_{\mathrm{s}}}

\global\long\def\Us{U_{\mathrm{s}}}

\global\long\def\Uc{U_{\mathrm{c}}}

\global\long\def\ex#1{\tilde{\epsilon}(#1)}

\global\long\def\epx#1{\tilde{\epsilon}'(#1)}

\global\long\def\eo{\epsilon_{0}^{*}}

\global\long\def\es{\epsilon_{\mathrm{s}}^{*}}

\global\long\def\ec{\epsilon_{\mathrm{c}}^{*}}

\global\long\def\est{\epsilon_{\mathrm{stab}}^{*}}

\global\long\def\emap{\epsilon^{\mathrm{MAP}}}

\global\long\def\emaxwell{\epsilon^{\mathrm{Max}}}

\global\long\def\sxmax{\tilde{x}_{\max}}

\global\long\def\sfs{\tilde{f}}

\global\long\def\sgs{\tilde{g}}

\global\long\def\sFs{\tilde{F}}

\global\long\def\sGs{\tilde{G}}

\global\long\def\sUs{\tilde{U}_{\mathrm{s}}}

\global\long\def\sfc{\tilde{\f}}

\global\long\def\sgc{\tilde{\g}}

\global\long\def\shc{\tilde{\h}}

\global\long\def\sUc{\tilde{U}_{\mathrm{c}}}

\global\long\def\q{\bm{q}}

\global\long\def\SS{\bm{S}}

\global\long\def\T{\bm{T}}

\global\long\def\D{\bm{D}}

\global\long\def\A{\bm{A}}

\global\long\def\X{\bm{X}}

\global\long\def\B{\bm{B}}

\global\long\def\t{\top}

\global\long\def\ti{i_{0}}

\global\long\def\tx{\tilde{\x}}

\global\long\def\Xc{\mathcal{X}}

\global\long\def\eps{\epsilon}

\global\long\def\vec{\operatorname{vec}}

\global\long\def\norm#1{\left\Vert #1\right\Vert }

\title{A Simple Proof of Maxwell Saturation \\
 for Coupled Scalar Recursions}

\author{Arvind Yedla~\IEEEmembership{Member,~IEEE}, Yung-Yih Jian, Phong S. Nguyen, and Henry D. Pfister~\IEEEmembership{Senior Member,~IEEE}%
\thanks{ This
work was done while the authors were with the Department of Electrical
and Computer Engineering at Texas A\&M University. The material is
based upon work supported by the National Science Foundation (NSF)
under Grants No. 0747470, No. 0802124, and No. 1320924. Any opinions,
findings, conclusions, and recommendations expressed in this material
are those of the authors and do not necessarily reflect the views
of these sponsors. This research was presented in part at the 2012
International Symposium on Turbo Codes and Iterative Information Processing
in Gothenburg, Sweden and at the 2013 Information Theory and Applications
Workshop in San Diego, CA~\cite{Yedla-istc12,Jian-ita13}.

Arvind Yedla is currently with Samsung Information Systems in San
Diego, CA (e-mail: arvind.yedla@gmail.com). Yung-Yih Jian is currently
with Qualcomm Inc. in Santa Clara, CA (e-mail: yungyih.jian@gmail.com).
Phong S. Nguyen is currently with Marvell Semiconductor in Santa Clara,
CA (email: phongece@gmail.com). Henry D. Pfister is now an Associate
Professor in the Department of Electrical and Computer Engineering
at Duke University (email: henry.pfister@duke.edu).%
} }
\maketitle
\begin{abstract}
Low-density parity-check (LDPC) convolutional codes (or spatially-coupled
codes) were recently shown to approach capacity on the binary erasure
channel (BEC) and binary-input memoryless symmetric channels. The
mechanism behind this spectacular performance is now called threshold
saturation via spatial coupling. This new phenomenon is characterized
by the belief-propagation threshold of the spatially-coupled ensemble
increasing to an intrinsic noise threshold defined by the uncoupled
system. 

In this paper, we present a simple proof of threshold saturation that
applies to a wide class of coupled scalar recursions. Our approach
is based on constructing potential functions for both the coupled
and uncoupled recursions. Our results actually show that the fixed
point of the coupled recursion is essentially determined by the minimum
of the uncoupled potential function and we refer to this phenomenon
as Maxwell saturation. 

A variety of examples are considered including the density-evolution
equations for: irregular LDPC codes on the BEC, irregular low-density generator matrix codes
on the BEC, a class of generalized LDPC codes with BCH component codes,
the joint iterative decoding of LDPC codes on intersymbol-interference
channels with erasure noise, and the compressed sensing of random vectors with i.i.d. components. \end{abstract}
\begin{IEEEkeywords}
convolutional LDPC codes, Maxwell conjecture, potential
functions, spatial coupling, threshold saturation 
\end{IEEEkeywords}

\section{Introduction}

Convolutional low-density parity-check (LDPC) codes, or spatially-coupled
(SC) LDPC codes, were introduced in \cite{Felstrom-it99} and shown
to have excellent belief-propagation (BP) thresholds in \cite{Sridharan-aller04,Lentmaier-isit05,Lentmaier-it10}.
Moreover, they have recently been observed to \emph{universally} approach
the capacity of various channels~\cite{Lentmaier-it10,Kudekar-istc10,Rathi-isit11,Yedla-isit11,Kudekar-isit11-DEC,Nguyen-arxiv11,Nguyen-icc12,Kudekar-it13}.

The fundamental mechanism behind this is explained well in \cite{Kudekar-it11},
where it is proven analytically for the binary erasure channel (BEC) that the BP threshold
of a particular SC ensemble converges to the maximum-a-posteriori
(MAP) threshold of the underlying ensemble. This phenomenon is now
called \emph{threshold saturation}. A similar result was also observed
independently in \cite{Lentmaier-isit10} and stated as a conjecture.
The same result for general binary memoryless symmetric (BMS) channels
was first empirically observed~\cite{Lentmaier-it10,Kudekar-istc10}
and recently proven analytically~\cite{Kudekar-it13}. This result
implies that one can achieve the capacity universally over BMS channels
because the MAP threshold of regular LDPC codes approaches the Shannon
limit universally, as the node degrees are increased. 

The underlying principle behind threshold saturation appears to be
very general and spatial coupling has now been applied, with much
success, to a variety of more general scenarios in information theory
and coding. In~\cite{Hassani-itw10,Hassani-jsp13}, the benefits
of spatial coupling are described for $K$-satisfiability, graph coloring,
and the Curie-Weiss model in statistical physics. SC codes are shown
to achieve the entire rate-equivocation region for the BEC wiretap
channel in~\cite{Rathi-isit11}. The authors observe in~\cite{Yedla-isit11}
that the phenomenon of threshold saturation extends to multi-terminal
problems (e.g., a noisy Slepian-Wolf problem) and can provide universality
over unknown channel parameters. Threshold saturation has also been
observed for the binary-adder channel~\cite{Kudekar-isit11-MAC},
for intersymbol-interference channels~\cite{Kudekar-isit11-DEC,Nguyen-arxiv11,Nguyen-icc12},
for message-passing decoding of code-division multiple access (CDMA)~\cite{Takeuchi-isit11,Schlegel-isit11,Truhachev-comlett12},
and for iterative hard-decision decoding of SC generalized LDPC codes~\cite{Jian-isit12}.
For compressive sensing, SC measurement matrices were investigated
with verification-based reconstruction in~\cite{Kudekar-aller10},
shown to have excellent performance with BP decoding in~\cite{Krzakala-physrevx},
and proven to achieve the information-theoretic limit in~\cite{Donoho-it13}.

In many of these papers it is conjectured, either implicitly or explicitly,
that threshold saturation occurs for the studied problem. A general
proof of threshold saturation (especially one where only a few details
must be verified for each system) would allow one to settle all of
these conjectures simultaneously. In this paper, we provide such a
proof for the case where the system of interest is characterized by
a coupled scalar recursion~\cite{Yedla-istc12}. Our results actually
go further and establish that the fixed point of the coupled recursion
is essentially determined by the minimum of a potential function associated
with the uncoupled recursion~\cite{Jian-ita13}. We call this phenomenon
\emph{Maxwell saturation} because the fixed point of the coupled recursion
saturates to a value that is closely related to the Maxwell curve
of the uncoupled system \cite{Measson-it08}. 

Our method is based on potential functions and was motivated mainly
by the approach taken in~\cite{Takeuchi-ieice12}. It turns out that
their approach is missing a few important elements and does not, as
far as we know, lead to a general proof of threshold saturation. Still,
it introduces the idea of using a potential function defined by an
integral of the DE recursion and this is an important element in our
approach. Elements of this approach have also appeared in some previous
proofs of threshold saturation (e.g., \cite{Donoho-it13,Truhachev-comlett12})
where the authors relied on a continuum approach to DE. A little while
after we posted the conference version of this article~\cite{Yedla-istc12},
Kudekar et al. posted their general proof of threshold saturation,
which was derived independently based on a continuum approach and
a slightly different potential function~\cite{Kudekar-unpub12}.
Later, we extended our approach to show Maxwell saturation in~\cite{Jian-ita13}
and they also extended their approach in~\cite{Kudekar-unpub13}.
The approach used in this paper has also been extended to vector recursions
in~\cite{Yedla-itw12} and BMS channels in \cite{Kumar-aller12,Kumar-itsub13}.
The connections between these approaches is discussed more thoroughly
in Section~\ref{sub:History}.

\subsection{Notation}

\label{sec:notation} The following notation is used throughout the
paper. We define the closed real intervals $\Xc=\left[0,\xmax\right]$
with $\Xcb\in(0,\infty)$, $\Yc=\left[0,\ymax\right]$ with $\ymax\in(0,\infty)$,
and $\Ec=[\Eca,\Ecb]$ with $\Ecb\in(0,\infty)$. Intervals of natural
numbers are denoted by $\ir mn\triangleq\{m,m+1,\ldots,n\}$. 

Vectors are denoted in boldface (e.g. $\x\in\Xc^{n}$), assumed to
be column vectors, and their elements are denoted by $[\x]_{i}$ or
$x_{i}$ for $i\in\ir 1n$. For vectors (e.g., $\x,\z\in\Xc^{n}$),
we use the partial order $\x\preceq\z$ defined by $x_{i}\leq z_{i}$
for $i\in\ir 1n$. A vector mapping (e.g., $\f(\x)$) that is non-decreasing
w.r.t. the partial order (i.e., $\x\preceq\z$ implies $\f(\x)\preceq\f(\z)$)
is called isotone. The symbols $\bm{0}_{n}$ and $\bm{1}_{n}$ denote
the all-zeros and all-ones column vectors of length $n$ and the subscript
is sometimes dropped when the length is apparent. 

We use standard-weight type (e.g., $f(x)$ and $F(\x$)) to denote
scalar-valued functions and boldface (e.g., $\f(\x)$ and $\g(\x)$)
to denote vector-valued functions. We say a function $f\colon\mathcal{Z}\to\mathbb{R}$
is $C^{k}$ if its $k$-th derivative exists and is continuous on
$\mathcal{Z}$ and we denote the supremum of $f$ over its domain
by $\left\Vert f\right\Vert _{\infty}\triangleq\sup_{z\in\mathcal{Z}}f(z)$. 

The gradient vector of a scalar-valued function is defined by $F'(\x)=\left[\partial F(\x)/\partial x_{1},\ldots,\partial F(\x)/\partial x_{n}\right]^{\t}$
and we use the convention that lowercase denotes the derivative (e.g.,
$f(x)=F'(x)$ and $\f(\x)=F'(\x)$) when both are defined. The Jacobian
matrix of a vector-valued function is defined to be
\[
\f'(\x)=\frac{\partial\f(\x)}{\partial\x}\triangleq\begin{bmatrix}\frac{\partial f_{1}(\x)}{\partial x_{1}} & \cdots & \frac{\partial f_{n}(\x)}{\partial x_{1}}\\
\vdots & \ddots & \vdots\\
\frac{\partial f_{1}(\x)}{\partial x_{n}} & \cdots & \frac{\partial f_{n}(\x)}{\partial x_{n}}
\end{bmatrix}.
\]
We note that the gradient and Jacobian definitions are transposed with respect to the standard convention.

\subsection{Outline}

Section~\ref{sub:problem_statement} describes the problem of interest
and states the main results for a fixed recursion. The recent history
of this problem is discussed in Section~\ref{sub:History}. Section~\ref{sub:proof_ub_theorem}
details the proof of Theorem~\ref{thm:main_scalar_ub} and Section~\ref{sec:proof_lb_theorem}
describes the proof of Theorem~\ref{thm:main_scalar_lb}. Section~\ref{sec:dependence}
extends these results to the case where the recursion depends on a
continuous parameter. Applications of these results are presented
in Section~\ref{sec:applications}.

\section{Maxwell Saturation}

\label{sec:MaxwellSat}

\subsection{Problem Statement and Main Results}

\label{sub:problem_statement} Let $f\fdef\Yc\to\Xc$ be a non-decreasing
$C^{1}$ function,  $g\fdef\Xc\to\Yc$ be a strictly increasing $C^{2}$
function, and assume $\ymax=g(\xmax)$. The main goal of this paper
is to analyze coupled recursions of the form
\begin{equation}
\begin{split}y_{i}^{(\ell+1)} & =g\left(x_{i}^{(\ell)}\right)\\
x_{i}^{(\ell+1)} & =\sum_{j=1}^{N}A_{j,i}\, f\left(\sum_{k=1}^{M}A_{j,k}\, y_{k}^{(\ell+1)}\right)
\end{split}
\label{eq:coupled_scalar_recursion}
\end{equation}
for $i\in\ir 1M$, where $M\triangleq N+w-1$ and $\A=\{A_{j,k}\}$
is the $N\times M$ matrix defined by
\begin{equation}
A_{j,k}=\left[\A\right]_{j,k}\triangleq\begin{cases}
\frac{1}{w} & \text{if }1\leq k-j+1\leq w\\
0 & \text{otherwise}.
\end{cases}\label{eq:Amatrix}
\end{equation}
It is important to observe that, while the inner sum for $x_{i}^{(\ell+1)}$
simply averages $w$ adjacent $y$-values, the outer sum implicitly
uses a boundary value of 0 because $\sum_{j=1}^{N}A_{j,i}<1$ for
$i\in\ir 1{w-1}\cup\ir{N+1}{N+w-1}$.  Under the conditions described
below, we show that the fixed point of the coupled recursion is intrinsically
connected to the dynamics of the uncoupled recursion
\begin{equation}
\begin{split}y^{(\ell+1)} & =g\left(x^{(\ell)}\right)\\
x^{(\ell+1)} & =f\left(y^{(\ell+1)}\right).
\end{split}
\label{eq:scalar_recursion}
\end{equation}
This problem is motivated by the density evolution (DE) recursions
that characterize the large-system performance of iterative demodulation
and decoding schemes with and without spatial coupling \cite{Luby-it01,Boutros-it02,Pfister-jsac08,Kudekar-it11,Schlegel-isit11,Nguyen-icc12,Jian-isit12}.

The recursions~\eqref{eq:coupled_scalar_recursion} and~\eqref{eq:scalar_recursion}
are initialized by choosing $x_{i}^{(0)}=\xmax$ for $i\in\ir 1M$
and $x^{(0)}=\xmax$, respectively. Since $f(g(\Xc))\subseteq\Xc$,
this initialization implies that $x^{(1)}\leq x^{(0)}$. Since $f,g$
are non-decreasing, $h(x)\triangleq f(g(x))$ is non-decreasing and
$x^{(\ell)}\leq x^{(\ell-1)}$ implies 
\[
x^{(\ell+1)}=f\left(g(x^{(\ell)})\right)\leq f\left(g(x^{(\ell-1)})\right)=x^{(\ell)}.
\]
Therefore, the sequence $\left(x^{(\ell)},y^{(\ell)}\right)$ converges
to a limit $\left(x^{(\infty)},y^{(\infty)}\right)$ and that limit
is a fixed point because $f,g$ are continuous. The set of all fixed
points will be important and is denoted by
\[
\mathcal{F}\triangleq\left\{ x\in\Xc\mid x=f(g(x))\right\} .
\]

For the coupled recursion, it is easy to verify that $A_{j,k}\geq0$
implies that
\[
x_{i}^{(\ell)}\leq x_{i}^{(\ell-1)}\;\forall i\in\ir 1M\,\implies\, x_{i}^{(\ell+1)}\leq x_{i}^{(\ell)}\;\forall i\in\ir 1M.
\]
Therefore, $x_{i}^{(\ell+1)}\leq x_{i}^{(\ell)}$ is given by induction
from the base case $x_{i}^{(1)}\leq x_{i}^{(0)}$, which follows from
$\sum_{k=1}^{M}A_{j,k}\leq1$ and $\sum_{j=1}^{N}A_{j,k}\leq1$. Hence,
for each $i\in\ir 1M$, the sequence $\big(x_{i}^{(\ell)},y_{i}^{(\ell)}\big)$
converges to a limit $\big(x_{i}^{(\infty)},y_{i}^{(\infty)}\big)$
as $\ell\to\infty$ and the limiting vector is a fixed point of the
coupled recursion because the implied vector update is continuous. 

Let the single-system (or uncoupled)\emph{ potential function} $\Us\fdef\Xc\to\mathbb{R}$
of the scalar recursion be 

\begin{equation}
\Us(x)\triangleq xg(x)-G(x)-F(g(x)),\label{eq:uni_scalar_pot}
\end{equation}
where $F(x)=\int_{0}^{x}f(z)\d z$ and $G(x)=\int_{0}^{x}g(z)\d z$.
The main results of this paper are encapsulated in the following two
theorems. The first theorem shows that the maximum element of the
fixed point vector for the coupled recursion is upper bounded by the
largest minimizer of the uncoupled potential function.
\begin{thm}
\label{thm:main_scalar_ub} For any $\delta>0$, there is $w_{0}<\infty$
such that, for all $w>w_{0}$ and all $N\in\ir 1{\infty}$, the fixed
point of the coupled recursion satisfies the upper bound 
\begin{equation}
\max_{i\in[1:M]}x_{i}^{(\infty)}-\delta\leq\overline{x}^{*}\triangleq\max\left(\argmin_{x\in\Xc}\Us(x)\right).\label{eq:scalar_theorem_ub}
\end{equation}
If $\overline{x}^{*}$ is not a limit point of $\mathcal{F}\cap[\overline{x}^{*},\xmax]$,
then $w_{0}<\infty$ for $\delta=0$.

\end{thm}
\begin{IEEEproof}
The proof is given in Section~\ref{sub:proof_ub_theorem}. 
\end{IEEEproof}

The second theorem shows that the maximum element of the fixed point
vector for the coupled recursion is lower bounded by the smallest
minimizer of the uncoupled potential function.
\begin{thm}
\label{thm:main_scalar_lb} For any fixed $w\geq1$ and $\eta>0$,
there is an $N_{0}<\infty$ such that, for all $N>N_{0}$, the fixed
point of the coupled recursion satisfies the lower bound 
\begin{equation}
\max_{i\in[1:M]}x_{i}^{(\infty)}+\eta\geq\underline{x}^{*}\triangleq\min\left(\argmin_{x\in\Xc}\Us(x)\right).\label{eq:scalar_theorem_lb}
\end{equation}
\end{thm}
\begin{IEEEproof}
The proof is given in Section~\ref{sec:proof_lb_theorem}.
\end{IEEEproof}
By combining Theorems~\ref{thm:main_scalar_ub} and ~\ref{thm:main_scalar_lb},
we see that the maximum element of the fixed-point vector for the
coupled recursion is essentially determined by the minimizer of uncoupled
potential. When the system also depends on a parameter (e.g., see
Section~\ref{sec:dependence}), this result is closely related to
the Maxwell construction in \cite{Measson-it08}. For that reason,
we refer to this phenomenon as \emph{Maxwell saturation}. In contrast,
\emph{threshold saturation} refers to the slightly weaker statement
that the iterative decoding threshold of the SC system (for reliable
communication) converges to the MAP threshold of the single system
for sufficiently large $w$.
\begin{rem}
It is natural to wonder how the choice of a uniform fixed-width coupling
(e.g., see \eqref{eq:Amatrix}) affects our results. Suppose instead
that $\left[\A\x\right]_{i}=\sum_{j=0}^{w-1}a_{j}x_{i+j}$ where $\bm{a}=(a_{0},\ldots,a_{w-1})$
is a symmetric non-negative vector that sums to 1. While it is possible
to extend Theorem~\ref{thm:main_scalar_ub} to this case (e.g., see
\cite{Donoho-it13,Kudekar-unpub13}), the resulting bounds depend
only on the maximum value $\max_{j\in\ir 0{w-1}}a_{j}$. In terms
of the bounds, the optimum $\bm{a}$ vector is $a_{j}=\frac{1}{w}$.
Hence, for the sake of simplicity, we consider only uniform fixed-width
coupling.
\end{rem}

\begin{rem}
\label{rem:w-limit cor of thm 1} The statement of Theorem~\ref{thm:main_scalar_ub}
is closely related to the definition of $\lim_{w\to\infty}$ and this
connection implies that
\[
\lim_{w\to\infty}\lim_{N\to\infty}\max_{i\in[1:N+w-1]}x_{i}^{(\infty)}\leq\overline{x}^{*}.
\]

\end{rem}
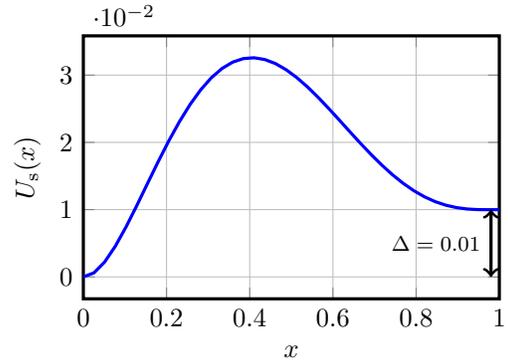
\begin{figure}[h]
\begin{center}
\ifextfig
\includegraphics{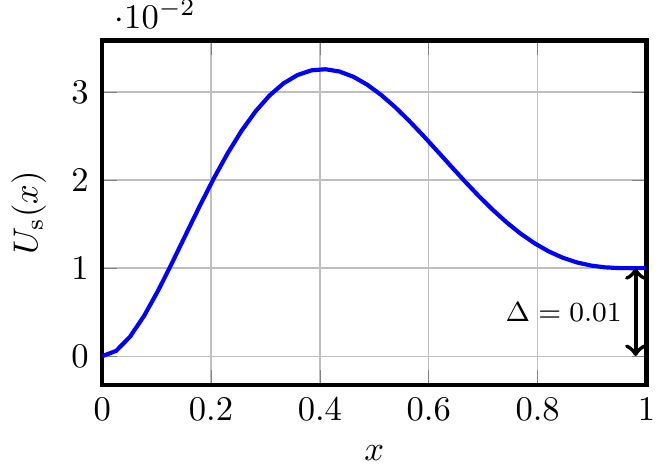}
\else
\pgfplotsset{every axis/.append style={very thick,enlarge x limits=false,scale=1.0,grid=major}}
\begin{tikzpicture}
\begin{axis}[
width=2.8in,
height=2in,
tick label style={font=\normalsize},
xmajorgrids,
ymajorgrids,
xlabel=$x$,ylabel={$U_\mathrm{s} (x)$}]
  \addplot[color=blue,domain=0:1,samples=40] {x*(1-(1-x)^2)- x-((1-x)^3-1)/3-97*((1-(1-x)^2)^3)/300};
\draw[<->,very thick] (axis cs:0.98,0) -- node[left=0.1pt]{\footnotesize{$\Delta=0.01$}} (axis cs:0.98,0.01);
\end{axis}
\end{tikzpicture}
\fi
\hspace*{7mm}
\end{center}
\vspace{-4mm}
\caption{The potential function for Example~\ref{ex:ldpc33}.}
\label{fig:ex_ldpc33}
\end{figure}
\begin{example}
\label{ex:ldpc33} Consider the recursion defined by $\xmax=\ymax=1$,
$f(x)=\frac{97}{100}x^{2}$ and $g(x)=1-(1-x)^{2}$. This recursion
can be seen as the DE recursion of a (3,3)-regular LDPC ensemble on
a BEC with erasure probability $\frac{97}{100}$. In this case, we
have $F(x)=\frac{97}{300}x^{3}$, $G(x)=x+\frac{1}{3}\left((1-x)^{3}-1\right)$,
and
\begin{align*}
\Us(x) & =x(1-(1-x)^{2})-\left(x+\tfrac{(1-x)^{3}-1}{3}\right)-\tfrac{97(1-(1-x)^{2})^{3}}{300}.
\end{align*}
In Fig.~\ref{fig:ex_ldpc33}, we see that the minimum of $\Us(x)$
is uniquely achieved at $\overline{x}^{*}=0$. Since $0$ is not a
limit point of $\mathcal{F}$ (e.g., see Proposition~\ref{prop:finitew}),
there is a $w_{0}<\infty$ such that the coupled recursion converges
to the $\bm{0}$ vector for all $w>w_{0}$. One can extract the upper
bound, $w_{0}\leq\frac{1}{2\Delta}K_{f,g}\xmax^{2}$, from the proof
of Theorem~\ref{thm:main_scalar_ub}. For this example, we find that
$w_{0}<600$ because straightforward computations show that $\Delta=0.01$
and $K_{f,g}<12$.
\end{example}

\begin{figure}[h]
\begin{center}
\pgfplotsset{every axis/.append style={very thick,enlarge x limits=false,scale=1.0,grid=major}}

\pgfmathdeclarefunction{ex4g}{2}{\pgfmathparse{1-(1-#2)*(2/45+(1-#1)*2/45+(1-#1)^2*21/45+(1-#1)^3*4/9*(1-#1)^4)}}
\pgfmathdeclarefunction{ex4gg}{2}{\pgfmathparse{#1-(1-#2)*(1/3-(1-#1)*2/45-(1-#1)^2*1/45-(1-#1)^3*7/45-(1-#1)^4*1/9)}}

\ifextfig
\includegraphics{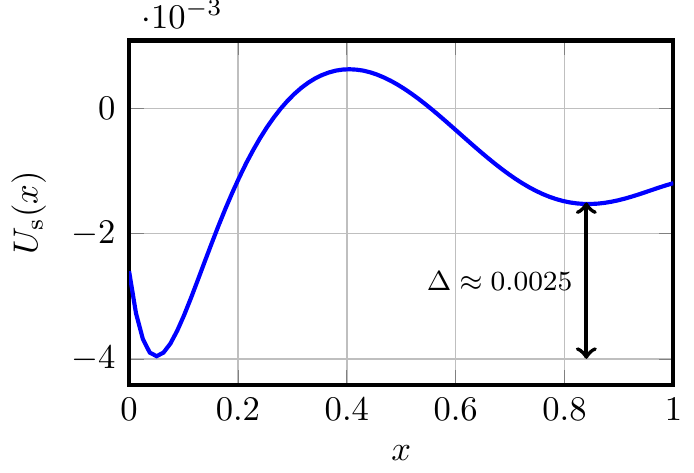}
\else
\begin{tikzpicture}
\begin{axis}[
width=2.8in,
height=2in,
tick label style={font=\normalsize},
xmajorgrids,
ymajorgrids,
xlabel=$x$,ylabel={$U_\mathrm{s} (x)$}]
  \addplot[color=blue,domain=0:1,samples=80] {x*ex4g(x,0.5) - ex4gg(x,0.5) - (1/6)*(ex4g(x,0.5))^6};
\draw[<->,very thick] (axis cs:0.84,-0.004) -- node[left=0.1pt]{\footnotesize{$\Delta\approx 0.0025$}} (axis cs:0.84,-0.0015);
\end{axis}
\end{tikzpicture}
\fi
\hspace*{7mm}
\end{center}
\vspace{-4mm}
\caption{The potential function for Example~\ref{ex:ldgm_irr1}.}
\label{fig:ex_ldgm_irr1}
\end{figure}
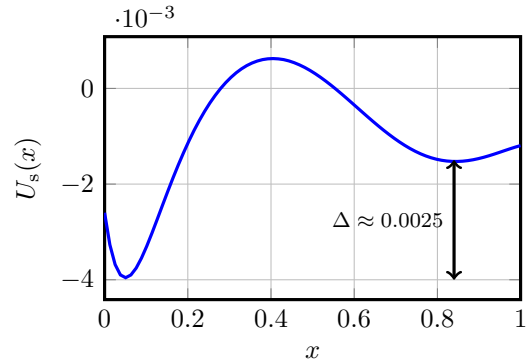
\begin{example}
\label{ex:ldgm_irr1} Consider the recursion defined by $\xmax=\ymax=1$,
$f(x)=x^{5}$ and $g(x)=1-\frac{1}{2}R'(1-x)/R'(1)$ where $R(x)=\tfrac{2}{15}x+\tfrac{1}{15}x^{2}+\tfrac{7}{15}x^{3}+\tfrac{1}{3}x^{4}$.
This recursion can be seen as the DE recursion of an irregular low-density generator matrix (LDGM)
ensemble on a BEC with erasure probability $\frac{1}{2}$. In this
case, we have $F(x)=\frac{1}{6}x^{6}$, $G(x)=x-\frac{1}{2}\left(1-R(1-x)\right)/R'(1)$,
and $\Us(x)$ is given by \eqref{eq:uni_scalar_pot}. In Fig.~\ref{fig:ex_ldgm_irr1},
we see that the function $\Us(x)$ has a unique global minima at $\overline{x}^{*}\approx0.05$.
Since $\overline{x}^{*}$ is not a limit point of $\mathcal{F}$ (e.g.,
see Proposition~\ref{prop:finitew}), there is a $w_{0}<\infty$
such that the maximum element of the coupled fixed-point vector is
upper bounded by $\overline{x}^{*}$ for all $w>w_{0}$. One can extract
the upper bound, $w_{0}\leq\frac{1}{2\Delta}K_{f,g}\xmax^{2}$, from
the proof of Theorem~\ref{thm:main_scalar_ub}. For this example,
we find that $w_{0}<2000$ because straightforward computations show
that $\Delta\geq0.0025$ and $K_{f,g}<10$.
\end{example}

\begin{figure}[h]
\begin{center}
\ifextfig
\includegraphics{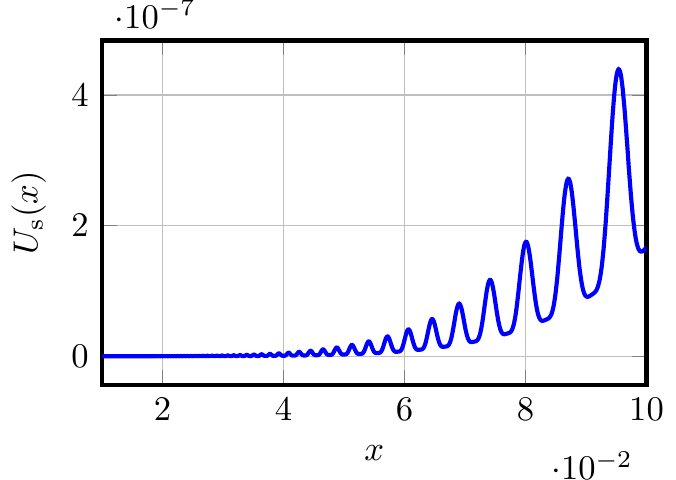}
\else
\pgfplotsset{every axis/.append style={very thick,enlarge x limits=false,scale=1.0,grid=major}}
\begin{tikzpicture}
\begin{axis}[
width=2.8in,
height=2in,
tick label style={font=\normalsize},
xmajorgrids,
ymajorgrids,
xlabel=$x$,ylabel={$U_\mathrm{s} (x)$}]
  \addplot[color=blue,domain=0.01:0.1,samples=800] {x^5*sin(180/x)^4/25+x^6/6};
\end{axis}
\end{tikzpicture}
\fi
\hspace*{7mm}
\end{center}
\vspace{-4mm}
\caption{The potential function for Example~\ref{ex:pathological}.}
\label{fig:ex_pathological}
\end{figure}
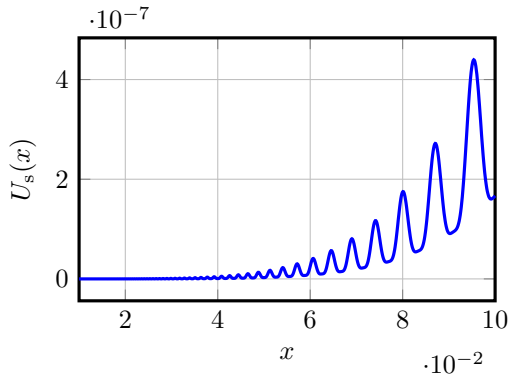
\begin{example}
\label{ex:pathological} Consider the following pathological example
defined by $\xmax=1$, $g(x)=x$, and $F(x)=\frac{1}{2}x^{2}-\frac{1}{25}x^{5}\sin^{4}(\frac{\pi}{x})-\frac{1}{30}x^{6}$.
With a little work, one can verify (e.g., numerically) that this system
satisfies the necessary conditions of Theorem~\ref{thm:main_scalar_ub}
(i.e., $F''(x)$ is continuous, bounded, and non-negative on $\Xc$).
The implied potential function, 
\[
\Us(x)=\tfrac{1}{25}x^{5}\sin^{4}(\tfrac{\pi}{x})+\tfrac{1}{30}x^{6},
\]
has a countably infinite set of local minima (e.g., see Fig.~\ref{fig:ex_pathological})
in the neighborhood of $x=0$ whose values approach 0 as $x\to0$.
Since all local minima are fixed points (e.g., see Lemma~\ref{lem:scalar_potential_minima_fp}),
this implies that $\mathcal{F}$ has a limit point at $x=0$. One
also finds that $\delta=0$ implies $w_{0}=\infty$. Thus, one cannot
guarantee that $\max_{i}x_{i}^{(\infty)}=0$ occurs for any finite
$w$. Still, Remark~\ref{rem:w-limit cor of thm 1} implies that
$\max_{i}x_{i}^{(\infty)}\to0$ as $w\to\infty$. \end{example}
\begin{rem}
While the stated result for Example~\ref{ex:ldpc33} is an instance
of threshold saturation (e.g., see \cite{Kudekar-it11,Yedla-istc12,Kudekar-unpub12}),
the result in Example~\ref{ex:ldgm_irr1} is an instance of Maxwell
saturation and does not follow from previously published proofs of
threshold saturation. Also, the bounds on $w_{0}$ computed in these
examples are quite loose and shown only for completeness. There is
some evidence that the necessary $w_{0}$ for Theorem~\ref{thm:main_scalar_ub}
could be as small as $O\left(\ln\frac{1}{\Delta}\right)$ if $f,g$
are analytic~\cite{Hassani-jsm12}.
\end{rem}

\subsection{\label{sub:History} History and Motivation}

The history of spatial coupling and threshold saturation starts with
the introduction of LDPC convolutional codes by Felstrom and Zigangirov
in 1999 \cite{Felstrom-it99}. After a few years, it became apparent
that these codes had significantly better performance (i.e., noise
thresholds close to capacity) when terminated \cite{Lentmaier-isit05}.
Without termination, however, these codes perform very similarly to
standard LDPC codes. The next big advance came when researchers realized
\cite{Kudekar-it11,Lentmaier-it10} that the noise threshold of the
$(j,k)$-regular convolutional LDPC ensemble was very close to the
MAP threshold of the standard $(j,k)$-regular LDPC ensemble and then
were able to prove it \cite{Kudekar-it11}. It is worth noting that
this connection would not have been possible without the calculation
of the MAP thresholds, via (G)EXIT functions, for standard LDPC ensembles
\cite{Measson-it08,Measson-it09}.

Kudekar, Richardson, and Urbanke showed that this observation was
part of a very general phenomenon which they named threshold saturation
via spatial coupling \cite{Kudekar-it11}. The term spatial coupling
refers to the fact that LDPC convolutional codes can be seen as a
spatial arrangement of standard LDPC ensembles that are locally connected
to each other along the spatial dimension. Their proof shows that
the BP noise threshold of the coupled ensemble must equal the MAP-decoding
noise threshold of the underlying LDPC ensemble by using the coupled
BP EXIT function to construct a spatial integration of the standard
BP EXIT function. Their result represents a major breakthrough in
coding theory and their paper introduces many ideas and techniques
(e.g., the $(\texttt{l},\texttt{r},L,w)$ ensemble and the modified
DE recursion) used in this paper. Still, the result was proven only
for regular LDPC ensembles on the BEC and the proof technique does
not generalize easily to other cases. For example, their extension
to BMS channels is a tour de force in analysis \cite{Kudekar-it13}.
Hassani, Macris, and Urbanke have also presented results for problems
in statistical physics including a rigorous analysis of the coupled
Curie-Weiss model \cite{Hassani-itw10,Hassani-jsm12}. Since then,
there has been mounting evidence that the phenomenon of threshold
saturation is indeed very general%
\footnote{For example, see \cite{Hagiwara-isit11,Rathi-isit11,Yedla-isit11,Kudekar-isit11-DEC,Kudekar-isit11-MAC,Yedla-aller11,Uchikawa-isit11,Schlegel-isit11,Takeuchi-isit11,Sekido-sita11,Donoho-it13,Sakaniwa-ieice11,Uchikawa-icnc12,Aref-isit12,Aref-arxiv13,Truhachev-comlett12,Nguyen-icc12,Kumar-aller12,Obata-isit13,Piemontese-icc13}.%
}. 

In \cite{Takeuchi-ieice12}, Takeuchi, Tanaka, and Kawabata present
a phenomenological description of threshold saturation that is based
on the single-system potential function
\begin{equation}
\int_{0}^{x}\left(z-f(g(z))\right)\d z.\label{eq:Takeuchi_wrong_potential}
\end{equation}
Our initial goal was to construct a rigorous proof based on the outline
presented in \cite{Takeuchi-ieice12}. After some time, we concluded
that this is not possible without two significant modifications. First,
their uncoupled potential does not correctly predict the MAP threshold
for regular LDPC codes on the BEC. Therefore, we use instead%
\footnote{ This change is inconsequential if $g(x)$ is linear. %
} the uncoupled potential function \eqref{eq:uni_scalar_pot} whose
integral form is given by
\begin{equation}
\Us(x)=\int_{0}^{x}\left(z-f(g(z))\right)g'(z)\d z+\mathrm{const}.\label{eq:scalar_potential_integral}
\end{equation}
For the special case of irregular LDPC codes, $\Us(x)$ is also a
scalar multiple of the trial entropy $P_{\epsilon}\left(\texttt{x},\texttt{y}(\texttt{x})\right)$
in \cite[Lem.~4]{Measson-it08}, which defines the conjectured MAP
(or Maxwell) threshold of these codes on the BEC \cite{Measson-it08}.
Second, we were unable to rigorously connect the coupled recursion
\eqref{eq:coupled_scalar_recursion} to their proposed free energy
functional for the coupled system \cite{Takeuchi-ieice12}. During
this process, however, we realized that the coupled system also has
a simple closed-form potential function~\eqref{eq:coupled_scalar_potential_scalar}.
We observed later that, for the special case of the coupled Curie-Weiss
model, a similar potential function was defined earlier in \cite[(18)]{Hassani-jsm12}.
Still, \cite{Takeuchi-ieice12} introduced us to the idea of defining
a potential function for a discrete-time recursion as the integral
of the current value minus the updated value. 

In 2011, rigorous proofs based on coupled potential functions in the
continuum limit were posted to arXiv by Truhachev, for iterative demodulation
of packet-based CDMA with outer codes, and by Donoho, Montanari, and
Javanmard, for compressed sensing \cite{Truhachev-comlett12,Donoho-it13}.
Our conference papers on this subject were submitted a few months
later~\cite{Yedla-istc12,Yedla-itw12}.

This paper is based on two conference papers by the same authors~\cite{Yedla-istc12,Yedla-itw12}.
These papers use coupled-system potentials for the discrete system
along with a Taylor series expansion to prove threshold saturation
for both coupled scalar and coupled vector recursions. For some coding
problems, these results imply that the iterative-decoding noise threshold
of the coupled system equals the conjectured MAP-decoding noise threshold.
Independently in \cite{Kudekar-unpub12}, Kudekar, Richardson, and
Urbanke develop a different approach based on potential functions
that proves threshold saturation for general scalar recursions. Their
approach is somewhat more complicated but also produces stronger performance
guarantees and requires only that $f$ and $g$ are non-decreasing.
 In this paper, the techniques from \cite{Yedla-istc12,Yedla-itw12}
are extended%
\footnote{The paper \cite{Yedla-istc12} also contains a small error that can
be exploited by pathological examples (e.g., see Example~\ref{ex:pathological}).
This error is corrected in \cite{Yedla-arxiv12}. %
} to show the coupled recursion essentially converges to a point determined
by the minimum of the single-system potential (e.g., the point given
by the Maxwell construction in \cite{Measson-it08}). On the other
hand, threshold saturation is only informative if this point is 0.
The approach taken in \cite{Yedla-istc12,Yedla-itw12} was also extended
to irregular LDPC codes on BMS channels in \cite{Kumar-aller12,Kumar-itsub13}.
In fact, this generalization highlights both the simplicity and generality
of our approach. 

Some of the aforementioned proofs (e.g., \cite{Truhachev-comlett12,Donoho-it13,Yedla-istc12,Yedla-itw12})
can also be seen as using Lyapunov techniques to prove the stability
of discrete-time dynamical systems (e.g., the scalar and coupled recursions).
For the single-system potential of LDPC codes on the BEC (i.e., the
Bethe free energy), this connection was observed in 2009 by Vontobel
\cite{Vontobel-acorn09}. More recently, it was discussed%
\footnote{While the observation, in \cite{Schlegel-arxiv13}, that the potential
functions in \cite{Yedla-istc12,Yedla-itw12} can be seen as Lyapunov
functions is certainly correct, the cited preprint also contains a
number of errors and incorrect statements.%
} in \cite{Schlegel-arxiv13}. However, we note that something much
more interesting is at work here. The single-system potential function
that we use is something more than just a Lyapunov function for the
single system. It is the thermodynamic potential associated with some
implicit thermodynamic system and therefore the global minimum is
intrinsically connected to some notion of a minimum energy state.
This is exactly why we can establish the lower bound%
\footnote{For the scalar case, this lower bound also settles a question posed
in \cite{Schlegel-arxiv13} by showing that the potential threshold
gives an asymptotically tight characterization of the coupled system. %
} in Theorem~\ref{thm:main_scalar_lb}. While many functions can be
used as a Lyapunov function for the single system, most of them will
not determine the fixed point of the coupled system. For example,
the choice of Takeuchi et al. in \eqref{eq:Takeuchi_wrong_potential}
is a Lyapunov function for the single system whose minimum does not
correctly predict the fixed point of the coupled system.

\subsection{Some Important Details}

Consider the integral form \eqref{eq:scalar_potential_integral} of
the potential \eqref{eq:uni_scalar_pot}. Taking the derivative of
\eqref{eq:uni_scalar_pot} gives 
\begin{align*}
\Us'(x) & =xg'(x)+g(x)-g(x)-f(g(x))g'(x)\\
 & =\left(x-f(g(x)\right)g'(x)
\end{align*}
and shows that \eqref{eq:uni_scalar_pot} is consistent with the integral
form \eqref{eq:scalar_potential_integral}. Calculating $\Us(0)$
in both equations shows that the constant in \eqref{eq:scalar_potential_integral}
is given by $-F(g(0))$. In our previous work \cite{Yedla-istc12,Yedla-itw12},
this constant was 0 due to the assumption that $g(0)=0$. In this
work, this choice simplifies some parts of Section~\ref{sec:dependence}.
Moreover, we have the following proposition.
\begin{prop}
\label{prop:fp0_U0} If $0$ is a fixed point of the uncoupled recursion
(i.e., $f(g(0))=0$), then $F(g(0))=0$ and $\Us(0)=0$.\end{prop}
\begin{IEEEproof}
Under this condition, we find $F(g(0))\!=\!0$ because $F$ is non-negative
and 
\[
F(g(0))=\int_{0}^{g(0)}f(x)\d x\leq g(0)f(g(0))=0,
\]
where the inequality holds because $f(x)$ is non-decreasing. Since
\eqref{eq:uni_scalar_pot} implies $\Us(0)=-F(g(0))$, the result
follows.
\end{IEEEproof}
Another important detail has to do with the case of $\delta=0$ in
Theorem~\ref{thm:main_scalar_ub}. Under the following conditions,
Theorem~\ref{thm:main_scalar_ub} predicts Maxwell saturation with
$w_{0}<\infty$ even for the case of $\delta=0$.
\begin{prop}
\label{prop:finitew} In Theorem~\ref{thm:main_scalar_ub}, $w_{0}<\infty$
for $\delta=0$ if any of the following hold:
\begin{enumerate}
\item the fixed-point set $\mathcal{F}$ is finite,
\item $f(x)$ and $g(x)$ are real analytic functions on $\mathcal{X}$
and $x-f(g(x))$ is not identically zero,
\item there exists a $\gamma>0$ such that $f(g(x))<x$ for all $x\in(\overline{x}^{*},\overline{x}^{*}+\gamma]$,
or
\item $f'(g(\overline{x}^{*}))g'(\overline{x}^{*})<1$.
\end{enumerate}
\end{prop}
\begin{IEEEproof}
See Appendix \ref{sec:app_maxsat_proof}.
\end{IEEEproof}

\subsection{A Half-Iteration Shift}

One can also shift the uncoupled recursion by half an iteration by
swapping $f$ and $g$. In this case, the recursion of interest becomes
$y^{(\ell+1)}=g(f(y^{(\ell)}))$, starting from $y^{(0)}=g(\xmax)$,
and the associated uncoupled potential function becomes 
\begin{equation}
\Vs(y)\triangleq yf(y)-F(y)-G(f(y)).\label{eq:V_half_iter_pot}
\end{equation}
While the uncoupled recursion only experiences a time shift, the change
in the coupled recursion is more complicated because of the 0-boundary
condition. Still, the coupled system associated with the shifted recursion
behaves very similar to the original system due to the following proposition.
\begin{lem}
\label{prop:half_iter_shift} If $f$ and $g$ are both strictly increasing
$C^{2}$ functions, then
\begin{enumerate}
\item the fixed points satisfy $\mathcal{F}'\!=\! g(\mathcal{F})$ and $\mathcal{F}\!=\! f(\mathcal{F}')$
with 
\[
\mathcal{F}'=\left\{ y\in\mathcal{Y}\mid y=g(f(y))\right\} .
\]

\item if $x\in\mathcal{F}$ (resp. $y\in\mathcal{F}'$), then $\Us(x)=\Vs(g(x))$
(resp. $\Vs(y)=\Us(f(y)))$, and
\item the minimizers satisfy $\mathcal{M}'=g(\mathcal{M})$ and $\mathcal{M}=f(\mathcal{M}')$
with 
\[
\mathcal{M}=\argmin_{x\in\Xc}\Us(x),\quad\mathcal{M}'=\argmin_{y\in\Yc}\Vs(y).
\]

\end{enumerate}
\end{lem}
\begin{IEEEproof}
See Appendix \ref{sec:app_maxsat_proof}.\end{IEEEproof}
\begin{rem}
The above proposition implies that all quantities computed from $\Us(x)$
in Theorem~\ref{thm:main_scalar_ub} can alternatively be computed
from $\Vs(y)$ (e.g., if it has a simpler functional form). Moreover,
the coupled systems based on the original and shifted recursions receive
the same guarantees (after the appropriate variable change) from Theorem~\ref{thm:main_scalar_ub}.
\end{rem}

\subsection{The Coupled Potential Function}

First, we observe that, using the vector notation $\x^{(\ell)}=\big(x_{1}^{(\ell)},\ldots,x_{M}^{(\ell)}\big)^{\t}$,
the recursion \eqref{eq:coupled_scalar_recursion} can be written
compactly as
\begin{equation}
\x^{(\ell+1)}=\h\left(\x^{(\ell)}\right)\triangleq\A^{\t}\f\left(\A\g(\x^{(\ell)})\right),\label{eq:coupled_scalar_recursion_vector}
\end{equation}
where $\f:\Yc^{N}\to\Xc^{N}$ and $\g:\Xc^{M}\to\Yc^{M}$ are defined
by $\left[\f(\x)\right]_{i}=f(x_{i})$, and $\left[\g(\x)\right]_{i}=g(x_{i})$.
The recursion starts from $\x^{(0)}=\x_{\max}\triangleq\xmax\cdot\bm{1}_{M}$
and it is easy to verify that the vector mappings $\f,\g$ are isotone,
which means that $\x\preceq\z$ implies $\f(\x)\preceq\f(\z)$ and
$\g(\x)\preceq\g(\z)$. Moreover, the linear transformations defined
by $\A,\A^{\t}$ are also isotone and $\x\preceq\z$ implies $\A\x\preceq\A\z$.
The mapping $\h$ is also isotone because it is the composition of
four isotone mappings. Finally, we say a length-$M$ vector $\x$
is \emph{symmetric} if $\left[\x\right]_{i}=\left[\x\right]_{M-i+1}$
and a symmetric vector is \emph{unimodal} if $\left[\x\right]_{i+1}\geq\left[\x\right]_{i}$
for $i<\left\lceil M/2\right\rceil $. We will see that each $\x^{(\ell)}$
is symmetric and unimodal.

Next, we extend the definition of the potential function to general
coupled recursions of the form \eqref{eq:coupled_scalar_recursion_vector}.
\begin{defn}
\label{def:coupledpotential} Let the potential function of the coupled
system be defined by 
\begin{align}
\!\!\!\Uc(\x) & \triangleq\sum_{i=1}^{M}\left(g(x_{i})x_{i}-G(x_{i})\right)-\!\sum_{i=1}^{N}F\!\left(\sum_{j=1}^{M}A_{i,j}g(x_{j})\right)\!\label{eq:coupled_scalar_potential_scalar}\\
 & =\g(\x)^{\t}\x-G(\x)-F(\A\g(\x)),\label{eq:coupled_scalar_potential_vector}
\end{align}
where $G(\x)=\sum_{i=1}^{M}G(x_{i})$ and $F(\y)=\sum_{i=1}^{N}F(y_{i})$.
One can verify that this is equivalent to 
\[
\Uc(\x)=\int_{\mathcal{C}}\g'(\z)(\z-\A^{\t}\f(\A\g(\z)))\cdot\mathrm{d}\z-F(\A\g(\bm{0})),
\]
where $\mathcal{C}$ is a smooth curve in $\Xc^{M}$ from $\bm{0}$
to $\x$, $G(\x)=\int_{\mathcal{C}}\g(\z)\cdot\mathrm{d}\z$, $F(\y)=\int_{\mathcal{C}'}\f(\z)\cdot\mathrm{d}\z$,
and $\mathcal{C}'$ is a smooth curve in $\Xc^{N}$ from $\bm{0}$
to $\bm{y}$. We note that the constant term in the integral $\Uc(\x)$
formula is chosen to be consistent with the scalar potential. \end{defn}
\begin{rem}
An important property of the potential function $\Us(x)$ is that
its derivative is closely related to the step taken by the recursion,
$x-f(g(x))$. A similar result holds for the coupled potential $\Uc(\x)$
and computing the derivative of \eqref{eq:coupled_scalar_potential_scalar}
shows that
\begin{align*}
\frac{\mathrm{d}}{\mathrm{d}x_{k}}\Uc(\x) & =\frac{\mathrm{d}}{\mathrm{d}x_{k}}\sum_{i=1}^{M}\left(g(x_{i})x_{i}-G(x_{i})\right)\\
 & \quad-\frac{\mathrm{d}}{\mathrm{d}x_{k}}\sum_{i=1}^{N}F\!\left(\sum_{j=1}^{M}A_{i,j}g(x_{j})\right)\\
 & =g'(x_{k})x_{k}+g(x_{k})-g(x_{k})\\
 & \quad-\sum_{i=1}^{N}f\!\left(\sum_{j=1}^{M}A_{i,j}g(x_{j})\right)\sum_{j=1}^{M}A_{i,j}\frac{\mathrm{d}}{\mathrm{d}x_{k}}g(x_{j})\\
 & =\left(x_{k}-\sum_{i=1}^{N}A_{i,k}f\!\left(\sum_{j=1}^{M}A_{i,j}g(x_{j})\right)\right)g'(x_{k})\\
 & =\left[\g'(\x)\left(\x-\A^{\t}\f(\A\g(\x))\right)\right]_{k}.
\end{align*}
A key observation in this work is that a potential function for general
coupled systems can be written in the simple form given by Def.~\ref{def:coupledpotential}.
Remarkably, this holds for general coupling coefficients because of
the reciprocity between $\A$ and $\A^{\t}$ that appears naturally
in spatial coupling.
\end{rem}

\begin{rem}
It is natural to wonder why the integral form of the single-system
potential needs a factor of $g'(x)$. Focusing instead on the coupled
potential function, one notices that the vector field $\x-\A^{\t}\f(\A\g(\x))$
is not conservative in general. Thus, there is no scalar function
whose gradient equals $\x-\A^{\t}\f(\A\g(\x))$. But, if one multiplies
by $\g'(\x)$ on the left, then $\g'(\x)\left(\x-\A^{\t}\f(\A\g(\x))\right)$
is conservative for all $\A,\f,\g$ (e.g., it is the gradient of the
coupled potential). Hence, we need the factor of $\g'(\x)$ in the
derivative of the coupled potential just to make it the gradient of
a scalar function and we need the $g'(x)$ in the derivative of the
single-system potential so that it matches the coupled potential.
\end{rem}

\subsection{A Convenient Change of Variables}

Now, we introduce a change of variables for the uncoupled recursion
that allows one to translate any fixed point to $x=0$. This result
is used in the proof of Theorem~\ref{thm:main_scalar_ub}.

For any fixed point $\xf\in\mathcal{F}$ of the uncoupled recursion,
the functions $\sfs(y)\triangleq f(y+g(\xf))-\xf$ and $\sgs(x)\triangleq g(x+\xf)-g(\xf)$
define a \emph{translated uncoupled recursion} that satisfies 
\[
\sfs(\sgs(x))=f(g(x+\xf))-\xf.
\]
This translation is given by the change of variable $x\mapsto x-\xf$
and, hence, the recursion operates on the space $\sXc\triangleq[0,\sxmax]$,
where $\sxmax\triangleq\xmax-\xf$. For the uncoupled system, this
change of variables has no effect other than mapping the value $\xf$
in the original uncoupled recursion to the value $0$ in the translated
uncoupled recursion (e.g., $\sfs(0)=\sgs(0)=0)$. For the translated
coupled system, $\sfc$ and $\sgc$ are defined by the pointwise application
of $\sfs$ and $\sgs$ to vectors of length $N$ and $M$. This implies
that $\sfc(\y)=\f(\y+g(\xf)\bm{1}_{N})-\xf\bm{1}_{N}$ and $\sgc(\x)=\g(\x+\xf\bm{1}_{M})-g(\xf)\bm{1}_{M}$.
It is important to note that the translation also affects implicitly
the boundary conditions introduced by coupling.
\begin{lem}
\label{lem:translated_recursion} The translated scalar system is
well defined and the overall update of the translated coupled system,
$\shc$, is 
\[
\shc(\x)=\A^{\t}\f\left(\A\g(\x\!+\!\xf\bm{1}_{M})\right)\!-\!\xf\bm{1}_{M}\!+\!\xf\left(\bm{1}_{M}\!-\!\A^{\t}\bm{1}_{N}\right).
\]
This differs from $\h(\x+\xf\bm{1}_{M})-\xf\bm{1}_{M}$ (i.e., translating
the coupled recursion) by the boundary condition term $\xf(\bm{1}_{M}-\A^{\t}\bm{1}_{N})$.
The potentials for the translated recursion are given by
\begin{enumerate}
\item $\sFs(y)=F(y+g(\xf))-y\xf-F(g(\xf))$
\item $\sGs(x)=G(x+\xf)-xg(\xf)-G(\xf)$
\item $\sUs(x)=x\sgs(x)-\sGs(x)-\sFs(\sgs(x))=\Us(x+\xf)-\Us(\xf)$ 
\end{enumerate}
\end{lem}
\begin{IEEEproof}
See Appendix \ref{sec:app_maxsat_proof}.\end{IEEEproof}
\begin{rem}
The term $\xf(\bm{1}_{M}-\A^{\t}\bm{1}_{N})$ in Lemma~\ref{lem:translated_recursion}
effectively changes the boundary values from 0 in original coupled
recursion to $\xf$ in the translated coupled recursion. However,
this is done by shifting the recursion down and keeping the same boundary
value at 0 rather than keeping the same recursion and changing the
boundary value. Therefore, if the translated coupled recursion converges
to the zero vector, then the original coupled recursion (with a modified
boundary value of $\xf$) converges to the all-$\xf$ vector.
\end{rem}

\section{Proof of Theorem~\ref{thm:main_scalar_ub}}

\label{sub:proof_ub_theorem}

Before we present the proof of Theorem~\ref{thm:main_scalar_ub},
we introduce some necessary definitions and lemmas. Proofs are relegated
to Appendix~\ref{sec:app_ub_proof} unless they reveal ideas central
to our approach.
\begin{lem}
\label{lem:scalar_potential_minima_fp} The potential value is non-increasing
with iteration (i.e., $\Us\left(f(g(x))\right)\leq\Us(x)$) and strictly
decreasing if $x$ is not a fixed point (i.e., $f(g(x))\neq x$).
Furthermore, all local minima of $\Us(x)$ occur at fixed points.\end{lem}
\begin{IEEEproof}
See Appendix \ref{sec:app_ub_proof}.\end{IEEEproof}
\begin{rem}
It turns out that the second part of Lemma~\ref{lem:scalar_potential_minima_fp}
depends crucially on the assumption that $g(x)$ is strictly increasing.
If $g(x)$ is allowed to be constant on an interval, then one can
construct recursions where a minimizer of the potential is not a fixed
point.
\end{rem}
Throughout, we will use $i_{0}\triangleq\left\lceil M/2\right\rceil $
to denote the spatial midpoint of the coupled system. The following
lemma shows that $\x^{(\ell)}$ is a non-increasing sequence of symmetric
unimodal vectors.
\begin{lem}
\label{lem:coupled_scalar_dec_sym} The coupled recursion \eqref{eq:coupled_scalar_recursion}
satisfies: (i) $\x{}^{(\ell+1)}\preceq\x{}^{(\ell)}$, (ii) $x_{i}^{(\ell)}=x_{M-i+1}^{(\ell)}$
and (iii) $x_{i+1}^{(\ell)}\geq x_{i}^{(\ell)}$ for $i<i_{0}$.\end{lem}
\begin{IEEEproof}
See Appendix \ref{sec:app_ub_proof}.\end{IEEEproof}
\begin{defn}
\label{def:mod_coupled_recursion} The \emph{modified coupled recursion}
is defined to be $\hat{\x}^{(\ell+1)}=\q\left(\h\left(\hat{\x}^{(\ell)}\right)\right)$,
starting from $\hat{\x}^{(0)}=\x_{\max}$, where
\[
\left[\q(\x)\right]_{i}\triangleq\begin{cases}
\left[\x\right]_{i_{0}} & \mathrm{if}\; i\in\ir{\ti+1}M\\
\left[\x\right]_{i} & \mathrm{otherwise}.
\end{cases}
\]
\end{defn}
\begin{rem}
The modified coupled recursion provides an upper bound on the coupled
recursion that has a few important properties which will be used in
the proof of Theorem~\ref{thm:main_scalar_ub}.\end{rem}
\begin{lem}
\label{lem:mod_coupled_scalar_dec_inc} The sequence generated by
the modified coupled recursion satisfies: \emph{(i)} $\hat{\x}^{(\ell+1)}\preceq\hat{\x}^{(\ell)}$,
\emph{(ii)} $\hat{\x}^{(\ell)}\succeq\x^{(\ell)}$, and \emph{(iii)}
$\hat{\x}^{(\ell)}$ satisfies $\left[\hat{\x}\right]_{i+1}\geq\left[\hat{\x}\right]_{i}$
for $i\in\ir 1{M-1}$.\end{lem}
\begin{IEEEproof}
See Appendix \ref{sec:app_ub_proof}.\end{IEEEproof}
\begin{lem}
\label{lem:coupled_scalar_shift_down} Consider the coupled system
potential $\Uc(\x)$ for any $\x\in\Xc^{M}$ satisfying $x_{i}=x_{i_{0}}$
for $i\in\ir{\ti}M$ and $x_{i}\leq x_{\ti}$ for $i\in\ir 1M$. Let
the shift operator $\SS$ be defined by $\left[\SS\x\right]_{i}=x_{i-1}$
for $i\in\ir 1M$ with $x_{0}=0$. Then, applying the shift to $\x$
changes the potential by $\Uc(\SS\x)-\Uc(\x)\leq\Us(0)-\Us(x_{i_{0}})$. \end{lem}
\begin{IEEEproof}
We compute $\Uc(\SS\x;\epsilon)-\Uc(\x;\epsilon)$ by treating the
three terms in~\eqref{eq:coupled_scalar_potential_vector} separately.
The first term gives 
\begin{align*}
\g(\SS\x)^{\t}(\SS\x)-\g(\x)^{\t}\x & =\!\sum_{i=0}^{M-1}\! g(x_{i})x_{i}-\!\sum_{i=1}^{M}\! g(x_{i})x_{i}\\
 & =g(x_{0})x_{0}-g(x_{M})x_{M}\\
 & =-g(x_{i_{0}})x_{i_{0}}.
\end{align*}
Similarly, the second term gives 
\begin{align*}
-G(\SS\x)+G(\x) & =-\sum_{i=0}^{M-1}G(x_{i})+\sum_{i=0}^{M}G(x_{i})\\
 & =-G(x_{0})+G(x_{M})\\
 & =G(x_{\ti}).
\end{align*}
For the third term, we consider $-F\left(\A\g(\SS\x)\right)=-\sum_{i=1}^{N}F\left([\A\g(\SS\x)]_{i}\right)$
and observe that $\left[\g(\SS\x)\right]_{1}=g(0)$ and $\left[\g(\SS\x)\right]_{i}=\left[\SS\g(\x)\right]_{i}$
for $i\in\ir 2N$. Since $\g$ is monotonic, this implies that $-F\left([\A\g(\SS\x)]_{1}\right)\leq-F\left(g(0)\right)$
and $-F\left([\A\g(\SS\x)]_{i}\right)=-F\left([\A\g(\x)]_{i-1}\right)$
for $i\in\ir 2N$. Therefore, we have the upper bound
\begin{align*}
-F & \left(\A\g(\SS\x)\right)+F\left(\A\g(\x)\right)\\
 & \leq-F\left(g(0)\right)-\sum_{i=2}^{N}F\left([\A\g(\x)]_{i-1}\right)+\sum_{i=1}^{N}F\left([\A\g(\x)]_{i}\right)\\
 & =-F\left(g(0)\right)\!+\! F\left([\A\g(\x)]_{N}\right)\leq-F\left(g(0)\right)\!+\! F\left(g(x_{\ti})\right),
\end{align*}
where $[\A\g(\x)]_{N}\leq x_{\ti}$ holds in the last step because
$x_{i}\leq x_{\ti}$ for $i\in\ir 1M$. Finally, summing these bounds
gives the stated result
\begin{align*}
\Uc(\SS\x)\,-\, & \Uc(\x)\leq-g(x_{\ti})x_{\ti}+G(x_{\ti})\\
 & -F\left(g(0)\right)+F\left(g(x_{\ti})\right)=\Us(0)-\Us(x_{\ti}).
\end{align*}
\end{IEEEproof}
\begin{lem}
\label{lem:modified_scalar_fp_dir_deriv} Let $\hat{\x}=\hat{\x}^{(\infty)}$
be the limit of the modified coupled recursion in Def.~\ref{def:mod_coupled_recursion}
and let $\SS$ be the shift operator defined in Lemma~\ref{lem:coupled_scalar_shift_down}.
 Then, the directional derivative of the coupled potential in the
direction $\SS\hat{\x}-\hat{\x}$ satisfies 
\[
\Uc'(\hat{\x})^{\t}(\SS\hat{\x}-\hat{\x})=0.
\]
\end{lem}
\begin{IEEEproof}
By Lemma~\ref{lem:mod_coupled_scalar_dec_inc}, the modified coupled
recursion generates a decreasing sequence that converges to a limit,
which is a fixed point because the update function is continuous.
From the definition of $\q(\cdot)$ and the fact that $\hat{\x}$
is non-decreasing, it follows that $\left[\hat{\x}\right]_{i}=\left[\h\left(\hat{\x}\right)\right]_{i}$
for $i\in[1:i_{0}]$, and $\left[\hat{\x}\right]_{i}=\hat{x}_{i_{0}}$
for $i\in\ir{\ti+1}M$. Since the last half of $\hat{\x}$ is constant,
we find that $\left[\SS\hat{\x}-\hat{\x}\right]_{i}=0$ for $i\in\ir{\ti+1}M$.
For the first half of the vector, the recursion is unaffected by $\q(\cdot)$
and this implies that $\left[\Uc'(\hat{\x})\right]_{i}=\left[\g'(\hat{\x})\left(\hat{\x}-\h(\hat{\x})\right)\right]_{i}=0$
for $i\in\ir 1{\ti}$. Therefore, we have $\Uc'(\hat{\x})^{\t}(\SS\hat{\x}-\hat{\x})=0$
because, in each position, one of the two vectors is zero.\end{IEEEproof}
\begin{lem}
\label{lem:coupled_scalar_hessian_bound} The Hessian matrix $\Uc''(\x)$
of $\Uc(\x)$ satisfies $\left\Vert \Uc''(\x)\right\Vert _{p}\leq K_{f,g}$,
where 
\begin{equation}
K_{f,g}\triangleq\left\Vert g''\right\Vert _{\infty}\xmax+\left\Vert g'\right\Vert _{\infty}+\left\Vert f'\right\Vert _{\infty}\left\Vert g'\right\Vert _{\infty}^{2}\label{eq:coupled_scalar_hessian_bound}
\end{equation}
for all $\x\in\Xc^{M}$ and $p\in\left\{ 1,2,\infty\right\} $.\end{lem}
\begin{IEEEproof}
See Appendix \ref{sec:app_ub_proof}.\end{IEEEproof}
\begin{lem}
\label{lem:Sx_minus_x} Let $\x=\q(\A^{\t}\z)$ for some $\z\in\Xc^{N}$
satisfying $z_{i+1}\geq z_{i}$ for $i\in\ir 1{M-1}$. Let the down-shift
operator $\SS:\Xc^{M}\to\Xc^{M}$ be defined by $\left[\SS\x\right]_{i}=x_{i-1}$
for $i\in\ir 1M$ with $x_{0}=0$. Then, we have the bounds $\norm{\SS\x-\x}_{\infty}\leq\frac{1}{w}z_{\ti}\leq\frac{1}{w}\xmax$
and $\norm{\SS\x-\x}_{1}=x_{\ti}-x_{0}\leq\xmax$. \end{lem}
\begin{IEEEproof}
For $i>\ti$, $x_{i}=x_{i-1}$ due to $\q(\cdot)$ and $\left|x_{i}-x_{i-1}\right|=0$.
For $i\leq\ti$, we have 
\begin{align*}
\left|x_{i}-x_{i-1}\right| & =\left|\tfrac{1}{w}\sum\nolimits _{k=0}^{w-1}z_{i-k}-\tfrac{1}{w}\sum\nolimits _{k=0}^{w-1}z_{i-k-1}\right|\\
 & \leq\tfrac{1}{w}z_{\ti}\leq\tfrac{1}{w}\xmax,
\end{align*}
where we assume $z_{i}=0$ for $i\notin\ir 1N$. Hence, $\norm{\SS\x-\x}_{\infty}\leq\frac{1}{w}z_{\ti}\leq\frac{1}{w}\xmax$.
By Lemma~\ref{lem:mod_coupled_scalar_dec_inc}(iii), $\x$ is non-decreasing
because $\z$ is non-decreasing. Therefore, we find that $\left[\SS\x-\x\right]_{i}=x_{i-1}-x_{i}\leq0$
and the $1$-norm sum telescopes to give 
\begin{align*}
\left\Vert \SS\x-\x\right\Vert _{1} & =\sum_{i=1}^{M}\left|x_{i-1}-x_{i}\right|=\sum_{i=1}^{\ti}\left(x_{i}-x_{i-1}\right)\\
 & =x_{\ti}-x_{0}\leq\xmax.
\end{align*}
\end{IEEEproof}
\begin{rem}
It is natural to wonder, ``Where in the proof of Theorem~\ref{thm:main_scalar_ub}
is the $0$-boundary of the SC system used?''. The best answer to
this question is probably Lemma~\ref{lem:Sx_minus_x}. This is
because the bound $\norm{\SS\x-\x}_{\infty}\leq\frac{1}{w}\xmax$
depends crucially on the fact that $\left|x_{0}-x_{1}\right|=\left|x_{1}\right|\leq\frac{1}{w}\xmax$
and this holds only because the $0$-boundary is implicit in the equation
$x_{1}=\left[\A^{\t}\z\right]_{1}=\frac{1}{w}z_{1}$.
\end{rem}
Now, we are ready to combine the previous results into a single lemma.
\begin{lem}
\label{lem:fg0_saturation} Consider an uncoupled recursion satisfying
$f(0)=g(0)=0$. For any $\delta\geq0$, if $\Delta_{0}\triangleq\inf\{\Us(x)\mid x\in\mathcal{F}\cap(\delta,\xmax]\}>0$
and $w>\frac{1}{2\Delta_{0}}K_{f,g}\xmax^{2}$ (where $K_{f,g}$ is
defined in \eqref{eq:coupled_scalar_hessian_bound}), then the fixed
point of the coupled recursion must satisfy $x_{i}^{(\infty)}\leq\delta$
for $i\in\ir 1M$.\end{lem}
\begin{IEEEproof}
First, we use Lemma~\ref{lem:mod_coupled_scalar_dec_inc} to see
that the modified coupled recursion converges to a fixed-point limit
denoted by $\hat{\x}$. If $\hat{x}_{i_{0}}\leq\delta$, then the
proof is complete because $\hat{x}_{i_{0}}$ is the maximum value
of $\hat{\x}$. Hence, we can assume $\delta\in[0,\xmax)$. Now, suppose
that $\hat{x}_{i_{0}}>\delta$. The remainder of the proof hinges
on calculating $\Uc(\SS\hat{\x})-\Uc(\hat{\x})$ using two different
approaches. 

The first approach is by applying Lemma~\ref{lem:coupled_scalar_shift_down}
and observing that
\[
\Uc(\SS\hat{\x})-\Uc(\hat{\x})\leq\Us(0)-\Us(\hat{x}_{i_{0}})=-\Us(\hat{x}_{i_{0}})
\]
because $f(0)=g(0)=0$ implies $\Us(0)=0$. In addition, we observe
that
\[
\hat{x}_{i_{0}}=\left[\q\left(\h\left(\hat{\x}\right)\right)\right]_{i_{0}}=\left[\h\left(\hat{\x}\right)\right]_{i_{0}}\leq\left[\h\left(\hat{x}_{i_{0}}\bm{1}\right)\right]_{i_{0}}\leq h\left(\hat{x}_{i_{0}}\right)
\]
holds because $\hat{\x}$ is a fixed point of $\q\left(\h\left(\cdot\right)\right)$
and $\hat{\x}\preceq\hat{x}_{i_{0}}\cdot\bm{1}$. This implies that
the recursion $z^{(\ell+1)}=f(g(z^{(\ell)}))$ from $z^{(0)}=\hat{x}_{i_{0}}$
satisfies $z^{(\ell+1)}\geq z^{(\ell)}$ and $\Us(z^{(\ell+1)})\leq\Us(z^{(\ell)})$
(by Lemma~\ref{lem:scalar_potential_minima_fp}). Therefore, $z^{(\ell)}$
increases to a stable fixed point $z^{(\infty)}\geq\hat{x}_{i_{0}}>\delta$
that satisfies $\Us(\hat{x}_{i_{0}})\geq\Us(z^{(\infty)})$ and
\[
\Us(z^{(\infty)})\geq\inf\left\{ \Us(x)\mid x\in(\delta,\xmax],\, x=f(g(x))\right\} .
\]
 Thus, $-\Us(\hat{x}_{i_{0}})\leq-\Delta_{0}$ and, if $\Delta_{0}>0$,
then this implies that $\left|\Uc(\SS\hat{\x})-\Uc(\hat{\x})\right|\geq\Delta_{0}$.

For the second approach, we evaluate $\Uc(\SS\hat{\x})$ using a 2nd-order
Taylor series expansion.  Let $\hat{\z}(t)=\hat{\x}+t(\SS\hat{\x}-\hat{\x})$
and $\phi(t)=\Uc(\hat{\z}(t))$. Then, Taylor's theorem says that,
for some $t_{0}\in[0,1]$, one has
\[
\phi(1)=\phi(0)+\phi'(0)+\frac{1}{2}\phi''(t_{0}).
\]
Using the chain rule to calculate the derivatives shows that
\begin{align*}
\Uc(\SS\hat{\x})-\Uc(\hat{\x}) & =\Uc'(\hat{\x})^{\t}(\SS\hat{\x}-\hat{\x})\\
 & \quad+\frac{1}{2}\left(\SS\hat{\x}-\hat{\x}\right)^{\t}\Uc''(\hat{\z}(t_{0}))\left(\SS\hat{\x}-\hat{\x}\right).
\end{align*}
Using Lemmas~\ref{lem:modified_scalar_fp_dir_deriv}~and~\ref{lem:coupled_scalar_hessian_bound},
this can be rewritten as
\begin{align*}
\Big|\Uc(\SS\hat{\x})- & \Uc(\hat{\x})\Big|=\left|\frac{1}{2}\left(\SS\hat{\x}-\hat{\x}\right)^{\t}\Uc''(\hat{\z}(t_{0}))\left(\SS\hat{\x}-\hat{\x}\right)\right|\\
 & \leq\frac{1}{2}\left\Vert \SS\hat{\x}-\hat{\x}\right\Vert _{1}\left\Vert \Uc''(\hat{\z}(t_{0}))\left(\SS\hat{\x}-\hat{\x}\right)\right\Vert _{\infty}\\
 & \leq\frac{1}{2}\left\Vert \SS\hat{\x}-\hat{\x}\right\Vert _{1}\left\Vert \Uc''(\hat{\z}(t_{0}))\right\Vert _{\infty}\left\Vert \SS\hat{\x}-\hat{\x}\right\Vert _{\infty}\\
 & \le\frac{1}{2}K_{f,g}\left\Vert \SS\hat{\x}-\hat{\x}\right\Vert _{1}\left\Vert \SS\hat{\x}-\hat{\x}\right\Vert _{\infty}\\
 & \le\frac{1}{2w}K_{f,g}\xmax^{2},
\end{align*}
where the norm bounds in the last step are computed in Lemma~\ref{lem:Sx_minus_x}.
This bound contradicts the result of the first approach if
\begin{equation}
w>\frac{1}{2\Delta_{0}}K_{f,g}\xmax^{2}.\label{eq:fg0_w_bound}
\end{equation}
Based on this contradiction, we conclude that~\eqref{eq:fg0_w_bound}
implies $\hat{x}_{i_{0}}\leq\delta$.
\end{IEEEproof}
Now, we are ready to prove Theorem~\ref{thm:main_scalar_ub}. This
theorem says that, for any $\delta>0$, there is $w_{0}<\infty$ such
that, for all $w>w_{0}$ and all $N\in [1\mathbin{:}\infty]$, we have $\max_{i\in[1:M]}x_{i}^{(\infty)}-\delta\leq\overline{x}^{*}$,
where $\overline{x}^{*}\triangleq\max\left(\argmin_{x\in\Xc}\Us(x)\right)$.
The primary  tools are the change of variables defined in Lemma~\ref{lem:translated_recursion}
and threshold saturation result in Lemma~\ref{lem:fg0_saturation}. 
\begin{IEEEproof}
[Proof of Theorem~\ref{thm:main_scalar_ub}] This theorem now follows
from combining Lemmas~\ref{lem:fg0_saturation} and~\ref{lem:translated_recursion}.
The idea is to first translate the uncoupled recursion by applying
Lemma~\ref{lem:translated_recursion} with $\xf=\overline{x}^{*}$
and then apply Lemma~\ref{lem:fg0_saturation} to show that coupled
translated recursion converges to the zero vector. 

First, we handle some special cases. If the uncoupled recursion has
no fixed points in $[\overline{x}^{*}+\delta,\xmax]$, then the promised
upper bound follows trivially because the coupled system is upper
bounded by the uncoupled system (i.e., $x_{i}^{(\infty)}\leq x^{(\infty)}$
for $i\in[1:M]$) and the uncoupled system satisfies $x^{(\infty)}\leq\overline{x}^{*}+\delta$.
Therefore, we consider only the case where the uncoupled system has
a fixed point in $[\overline{x}^{*}+\delta,\xmax]$ and, hence,  $\delta\in[0,\xmax-\overline{x}^{*}]$.

Next, we define $w_{0}\triangleq\frac{1}{2\Delta}K_{f,g}\xmax^{2},$
where $K_{f,g}$ is defined in \eqref{eq:coupled_scalar_hessian_bound},
and 
\[
\Delta\triangleq\inf\left\{ \Us(x)-\Us(\overline{x}^{*})\,|\, x\in\mathcal{F}\cap(\overline{x}^{*}+\delta,\xmax]\right\} .
\]
To show that $w_{0}<\infty$, we will show that $\Delta>0$ for any
$\delta\in(0,\xmax-\overline{x}^{*})$. Observe that the set of fixed
points $\mathcal{F}=\{x\in\Xc\mid x=f(g(x))\}$ is compact because
it a closed subset (i.e., it contains all its limit points because
it is defined by a continuous equality) of a compact set. From this,
we see that the set $\mathcal{F}\cap[\overline{x}^{*}+\delta,\xmax]$
is compact (i.e., it is intersection of compact sets) and non-empty
(i.e., there is a fixed point in $[\overline{x}^{*}+\delta,\xmax]$).
Since $\overline{x}^{*}$ is the largest value that minimizes $\Us(x)$,
it follows that $\Us(x)>\Us(\overline{x}^{*})$ for all $x\in(\overline{x}^{*},\xmax]$.
Hence, for any $\delta\in(0,\xmax-\overline{x}^{*})$, we have
\begin{align*}
\Delta & \geq\min\left\{ \Us(x)-\Us(\overline{x}^{*})\,|\, x\in\mathcal{F\cap}[\overline{x}^{*}\!+\!\delta,\xmax]\right\} \\
 & >0.
\end{align*}
Except for pathological cases (e.g., where $\overline{x}^{*}$ is
a limit point of $\mathcal{F}\cap[\overline{x}^{*},\xmax]$), the
infimum must equal the minimum. This implies that, if $\overline{x}^{*}$
is not a limit point of $\mathcal{F}\cap[\overline{x}^{*},\xmax]$,
then $\Delta>0$ and $w_{0}<\infty$ for all $\delta\in[0,\xmax-\overline{x}^{*})$. 

Now, we focus on the translated uncoupled recursion defined by Lemma~\ref{lem:translated_recursion}
with $\xf=\overline{x}^{*}$. For the translated system, one finds
that $\sfs(0)=\sgs(0)=0$, $\sxmax=\xmax-\overline{x}^{*}$, and $\sUs(x)=\Us(x+\overline{x}^{*})-\Us(\overline{x}^{*})$
for all $x\in(0,\sxmax]$. To apply Lemma~\ref{lem:translated_recursion}
to the translated system, we start by computing
\begin{align*}
\!\Delta_{0} & =\inf\left\{ \sUs(x)\,|\, x\in(\delta,\sxmax],x\!=\!\sfs(\sgs(x))\right\} \\
 & =\inf\left\{ \Us(x\!+\!\overline{x}^{*})\!-\!\Us(\overline{x}^{*})\,|\, x\in(\delta,\sxmax],x\!=\!\sfs(\sgs(x))\right\} \\
 & =\inf\left\{ \Us(x)\!-\!\Us(\overline{x}^{*})\,|\, x\in(\overline{x}^{*}\!+\!\delta,\xmax],x\!=\! f(g(x))\right\} \\
 & =\Delta>0.
\end{align*}
Therefore, Lemma~\ref{lem:fg0_saturation} implies that the fixed
point of the translated coupled system, $\tilde{x}_{i}^{(\infty)}$,
must satisfy $\tilde{x}_{i}^{(\infty)}\leq\delta$ for $i\in\ir 1M$.
The proof is completed by noting that the fixed point of the original
coupled system, $x_{i}^{(\infty)}$, is upper bounded by $\tilde{x}_{i}^{(\infty)}+\overline{x}^{*}$
because the two coupled recursions are identical except for the translation
and the fact that the translated system uses the larger boundary value
$\overline{x}^{*}\geq0$.
\end{IEEEproof}

\section{Proof of Theorem~\ref{thm:main_scalar_lb}}

\label{sec:proof_lb_theorem}

Before we present the proof of Theorem~\ref{thm:main_scalar_lb},
we introduce the following lemma.
\begin{lem}
\label{lem:U_vec_bound} The functions $\Us(x)$ and $\Uc(\x)$ satisfy
\begin{enumerate}
\item $\Us(x)>\Us(\underline{x}^{*})$ for all $x\in[0,\underline{x}^{*})$
\item If $\x\!\succeq\!\A^{\t}\f(\A\g(\x))$, then $\Uc(\A^{\t}\f(\A\g(\x)))\!\leq\!\Uc(\x)$
\item For an arbitrary vector $\x$, we have 
\[
\Uc(\x)\geq\sum_{i=1}^{M}\Us(x_{i})
\]
 
\item For a constant vector $\x=(x,\ldots,x)^{\t}$, we have 
\[
\Uc(\x)=M\,\Us(x)+(w-1)F(g(x))
\]

\end{enumerate}
\end{lem}
\begin{IEEEproof}
For (i), we observe that $\underline{x}^{*}$ is precisely the smallest
value of $x$ that achieves the minimum value of $\Us(x)$. It follows
easily that $\Us(x)>\Us(\underline{x}^{*})$ for $x\in[0,\underline{x}^{*})$.
For (ii), we let $\z(t)=\x+t(\A^{\t}\f(\A\g(\x))-\x)$ and observe
that $\x\succeq\z(t)\succeq\A^{\t}\f(\A\g(\x))$ holds because $\z(t)$
is a convex combination of the endpoints $\x\succeq\A^{\t}\f(\A\g(\x))$.
Also, $\A^{\t}\f(\A\g(\x))\succeq\A^{\t}\f(\A\g(\z(t)))$ is implied
by $\x\succeq\z(t)$ because $\h(\x)=\A^{\t}\f(\A\g(\x))$ is isotone.
Next, we use~\eqref{eq:coupled_scalar_potential_vector} to write
$B=\Uc\left(\A^{\t}\f(\A\g(\x))\right)-\Uc(\x)$ as
\[
B=\int_{0}^{1}\g'(\z(t))\underbrace{\left(\z(t)-\A^{\t}\f(\A\g(\z(t)))\right)}_{\succeq\bm{0}}\cdot\underbrace{\z'(t)}_{\preceq\bm{0}}\d t.
\]
Since $\g'(\z(t))$ is an $M\times M$ non-negative matrix (e.g.,
$g$ is non-decreasing), it follows that $B\leq0$.

For (iii), consider any vector $\x$ and observe that
\begin{align*}
\Uc(\x) & =\x^{T}\g(\x)-G(\x)-F(\A\g(\x))\\
 & =\sum_{i=1}^{M}x_{i}g(x_{i})-\sum_{i=1}^{M}G(x_{i})-\sum_{j=1}^{N}F\left(\sum_{i=1}^{M}A_{j,i}g(x_{i})\right)\\
 & \stackrel{(a)}{\geq}\sum_{i=1}^{M}x_{i}g(x_{i})-\sum_{i=1}^{M}G(x_{i})-\sum_{i=1}^{M}\sum_{j=1}^{N}A_{j,i}F(g(x_{i}))\\
 & \stackrel{(b)}{\geq}\sum_{i=1}^{M}x_{i}g(x_{i})-\sum_{i=1}^{M}G(x_{i})-\sum_{i=1}^{M}F(g(x_{i}))\\
 & =\sum_{i=1}^{M}\Us(x_{i}),
\end{align*}
where $(a)$ $F\left(\sum_{i=1}^{M}A_{j,i}g(x_{i})\right)\leq\sum_{i=1}^{M}A_{j,i}F(g(x_{i}))$
holds because $F$ is convex and $\sum_{i=1}^{M}A_{j,i}=1$ and $(b)$
follows from $\sum_{j=1}^{N}A_{j,i}\leq1$.

For (iv), $\x=(x,\ldots,x)^{\t}$ is a constant vector and we have
\begin{align*}
\Uc(\x) & =\x^{T}\g(\x)-G(\x)-F(\A\g(\x))\\
 & =\sum_{i=1}^{M}xg(x)-\sum_{i=1}^{M}G(x)-\sum_{j=1}^{N}F\left(\sum_{i=1}^{M}A_{j,i}g(x_{i})\right)\\
 & =\sum_{i=1}^{M}xg(x)-\sum_{i=1}^{M}G(x)-\sum_{i=1}^{N}F(g(x))\\
 & =M\,\Us(x)+(M-N)F(g(x))\\
 & =M\,\Us(x)+(w-1)F(g(x)).
\end{align*}

\vspace*{-6mm}
\end{IEEEproof}
Now, we are ready to prove Theorem~\ref{thm:main_scalar_lb}. This
theorem says that, for any fixed $w\geq1$ and $\eta>0$, there is
an $N_{0}<\infty$ such that, for all $N\geq N_{0}$, we have $\max_{i\in\ir 1M}x_{i}^{(\infty)}\geq\underline{x}^{*}-\eta$,
where $\underline{x}^{*}\triangleq\min\left(\arg\min_{x\in\Xc}\Us(x)\right)$. 
\begin{IEEEproof}
[Proof of Theorem~\ref{thm:main_scalar_lb}] The fixed point of the
vector recursion is given by $\x^{(\infty)}$. The idea is to lower
bound this fixed point by starting the vector recursion $\z^{(\ell+1)}=\A^{\t}\f(\A\g(\z^{(\ell)}))$
from $\z^{(0)}=\underline{x}^{*}\bm{1}_{M}$. Since $\underline{x}^{*}$
is a fixed point of the scalar recursion and the coupled recursion
has 0's at the boundary, $\z^{(0)}\succeq\z^{(1)}$ and the vector
sequence $\z^{(0)}\succeq\z^{(1)}\succeq\cdots$ is non-increasing.
This implies that $\z^{(\ell)}$ converges to a limit $\z^{(\infty)}$
and, using the fact that $\z^{(0)}\preceq\x^{(0)}$, one finds that
$\z^{(\infty)}\preceq\x^{(\infty)}$. From Lemma~\ref{lem:U_vec_bound},
it follows that 
\begin{equation}
\begin{split}\sum_{i=1}^{M} & \Us\left(z_{i}^{(\infty)}\right)\stackrel{\text{part (iii)}}{\leq}\Uc(\z^{(\infty)})\stackrel{\text{part (ii)}}{\leq}\Uc(\z^{(0)})\\
 & =\Uc(\underline{x}^{*}\bm{1})\stackrel{\text{part (iv)}}{=}M\,\Us(\underline{x}^{*})+(w-1)F(g(\underline{x}^{*})).
\end{split}
\label{eq:scalar_converse_bound}
\end{equation}

The proof continues by contradiction. First, we define $\gamma=\min_{x\in[0,\underline{x}^{*}-\eta]}\Us\left(x\right)-\Us(\underline{x}^{*})$
and observe that $\gamma>0$ is implied by Lemma~\ref{lem:U_vec_bound}(i).
Next, we choose 
\[
N>N_{0}\triangleq{\textstyle \frac{(w-1)F(g(\underline{x}^{*}))}{\gamma}}-(w-1)
\]
 and suppose that $\max_{i\in\ir 1M}z_{i}^{(\infty)}<\underline{x}^{*}-\eta$.
This implies $z_{i}^{(\infty)}<\underline{x}^{*}-\eta$ for all $i\in\ir 1M$
and, hence, that
\[
\sum_{i=1}^{M}\Us\left(z_{i}^{(\infty)}\right)\geq\sum_{i=1}^{M}\min_{x\in[0,\underline{x}^{*}-\eta]}\Us\left(x\right)=M\left(\Us(\underline{x}^{*})+\gamma\right).
\]
Next, we observe that $N>N_{0}$ implies 
\[
M>N_{0}+(w-1)>{\textstyle \frac{(w-1)F(g(\underline{x}^{*}))}{\gamma}}
\]
 and 
\[
M\left(\Us(\underline{x}^{*})+\gamma\right)>M\,\Us(\underline{x}^{*})+(w-1)F(g(\underline{x}^{*})).
\]
But, this contradicts~\eqref{eq:scalar_converse_bound} and implies
that $\max_{i\in\ir 1M}z_{i}^{(\infty)}\geq\underline{x}^{*}-\eta$.
The stated result follows from $\x^{(\infty)}\succeq\z^{(\infty)}$.
\end{IEEEproof}

\section{Dependence on a Parameter}

\label{sec:dependence} In many applications, the recursion of interest
also depends on an additional parameter. If the uncoupled recursion
has a stable fixed point at 0 when this parameter is sufficiently
small, then one is often interested in the largest parameter value
such that the coupled recursion converges to the zero vector. In this
section, we characterize this threshold and show that, under some
conditions, it is equal to the natural generalization of the threshold
associated with the Maxwell conjecture for LDPC codes~\cite[Conj.~1]{Measson-it08}.

\subsection{Admissible Systems}

\label{sub:dependence_scalar}

For $\emax\in(0,\infty)$, let $\Ec\triangleq[0,\emax]$ and suppose
that the system of interest depends on a parameter $\epsilon\in\mathcal{E}$.
In this case, the recursion is defined by the bivariate functions
\[
f\colon\Yc\times\Ec\to\Xc\quad\mbox{and}\quad g\colon\Xc\times\Ec\to\Yc.
\]
For bivariate functions with dependence on $x$ and $\epsilon$, we
use the notation $f(x;\epsilon)$ and the first two $x$-derivatives
are denoted by $f'(x;\epsilon)$ and $f''(x;\epsilon)$. Derivatives
in $\epsilon$ (or mixed partial derivatives) are denoted by $f^{(m,n)}(x;\epsilon)\triangleq\frac{\d^{m}}{\d x^{m}}\frac{\d^{n}}{\d\epsilon^{n}}f(x;\epsilon)$.
For convenience, we also define $h(x;\epsilon)\triangleq f(g(x;\epsilon);\epsilon)$
and $\Xm\triangleq(0,\xmax]$.
\begin{defn}
\label{def:scalar_admissible} An \emph{ admissible system} is a system
where the functions $f(x;\epsilon)$ and $g(x;\epsilon$) satisfy:
\begin{enumerate}
\item $f(x;\epsilon)$ and $g(x;\epsilon)$ are $C^{1}$ functions on $\Xc\times\Ec$,
\item $f(x;\epsilon)$ and $g(x;\epsilon)$ are non-decreasing in both $x$
and $\epsilon$,
\item $g(x;\epsilon)$ is strictly increasing in $x$, and
\item $g''(x;\epsilon)$ exists and is jointly continuous on $\Xc\times\Ec$.
\end{enumerate}

In addition, we say that an admissible system is \emph{proper} if
$h^{(0,1)}(x;\epsilon)>0$ for all $(x,\epsilon)\in\Xm\times\Ec$.

\end{defn}
Based on the above conditions, it is easy to verify that the results
of Theorem~\ref{thm:main_scalar_ub} and Theorem~\ref{thm:main_scalar_lb}
hold for each $\epsilon\in\Ec$. For admissible systems, the uncoupled
potential function, $\Us:\Xc\times\Ec\to\mathbb{R}$, is defined to
be 
\begin{equation}
\Us(x;\epsilon)\triangleq xg(x;\epsilon)-G(x;\epsilon)-F(g(x;\epsilon);\epsilon),\label{eq:univariate_scalar_potential_parameter}
\end{equation}
where $F(x;\epsilon)=\int_{0}^{x}f(z;\epsilon)\d z$ and $G(x;\epsilon)=\int_{0}^{x}g(z;\epsilon)\d z$.

\begin{defn}
Applying Theorem~\ref{thm:main_scalar_ub} to a system that depends
on a parameter naturally leads to the following definitions:
\begin{align*}
\Psi(\epsilon) & \triangleq\min_{x\in\Xc}\Us(x;\epsilon),\\
X^{*}(\epsilon) & \triangleq\left\{ x\in\Xc\mid\Us(x;\epsilon)=\Psi(\epsilon)\right\} ,\mbox{ and}\\
\overline{x}^{*}(\epsilon) & \triangleq\max\, X^{*}(\epsilon).
\end{align*}

\end{defn}
In many cases, one is interested in the $\epsilon$-threshold below
which the uncoupled (resp. coupled) recursion converges to 0 (resp.
the $\bm{0}$ vector). This leads us to the following definitions.
\begin{defn}
Let the \emph{single-system threshold} $\es$ be 
\begin{equation}
\es\triangleq\sup\left\{ \epsilon\in\Ec\,\bigg|\, h(x;\epsilon)<x,\, x\in\Xm\right\} ,\label{eq:single_system_threshold}
\end{equation}
which is well-defined as long as $h(x;0)<x$ for $x\in\Xm$.
\end{defn}

\begin{defn}
Let the \emph{stability threshold} $\est$ be 
\begin{equation}
\est\triangleq\sup\left\{ \epsilon\in\Ec\,\bigg|\,\exists\delta\!>\!0,\,\forall x\in(0,\delta],h(x;\epsilon)\!<\! x\right\} ,\!\label{eq:stability_threshold}
\end{equation}
which is well-defined as long as $0$ is a stable fixed point of the
uncoupled recursion for $\epsilon=0$. We say that a system has a
\emph{strict stability threshold} if, for all $\epsilon\in(\est,\emax]$,
there is a $\delta>0$ such that $h(x;\epsilon)>x$ for all $x\in(0,\delta]$.
Likewise, a system is called \emph{unconditionally stable} if there
is a $\delta>0$ such that $h(x;\emax)<x$ for all $x\in(0,\delta]$. 
\end{defn}
For the coupled recursion, Theorem~\ref{thm:main_scalar_ub} implies
a simple threshold for convergence to $\bm{0}$ as $w\to\infty$.
This expression is stated precisely below. Lemma~\ref{lem:alt_thresh}
also gives a number of equivalent expressions.
\begin{defn}
Let the \emph{coupled threshold }(or \emph{potential threshold)} $\ec$
be defined by 
\begin{equation}
\ec\triangleq\sup\left\{ \epsilon\in\Ec\mid\overline{x}^{*}(\epsilon)=0\right\} ,\label{eq:potential_threshold_0}
\end{equation}
which is well-defined as long as $\overline{x}^{*}(0)=0$. \end{defn}
\begin{lem}
\label{lem:threshold_lemma} If the above thresholds are well-defined,
then we have the following:
\begin{enumerate}
\item If $\epsilon<\es$, then the uncoupled recursion converges to a fixed
point at $0$. If $\epsilon>\es$, then it converges to a non-zero
fixed point.
\item If $\epsilon<\est$, then $0$ is a stable fixed point of the uncoupled
recursion. If $\epsilon>\est$, then $0$ is not a stable fixed point.
\item If\textup{ $\epsilon<\ec$}, then $\overline{x}^{*}(\epsilon)=0$
and the fixed point of the coupled recursion converges to $\bm{0}$
as $w\to\infty$. 
\item If\textup{ $\epsilon<\ec\leq\est$}, then there is a $w_{0}<\infty$
such that the coupled recursion converges to $\bm{0}$ for all $w>w_{0}$. 
\end{enumerate}
\end{lem}
\begin{IEEEproof}
The statement of (i) follows from well-known properties of continuous
scalar recursions~\cite[p.~96]{RU-2008}. A fixed point of a recursion
is called \emph{stable} if all sufficiently small perturbations return
to that fixed point. Using this definition, we see that (ii) is true
by definition. Statement (iii) follows from the definition of $\ec$
and Theorem~\ref{thm:main_scalar_ub}. 

For (iv), we note that $\epsilon<\ec\leq\est$ implies that $\overline{x}^{*}(\epsilon)=0$
and that there is a $\delta>0$ such that $h(x;\epsilon)<x$ for all
$x\in(0,\delta]$. Therefore, Proposition~\ref{prop:finitew}(iii)
implies that there is a $w_{0}<\infty$ such that the coupled recursion
converges to $\bm{0}$ for all $w>w_{0}$. 
\end{IEEEproof}

\subsection{Connections with EXIT Functions}

Recursions that depend on a parameter also have an extra degree of
freedom that allows some quantities to be computed in creative ways.
For example, the DE equations for LDPC codes on a BEC have a deep
structure and the Maxwell construction in statistical physics can
be applied to their EXIT functions to bound the performance of MAP
detectors~\cite{Measson-it08}. Some elements of that theory (e.g.,
parts that depend only on the mathematical structure of the problem)
can be extended to general recursions. The remainder of this section
describes these extensions.
\begin{example}
\label{ex:irregular_ldpc} For the case of irregular LDPC codes on
the BEC~\cite{RU-2008}, we have $\emax=\xmax=\ymax=1$ and 
\begin{align*}
f(x;\epsilon) & =\epsilon\lambda(x)\\
g(x;\epsilon) & =1-\rho(1-x).
\end{align*}
The polynomials $L(x)$ (resp. $R(x)$) define the bit (resp. check)
node degree distributions and we use the standard convention that
$\lambda(x)=L'(x)/L'(1)$ (resp. $\rho(x)=R'(x)/R'(1)$)~\cite{RU-2008}.
In terms of these functions, we find that 
\begin{equation}
\begin{split}F(x;\epsilon) & =\epsilon L(x)/L'(1)\\
G(x;\epsilon) & =x-\left(1-R(1-x)\right)/R'(1).
\end{split}
\label{eq:irr_ldpc_FG}
\end{equation}
The trial entropy%
\footnote{One source of confusion is that this term is typically used to describe
essentially the same quantity in a number of functional forms (e.g.,
as a function of $x$ in \cite[p.~124]{RU-2008}, of $(x,y)$ in \cite[Lem.~4]{Measson-it08},
and of $(x,\epsilon)$ in $\Us$). In general, the extra parameters
can be suppressed by requiring that the function be evaluated at a
DE fixed point.%
} for this problem is defined by $\phi(x,1-\rho(1-x))$ in \cite[(9.10)]{Montanari-it05}
and $P_{\epsilon}(\texttt{x},\texttt{y}(\texttt{x}))$ in~\cite[Lem.~4]{Measson-it08}.
Translating notation between these papers shows that the uncoupled
potential $\Us(x;\epsilon)$ is exactly equal to $-1/L'(1)$ times
the trial entropy. Conjectured values for the conditional entropy
and the MAP erasure rate are also given explicitly in \cite[Sec.~X]{Montanari-it05}.
The Maxwell construction in \cite{Measson-it08} also implies conjectured
values for conditional entropy, the MAP EXIT function, and erasure
rate of the MAP decoder. The conjectured%
\footnote{Based on the results in \cite{Montanari-it05,Measson-it08,Giurgiu-isit13,Giurgiu-arxiv13},
this conjecture is largely resolved but there are still some exceptional
cases that remain unproven (e.g., see \cite[Thm.~2]{Montanari-it05},
\cite[Sec.~VII]{Measson-it08}, and \cite{Measson-isit07}).%
} value of the conditional entropy of the code bits given the observations
is given by $\max_{x\in[0,1]}\phi(x,1-\rho(1-x))=-L'(1)\Psi(\epsilon)$
and this was first proven to be an upper bound (under some restrictions)
in~\cite[(9.10)]{Montanari-it05}. The conjectured value of the MAP
EXIT function is given by $-L'(1)\Psi'(\epsilon)$. The conjectured
value of the MAP erasure rate is given by $\epsilon L(\overline{x}^{*}(\epsilon))=L'(1)F(\overline{x}^{*}(\epsilon);\epsilon)$,
where $\overline{x}^{*}(\epsilon)$ is a DE fixed point because Lemma~\ref{lem:scalar_potential_minima_fp}
shows that all minima of $\Us(x;\epsilon)$ occur at fixed points.
We note that the results of this paper show that SC ensembles achieve
these conjectured values.
\end{example}
The perspective in this section provides generalizations of the quantities
in Example~\ref{ex:irregular_ldpc} for general $f(x;\epsilon)$
and $g(x;\epsilon)$. An important part of this is the result that
$\Psi(\epsilon)$ has a simple integral expression. This follows from
the fact that $x\in X^{*}(\epsilon)$ implies that $x$ is a fixed
point (i.e., $h(x;\epsilon)=x$) and, hence, $\Us^{(1,0)}(x;\epsilon)=0$
and $\Us^{(0,1)}(x;\epsilon)$ has a simple expression (e.g., see
\eqref{eq:dUde_simple}). If the system satisfies a few additional
constraints, then this simple expression can also be used to compute
$\Us(x;\epsilon)$ at any fixed point. The following theorem forms
a basis for these connections. 
\begin{thm}
\label{thm:scalar_min_potential_integral} The function $\Psi(\epsilon)$
is non-increasing, and satisfies $\Psi(\epsilon)=\int_{0}^{\epsilon}\psi(t)\,\d t$
with 
\begin{equation}
\psi(t)\triangleq-G^{(0,1)}\left(\overline{x}^{*}(t);t\right)-F^{(0,1)}\left(g(\overline{x}^{*}(t);t);t\right).\label{eq:psi_def}
\end{equation}
For a proper admissible system, $\Psi(\epsilon)$ is strictly decreasing
on $\epsilon\in[\ec,\emax]$ \textup{or, similarly, if $\overline{x}^{*}(\epsilon)>0$.}
\end{thm}
The following Lemma, whose proof appears in Appendix \ref{sec:app_dep_proof},
is used in the proof of Theorem~\ref{thm:scalar_min_potential_integral}.
\begin{lem}
\label{lem:lipshcitz_FGUP} For admissible systems, the function $\Us(x;\epsilon)$
(resp. $\Psi(\epsilon)$) is (resp. uniformly) Lipschitz continuous
in $\epsilon$. Moreover, $F^{(0,1)}(x;\epsilon)$ and $G^{(0,1)}(x;\epsilon)$
are non-negative and non-decreasing in $x$.
\end{lem}

\begin{IEEEproof}
[Proof of Theorem~\ref{thm:scalar_min_potential_integral}] Consider
a continuously differentiable bivariate function (e.g., $\Us(x;\epsilon)$)
and its $x$-minimum over a compact set (e.g., $\Psi(\epsilon)=\min_{x\in\Xc}\Us(x;\epsilon)$).
The standard \emph{envelope theorem} states that, if the minimum is
achieved uniquely, then the derivative of $\Psi$ equals $\Us^{(0,1)}(x;\epsilon)$
evaluated at the $x$-minimizer (e.g., $x=\overline{x}^{*}(\epsilon)$).
In this proof, we make use of an elegant generalization of the envelope
theorem to the case where the minimum is not necessarily unique \cite{Milgrom-ecmet02}. 

Since $\Psi(\epsilon)$ is Lipschitz continuous by Lemma~\ref{lem:lipshcitz_FGUP},
it follows that $\Psi(\epsilon)$ is differentiable almost everywhere
and satisfies the fundamental theorem of calculus~\cite[pp.~106-108]{Folland-1999}.
Therefore, there is a Lebesgue integrable function $\psi(t)=\Psi'(t)$
such that
\[
\Psi(\epsilon)=\int_{0}^{t}\psi(t)\mathrm{d}t.
\]
Based on this, the envelope theorem in \cite{Milgrom-ecmet02} shows
that $\psi(t)=\Us^{(0,1)}\left(\overline{x}^{*}(t);t\right)$ almost
everywhere. Using \eqref{eq:univariate_scalar_potential_parameter},
we can compute
\begin{align}
\Us^{(0,1)}\left(x;\epsilon\right) & =\frac{\d}{\d\epsilon}\Us(x;\epsilon)\nonumber \\
 & =\frac{\d}{\d\epsilon}\left(xg(x;\epsilon)-G(x;\epsilon)-F(g(x;\epsilon);\epsilon)\right)\nonumber \\
 & =\left(x-f(g(x;\epsilon);\epsilon)\right)g^{(0,1)}(x;\epsilon)-G^{(0,1)}(x;\epsilon)\nonumber \\
 & \quad\quad-F^{(0,1)}(g(x;\epsilon);\epsilon).\label{eq:dUde_simple}
\end{align}
Since Lemma~\ref{lem:scalar_potential_minima_fp} shows that all
minima of $\Us(x;\epsilon)$ occur at fixed points, we see that
\[
\Us^{(0,1)}\left(\overline{x}^{*}(\epsilon);\epsilon\right)=-G^{(0,1)}(\overline{x}^{*}(\epsilon);\epsilon)-F^{(0,1)}(g(\overline{x}^{*}(\epsilon);\epsilon);\epsilon).
\]
Finally, we note that Lemma~\ref{lem:lipshcitz_FGUP} also shows
that $F^{(0,1)}(x;\epsilon)$ and $G^{(0,1)}(x;\epsilon)$ are non-negative
on $\Xc\times\Ec$. From this, we see that $\psi(\epsilon)$ is almost
everywhere non-positive and conclude that $\Psi(\epsilon)$ is non-increasing. 

For a proper admissible system, 
\[
h^{(0,1)}(x;\epsilon)\!=\! f^{(1,0)}(g(x;\epsilon);\epsilon)g^{(0,1)}(x;\epsilon)\!+\! f^{(0,1)}(g(x;\epsilon);\epsilon)
\]
is positive on $\Xm\times\Ec$ and this implies either $g^{(0,1)}(x;\epsilon)>0$
or $f^{(0,1)}(g(x;\epsilon);\epsilon)>0$ for each $(x,\epsilon)\in\Xm\times\Ec$
. By integrating, we see that, if $\overline{x}^{*}(\epsilon)>0$,
then either $G^{(0,1)}(\overline{x}^{*}(\epsilon);\epsilon)>0$ or
$F^{(0,1)}(g(\overline{x}^{*}(\epsilon);\epsilon);\epsilon)>0$. Since
the definition of $\ec$ implies that $\overline{x}^{*}(\epsilon)>0$
for $\epsilon>\ec$, it follows that $\Psi(\epsilon)$ is strictly
decreasing on $\epsilon\in[\ec,\emax]$.\end{IEEEproof}
\begin{rem}
The function $\psi(t)$ is analogous to the MAP EXIT function and
is computed by evaluating a function, which is closely related to
the recursion, along a fixed-point curve (e.g., $\overline{x}^{*}(t)$)
of the recursion. The difference between the MAP EXIT function and
the EBP EXIT function is the fixed-point curve along which the integral
is taken. For the DE recursion of irregular LDPC codes with degree
profile $(\lambda,\rho)$, the EBP EXIT function associated with a
fixed point $\overline{x}^{*}(t)$ is $L(1-\rho(1-\overline{x}^{*}(t)))$.
Likewise, we find that $G^{(0,1)}(x;\epsilon)=0$ and $F^{(0,1)}(x;\epsilon)=L(1-\rho(1-x))/L'(1)$
implies $\psi(t)=-L(1-\rho(1-\overline{x}^{*}(t)))/L'(1)$.
\end{rem}

\begin{defn}
\label{def:Xf_ex_Q} For admissible systems, the following observations
and definitions will be useful:
\begin{enumerate}
\item Since $h(x;\epsilon)$ is continuous and non-decreasing in $\epsilon$,
the subset of $\Xm$ that supports a fixed point is
\[
\Xc_{f}\triangleq\left\{ x\in\Xm\mid h(x;0)\leq x,h(x;\emax)\geq x\right\} .
\]

\item For each $x\in\Xc_{f}$, let $\ex x\triangleq\min\left\{ \epsilon\in\Ec\,|\, h(x;\epsilon)=x\right\} $
be the smallest $\epsilon$ that supports a fixed point at $x$.
\item The \emph{fixed-point potential}, $Q\colon\Xc_{f}\to\R$, is defined
by $Q(x)\triangleq\Us(x;\ex x)$. 
\end{enumerate}
\end{defn}
\begin{lem}
\label{lem:e(x)_derivative} For a proper admissible system, $\ex x$
maps $x\in\Xc_{f}$ to the unique $\epsilon\in\Ec$ such that $h(x;\epsilon)=x$.
Moreover, each $x_{0}\in\Xc_{f}$ with $\ex{x_{0}}\in(0,\emax)$ lies
in the interior of $\Xc_{f}$ and $\ex x$ is $C^{1}$ on any interval
in $\Xc_{f}$.\end{lem}
\begin{IEEEproof}
See Appendix~\ref{sec:app_dep_proof}.
\end{IEEEproof}
The following lemma is the natural generalization of \cite[Lem.~4]{Measson-it08}
to the more general setup considered in this paper. 
\begin{lem}
\label{lem:Q_ebp_integral} For a proper admissible system, if the
interval $[x_{1},x_{2}]\subset\Xc_{f}$, then $Q(x_{2})-Q(x_{1})$
is given by
\[
-\int_{x_{1}}^{x_{2}}\left(G^{(0,1)}\left(x;\ex x\right)+F^{(0,1)}\left(x;\ex x\right)\right)\epx x\d x.
\]
\end{lem}
\begin{IEEEproof}
See Appendix \ref{sec:app_dep_proof}.\end{IEEEproof}
\begin{rem}
It is very likely that the precise connection between the Maxwell
construction and the $\Psi(\epsilon)$ function is still rather opaque.
The key to understanding this connection is to note that $\overline{x}^{*}(\epsilon)$
and $\psi(\epsilon)$ can only jump when the minimum of $\Us(x;\epsilon)$
is achieved at multiple $x$-values. In this case, there exist $x_{1},x_{2}\in X^{*}(\eps)$
such that $\Us(x_{1};\epsilon)=\Us(x_{2};\epsilon)=\Psi(\epsilon)$
and $\ex{x_{1}}=\ex{x_{2}}=\epsilon$. Therefore, if $[x_{1},x_{2}]\in\Xc_{f}$,
then Lemma~\ref{lem:Q_ebp_integral} shows that the natural analogue
of the EBP EXIT area integral is given by 
\begin{align*}
0 & =\Us(x_{2};\epsilon)-\Us(x_{1};\epsilon)\\
 & =Q(x_{2})-Q(x_{1})\\
 & =\int_{x_{1}}^{x_{2}}\Us^{(0,1)}(x;\ex x)\epx x\d x.
\end{align*}
Hence, the positive and negative portions of the parametric area computation
(e.g., EBP EXIT integral) must balance. 

We note that this connection between the Maxwell construction and
minimizers of thermodynamic potential functions is certainly known
by many physicists. It can be seen implicitly in the statistical physics
literature (via Legendre transforms) but is rarely discussed explicitly.\end{rem}
\begin{lem}
\label{lem:e(x)_properties} The following properties of $\Xc_{f}$
and $\ex x$ will be useful:
\begin{enumerate}
\item The set $\Xc_{f}$ is either closed or it is missing a single limit
point at $x=0$.
\item If $\est\in[0,\emax)$, then there exists a sequence $x_{n}\in\Xc_{f}$
such that $x_{n}\to0$ and $\lim_{n}\ex{x_{n}}=\est$.
\item If the system is unconditionally stable, then $\Xc_{f}$ is closed
and $\ec\leq\est$.
\item If the system has a strict stability threshold $\est\in[0,\emax)$,
then $\ec\leq\est$ and $\lim_{n}\ex{x_{n}}=\est$ for any sequence
$x_{n}\in\Xc_{f}$ such that $x_{n}\to0$. 
\end{enumerate}
\end{lem}
\begin{IEEEproof}
See Appendix \ref{sec:app_dep_proof}.\end{IEEEproof}
\begin{lem}
\label{lem:alt_thresh} The following expressions will also be useful:
\begin{enumerate}
\item For any admissible system, $\Psi(\epsilon)=0$ for $\epsilon\leq\ec$
and $\Psi(\epsilon)<0$ implies $\overline{x}^{*}(\epsilon)>0$.
\item For a proper admissible system, 
\begin{equation}
\ec=\sup\left\{ \epsilon\in\Ec\mid{\textstyle \min_{x\in\Xc}}\Us(x;\epsilon)\geq0\right\} .\label{eq:istc_pot_thresh}
\end{equation}
Also, if $Q(x)<0$ for $x\in\Xc_{f}$, then $\ec<\ex x$.
\item Consider a proper admissible system where either $\Xc_{f}$ is closed
or the system has a strict stability threshold $\est\in[0,\emax]$.
If $\ec\in[0,\emax)$, then
\begin{equation}
\ec=\min\left\{ \ex x\mid x\in\overline{\Xc_{f}},Q(x)=0\right\} .\label{eq:ec_eq_Maxwell}
\end{equation}

\end{enumerate}
\end{lem}
\begin{IEEEproof}
See Appendix \ref{sec:app_dep_proof}.\end{IEEEproof}
\begin{rem}
We note that the definition of the potential threshold in \eqref{eq:istc_pot_thresh}
is exactly the same as the definition given in \cite{Yedla-istc12}
but there is one caveat. In \cite{Yedla-istc12}, the energy gap is
defined as the minimum of the potential over all $x$-values not less
than the first unstable fixed point. It is stated incorrectly in that
this value will always be positive. Instead, this value is strictly
positive only if there are no unstable fixed points arbitrarily close
to $x=0$. For some pathological cases (e.g., see Example~\ref{ex:pathological}),
the energy gap can be zero. Remark~\ref{rem:w-limit cor of thm 1}
shows that threshold of the SC system still converges to $\ec$, but
only as $w\to\infty$.
\end{rem}

\section{Applications}

\label{sec:applications}

\subsection{Application to Irregular LDPC Codes}

\label{sec:irregular-ldpc-codes}

Consider the ensemble LDPC$(\lambda,\rho)$ of irregular LDPC codes
and assume transmission takes place over an erasure channel with parameter
$\epsilon$ \cite{RU-2008}. For this ensemble, the node degree distributions
are given by $L(z)=\sum_{i=2}^{d_{v}}L_{i}z^{i}$ (resp. $R(z)=\sum_{i=1}^{d_{c}}R_{i}z^{i}$)
where $L_{i}$ (resp. $R_{i}$) is a rational number that defines
the fraction of variable (resp. check) nodes with degree $i$. The
edge degree distributions are given by $\lambda(z)=L'(z)/L'(1)=\sum_{i=2}^{d_{v}}\lambda_{i}z^{i-1}$
(resp. $\rho(z)=R'(z)/R'(1)=\sum_{i=2}^{d_{c}}\rho_{i}z^{i-1}$) where
$\lambda_{i}$ (resp. $\rho_{i}$) is the fraction of edges attached
to variable (resp. check) nodes with degree $i$. 

Let $x^{(\ell)}$ be the fraction of erasure messages sent from variable
to check nodes during iteration $\ell$. Then, the DE equation can
be written in the form of~\eqref{eq:scalar_recursion}, where $f(x;\epsilon)=\epsilon\lambda(x)$
and $g(x;\epsilon)=1-\rho(1-x)$ \cite{RU-2008}. This example was
discussed in Example~\ref{ex:irregular_ldpc} and it is easy to verify
that $f$ and $g$ describe a proper admissible system with $\emax=\xmax=\ymax=1$.
\begin{defn}
\label{def:sc_irr_ldpc} The ensemble of SC irregular LDPC$(\lambda,\rho)$
codes with length $\kappa N$ is defined as follows. Assume that $\kappa$
is chosen such that $\kappa L_{i}$, $\kappa R_{j}L'(1)/R'(1)$, and
$kL'(1)/w$ are integers. A collection of $N$ variable-node groups
are placed at positions labeled $\left\{ 1,2,\ldots,N\right\} $ and
a collection of $M=N+w-1$ check-node groups are placed at positions
labeled $\left\{ 1,2,\ldots,M\right\} $. In each variable-node group,
$\kappa L_{i}$ nodes of degree $i$ are placed for $i\in\ir 2{d_{v}}$.
Similarly, in each check-node group, $\kappa R_{j}L'(1)/R'(1)$ nodes
of degree $j$ are placed for $j\in\ir 1{d_{c}}$. Next, the $\kappa L'(1)$
edge sockets in each group of variable and check nodes are partitioned
in $w$ groups using a uniform random permutation. The set of variable-node
(resp. check-node) sockets in the $k$-th group of position $i$ is
denoted by $\mathcal{P}_{i,k}^{v}$ (resp. $\mathcal{P}_{i,k}^{c}$)
for $k=0,1,\ldots,w-1$. The SC code is constructed by defining edges
that connect the sockets in $\mathcal{P}_{i,k}^{v}$ to the sockets
in $\mathcal{P}_{i+k-1,k}^{c}$ in some fixed manner. This construction
leaves some sockets of the check-node groups at the boundaries unconnected
and these sockets are removed (i.e., the bit value is assumed to 0).
It is easy to verify that the DE analysis equations for this ensemble
are given by \eqref{eq:coupled_scalar_recursion} as $\kappa\to\infty$.
This construction can be seen as a variation on the original $(\texttt{l},\texttt{r},L,w)$
SC ensemble defined in~\cite{Kudekar-it11}.
\end{defn}
The results of this paper allow us to analyze the performance of the
above SC LDPC ensemble. In particular, the uncoupled potential function
can be computed easily from~\eqref{eq:irr_ldpc_FG} and we note that
it equals $-\phi(x,1-\rho(1-x))/L'(1)$, where $\phi$ is given by
\cite[(9.10)]{Montanari-it05}. For the uncoupled ensemble, this implies
that, under mild conditions on $R(x)$, the conditional entropy of
the code bits given the received sequence is upper bounded by $-L'(1)\Psi(\epsilon)$
\cite[Thm.~2]{Montanari-it05}. 

\begin{figure}[t]
\begin{center}
\ifextfig
\includegraphics{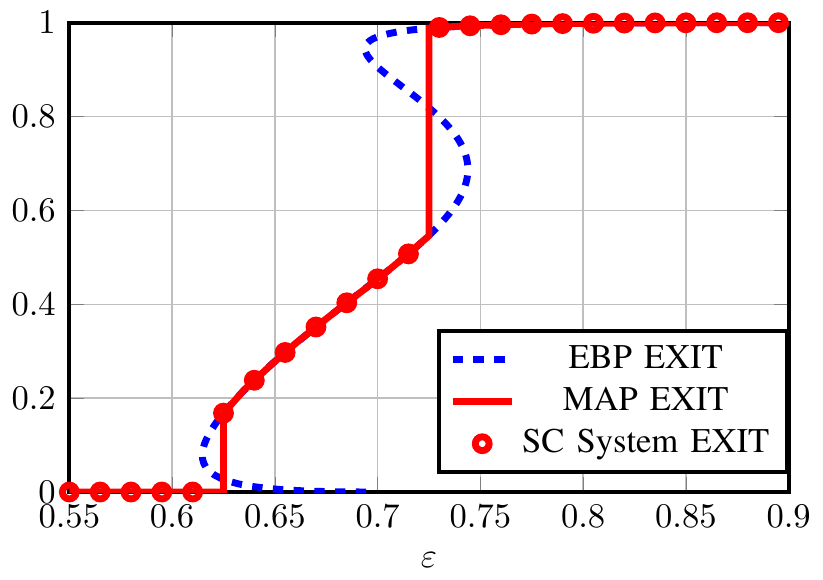}
\else
\input{sc_ebp_ldpc}
\fi
\vspace{-4mm}
\caption{Graphical depiction of Maxwell saturation for the irregular LDPC ensemble in Example~\ref{ex:maxwell_ex_ldpc}.}
\label{fig:maxwell_ex_ldpc}
\end{center}
\end{figure}

For this system, the unique $\epsilon$ associated with a fixed point
$x$ is well-known and given by $\epsilon(x)\triangleq x/\lambda(1-\rho(1-x))$.
For consistency with Definition~\ref{def:Xf_ex_Q}, we use $\ex x$
to denote the restriction of $\epsilon(x)$ to the domain $\Xc_{f}=\left\{ x\in(0,1]\mid\epsilon(x)\in[0,1]\right\} $.
For this setup, the conjectured MAP decoding threshold is called the
\emph{Maxwell threshold} \cite[Conj.~1]{Measson-it08} and is defined
in terms of the \emph{trial entropy}
\begin{align*}
P & (x)=\int_{0}^{x}L\left(1-\rho(1-z)\right)\epsilon'(z)\d z\\
 & \stackrel{(a)}{=}-L'(1)x\left(1-\rho(1-x)\right)+L'(1)\left(x+\frac{1-R(1-x)}{R'(1)}\right)\\
 & \quad+\epsilon(x)L\left(1-\rho(1-x)\right)\\
 & =-L'(1)Q(x)\;\mathrm{for}\, x\in\Xc_{f},
\end{align*}
where $(a)$ comes from~\cite[p.~124]{RU-2008} and the relationship
to $Q(x)$ can be verified directly by computing $Q(x)$ according
to Definition~\ref{def:Xf_ex_Q}. In particular, the Maxwell threshold
\cite[Conj.~1]{Measson-it08} is given by%
\footnote{Their statement actually neglects the stability condition and is incorrect
if the threshold is determined by stability. Also, there are pathological
cases where this definition implies $\emaxwell=\infty$ (e.g., when
the design rate $1-L'(1)/R'(1)$ is negative).%
}
\begin{align*}
\emaxwell\! & \triangleq\!\min\left\{ {\textstyle \frac{1}{\lambda'(0)\rho'(1)}},\min\left\{ \epsilon(x)\,|\, P(x)\!=\!0,x\!\in\!(0,1]\right\} \right\} \\
 & =\min\left\{ \epsilon(x)\,|\, P(x)=0,x\in[0,1]\right\} ,
\end{align*}
where $P(0)\triangleq\lim_{x\to0}P(x)=0$ and 
\[
\epsilon(0)\triangleq\lim_{x\to0}\epsilon(x)=\begin{cases}
{\textstyle \frac{1}{\lambda'(0)\rho'(1)}} & \mathrm{if}\,\lambda'(0)>0\\
\infty & \mathrm{otherwise.}
\end{cases}
\]
In many cases, the results of \cite{Montanari-it05,Measson-it08,Kudekar-it09}
can be used to show that $\emap\leq\emaxwell$, where $\emap$ is
noise threshold of the MAP decoder. We also note that the potential
$\Us(x;\epsilon)$ is the same as the pseudo-dual of the average Bethe
variational entropy (e.g., see \cite{Walsh-com10}, \cite[Part 2, pp.~62-65]{Vontobel-acorn09}). 
\begin{lem}
\label{lem:maxwell_irr_ldpc_bec} For the ensemble LDPC$(\lambda,\rho)$,
the potential threshold \eqref{eq:potential_threshold_0} equals the
Maxwell threshold if the code rate $r=1-L'(1)/R'(1)$ satisfies $r>0$. \end{lem}
\begin{IEEEproof}
Based on well-known properties of LDPC DE equations, it is easy to
verify that they define an admissible system. It is a proper admissible
system because $h^{(0,1)}(x;\epsilon)=\lambda(1-\rho(1-x))>0$ for
$x\in(0,1]$. Also, the expansion $h(x;\epsilon)=\epsilon\lambda'(0)\rho'(1)x+O(x^{2})$
shows that the system has a strict stability threshold $\est\in(0,1]$.
If $r>0$, then we can compute directly $L'(1)\Us(1;1)=L'(1)/R'(1)-1<0$.
Since $\Us(0,1)=0$, this implies that $\overline{x}^{*}(1)>0$ and,
hence, that $\ec\in[0,1)$. Along with the strict stability threshold,
this allows us to apply Lemma~\ref{lem:alt_thresh}(iii) to see that
\eqref{eq:ec_eq_Maxwell} holds. This also implies that there is an
$x^{*}\in\overline{\Xc_{f}}$ such that $\ex{x^{*}}<1$ and $Q(x^{*})=0$.
From this, we see that $\emaxwell<1$ and that
\begin{alignat*}{1}
\ec & =\inf\left\{ \ex x\,|\, x\in\overline{\Xc_{f}},Q(x)=0\right\} \\
 & =\inf\left\{ \epsilon(x)\,|\, x\in[0,1],P(x)=0\right\} =\emaxwell
\end{alignat*}
because $P(x)=-L'(1)Q(x)$ for $x\in\Xc_{f}$ and $\epsilon(x)>1$
for $x\in[0,1]\backslash\overline{\Xc_{f}}$.\end{IEEEproof}
\begin{cor}
If the conditions of Lemma~\ref{lem:maxwell_irr_ldpc_bec} hold and
$\epsilon<\emaxwell$, then there is a $w_{0}<\infty$ such that the
SC DE recursion converges to the zero vector for all $w>w_{0}$.\end{cor}
\begin{IEEEproof}
Since $\emaxwell=\ec\leq\est$ by Lemma~\ref{lem:maxwell_irr_ldpc_bec}
and $\epsilon<\emaxwell$, we can apply Lemma~\ref{lem:threshold_lemma}(iv)
to see that there is a $w_{0}<\infty$ such that, for all $w<w_{0}$,
the SC DE recursion will converge to the zero vector.\end{IEEEproof}
\begin{example}
\label{ex:maxwell_ex_ldpc}Consider the irregular LDPC ensemble with
degree distribution 
\[
\begin{split}\lambda(x) & =\tfrac{4}{20}x+\tfrac{5}{20}x^{2}+\tfrac{2}{20}x^{6}+\tfrac{9}{20}x^{20}\\
\rho(x) & =\tfrac{6}{10}x^{4}+\tfrac{4}{10}x^{12}.
\end{split}
\]
In this case, $\ec$ equals the conjectured MAP threshold and we can
compute $\ec\approx0.625$. 

Recall that the parametric EBP EXIT curve~\cite[Ch.~3]{RU-2008}
is given by $\left(\epsilon(x);L(1-\rho(1-x))\right)$ and notice
that, up to a scale factor, this equals $\Us^{(0,1)}(x;\epsilon(x))$.
Likewise, the conjectured MAP EXIT curve is given by $L(1-\rho(1-\overline{x}^{*}(\epsilon)))$
and, up to a scale factor, this equals $\psi(\epsilon)$. We can also
compute $L(1-\rho(1-x_{406}^{(\infty)}(\epsilon)))$, where $x_{406}^{(\infty)}(\epsilon)$
is the fixed-point erasure rate for the worst-case position of a finite
SC system with $w=11$ and $N=800$. Fig.~\ref{fig:maxwell_ex_ldpc}
shows the EBP EXIT curve, the conjectured MAP EXIT curve, and the
performance of the finite SC system. Notice that the finite SC system
matches the asymptotic prediction almost exactly. 
\end{example}

\subsection{Application to Irregular LDGM Codes}

\label{sec:irregular_ldgm_codes}

Any code in the irregular LDPC ensemble, LDPC$(\lambda,\rho)$, can
be converted into an LDGM code by adding a degree-1 variable node
to each check node. During transmission, all degree-1 variable nodes
are transmitted and all other variable nodes are punctured. Let LDGM$(\lambda,\rho)$
denote the standard irregular LDGM ensemble formed by converting each
code in LDPC$(\lambda,\rho)$ to an LDGM code. The degree distributions
are defined identically to the LDPC case in Section~\ref{sec:irregular_ldgm_codes}
except that the new degree-1 variable nodes are not counted.

Consider the iterative decoding of a code in LDGM$(\lambda,\rho)$
assuming transmission takes place over an erasure channel with parameter
$\epsilon$ \cite{RU-2008}. Let $x^{(\ell)}$ be the fraction of
erasure messages sent from variable to check nodes during iteration
$\ell$. Then, the DE equation can be written in the form of~\eqref{eq:scalar_recursion},
where $f(x;\epsilon)=\lambda(x)$ and $g(x;\epsilon)=1-(1-\epsilon)\rho(1-x)$
\cite{RU-2008}. This example was discussed in Example~\ref{ex:ldgm_irr1}
and it is easy to verify that $f$ and $g$ describe an proper admissible
system with $\emax=\xmax=\ymax=1$. The SC LDGM ensemble is essentially
identical to the SC LDPC ensemble in Definition~\ref{def:sc_irr_ldpc}
except that degree-1 variable nodes are added to each check node.

The results of this paper also allow us to analyze the performance
of the SC LDGM ensemble. This is in contrast with our previous results
in \cite{Yedla-istc12,Yedla-itw12} because $g(0;\epsilon)=1-(1-\epsilon)\rho(1)>\epsilon$
and $h(0;\epsilon)>\lambda(\epsilon)$ do not satisfy the necessary
conditions in those papers. This is related to the well-known fact
that LDGM codes do not have a perfect-decoding fixed point. Still,
the uncoupled potential function can be computed easily from~\eqref{eq:irr_ldpc_FG}
using the fact that $F(x;\epsilon)=L(x)/L'(1)$ and $G(x;\epsilon)=x-(1-\epsilon)\left(1-R(1-x)\right)/R'(1)$.
In this case, $\Us(x;\epsilon)$ is the same as $-\phi(x,1-(1-\epsilon)\rho(1-x))/L'(1)$,
where $\phi$ is given by \cite[(9.16)]{Montanari-it05}. For the
uncoupled ensemble, \cite[Thm.~2]{Montanari-it05} states that, under
mild conditions on $R(x)$, the conditional entropy per information
bit%
\footnote{This normalization is somewhat unconventional but simplifies their
unified treatment of LDPC and LDGM ensembles. %
} is upper bounded by $\max_{x\in[0,1]}\phi(x,1-(1-\epsilon)\rho(1-x))=-L'(1)\Psi(\epsilon)$.
Normalizing instead by the number of code bits, we find that the conditional
entropy per code bit is upper bounded by $-R'(1)\Psi(\epsilon)$.

\begin{figure}[t]
\begin{center}
\ifextfig
\includegraphics{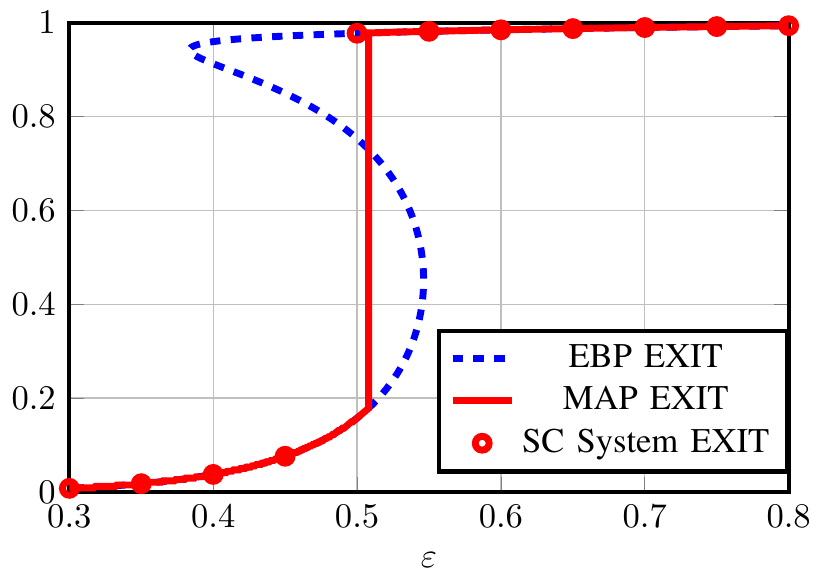}
\else
\input{sc_ebp_ldgm}
\fi
\vspace{-4mm}
\caption{Graphical depiction of Maxwell saturation for the irregular LDGM ensemble in Example~\ref{ex:maxwell_ex_ldgm}. }
\label{fig:maxwell_ex_ldgm}
\end{center}
\end{figure}

Likewise, one can solve for $\epsilon$ in the fixed point equation
to see that
\[
\epsilon(x)=1-\frac{1-\lambda^{-1}(x)}{\rho(1-x)}
\]
and $\Xc_{f}=\left\{ x\in(0,1]\mid\epsilon(x)\in[0,1]\right\} $.
If there are no degree-1 checks (i.e., $\rho(0)=0$), then $x=1$
is a fixed point for all $\epsilon\in\Ec$ and the formula for $\epsilon(x)$
must be treated carefully at this point. Regardless, we can write
the fixed-point potential as
\begin{align*}
Q(x) & =x\left(1-(1-\epsilon(x))\rho(1-x)\right)\\
 & \quad\quad-\left(x-(1-\epsilon(x))\frac{1-R(1-x)}{R'(1)}\right)\\
 & \quad\quad-\frac{1}{L'(1)}L\left(1-(1-\epsilon(x))\rho(1-x)\right).
\end{align*}
We note that, since the DE recursion for LDGM codes does not converge
to 0 for any $\epsilon>0$, the previous threshold results do not
apply. 
\begin{rem}
Although we cannot use the previous threshold results for this system,
the following observations provide something similar. Notice that
$\Psi(\epsilon(\overline{x}^{*}(\epsilon)))=Q(\overline{x}^{*}(\epsilon))$
implies $\epsilon(\overline{x}^{*}(\epsilon))=\Psi^{-1}(Q(\overline{x}^{*}(\epsilon)))$,
where the inverse exists because $h^{(0,1)}(x;\epsilon)>0$ for $\epsilon\in[0,1)$
implies $\Psi$ is strictly decreasing for $\epsilon\in[0,1]$. From
this, we see that, for any $x\in\Xc_{f}$, $\Psi^{-1}(Q(x))$ can
be seen as the $\epsilon$-threshold below which $\overline{x}^{*}(\epsilon)\leq x$.\end{rem}
\begin{example}
\label{ex:maxwell_ex_ldgm} Consider the irregular LDGM ensemble with
degree distribution 
\[
\begin{split}\lambda(x) & =x^{5}\\
\rho(x) & =\tfrac{2}{45}+\tfrac{2}{45}x+\tfrac{7}{15}x^{2}+\tfrac{4}{9}x^{3}.
\end{split}
\]
In this case, the Maxwell curve has a single discontinuity at $\epsilon_{0}\approx0.508$.

Applying the techniques in~\cite[Ch.~3]{RU-2008} to the LDGM ensemble
(i.e., differentiating the conditional entropy) shows that the parametric
EBP EXIT curve is given by
\[
\left(\epsilon(x);1-R(1-x)\right)=\left(\epsilon(x),R'(1)G^{(0,1)}(x;\epsilon(x))\right)
\]
and notice that, up to a scale factor, this equals $\Us^{(0,1)}(x;\epsilon(x))$.
Likewise, the conjectured MAP EXIT curve is given by 
\[
1-R(1-\overline{x}^{*}(\epsilon))=R'(1)G^{(0,1)}(\overline{x}^{*}(\epsilon);\epsilon)
\]
and, up to a scale factor, this equals $\psi(\epsilon)$. We can also
compute $1-R(1-x_{406}^{(\infty)}(\epsilon))$, where $x_{406}^{(\infty)}(\epsilon)$
is the fixed-point erasure rate for the worst-case position of a finite
SC system with $w=11$ and $N=800$. Fig.~\ref{fig:maxwell_ex_ldgm}
shows the parametric EBP EXIT, the conjectured MAP EXIT curve, and
the performance of the finite SC system. Notice that the finite SC
system has a slight overhang at $\epsilon_{0}$ due to finite $w$
but otherwise matches the asymptotic prediction very well. 
\end{example}

\subsection{Application to Generalized LDPC Codes}

\label{sec:gldpc-codes}

Consider a generalized LDPC (GLDPC) code with degree-2 bits and generalized
check constraints given by a primitive BCH code of block-length $n$.
For an iterative decoder based on bounded-distance decoding of the
BCH code, the DE recursions can be derived for both the BEC and binary
symmetric channel (BSC)~\cite{Jian-isit12}. On the BEC, the code
is chosen to correct all patterns of at most $t$ erasures. On the
BSC, the code is chosen to correct all error patterns of weight at
most $t$ and it is assumed that miscorrections can be ignored. 

For fixed $t$ and large $n$, the rate of a $t$ error-correcting
primitive BCH code is given by $r_{\mathrm{BCH}}=1-t\log_{2}(n+1)/n$
and the same code corrects $2t$ erasures. For the GLDPC code, the
overall design rate is $r_{\mathrm{GLDPC}}=1-2(1-r_{\mathrm{BCH}})$
\cite{Jian-isit12}. Therefore, the GLDPC rates for the BSC and BEC
constructions are $r_{\mathrm{BSC}}=1-2t\log_{2}(n+1)/n$ and $r_{\mathrm{BEC}}=1-t\log_{2}(n+1)/n$. 

For both cases, the iterative decoding performance of this ensemble
is characterized by a DE recursion of the form~(\ref{eq:scalar_recursion}),
where $\epsilon$ denotes the channel parameter. In this case, the
recursion is defined by $\emax=\xmax=\ymax=1$, $f(x;\epsilon)\triangleq\epsilon x$,
and $g(x;\epsilon)\triangleq\sum_{i=t}^{n-1}\binom{n-1}{i}x^{i}(1-x)^{n-1-i}$
\cite{Jian-isit12}. Here, $x$ denotes the erasure (resp. error)
probability of bit-to-check messages for the BEC (resp. BSC) case.
We note that $g(0;\epsilon)=0$ and $g(1;\epsilon)=1$. To highlight
the fact that $g(x;\epsilon)$ is independent of $\epsilon$, we also
write $g(x)$ instead of $g(x;\epsilon)$. 

Let $B(a,b)\triangleq\frac{(a-1)!(b-1)!}{(a+b-1)!}$ denote the Beta
function and 
\[
I_{x}(t,n-t)\triangleq\frac{1}{B(t,n-t)}\int_{0}^{x}z^{t-1}(1-z)^{n-t-1}dz
\]
denote the regularized incomplete Beta function \cite[Sec.~8.17]{Olver-2010}.
Using the fact that $g(x)=I_{x}(t,n-t)$ \cite[Sec.~8.17]{Olver-2010},
we can verify that, for $x\in(0,1)$,
\begin{align*}
g'(x) & =\frac{\d}{\d x}I_{x}(t,n-t)\\
 & =\frac{x^{t-1}(1-x)^{n-t-1}}{B(t,n-t)}>0.
\end{align*}
Therefore, the functions $f,g$ define an admissible system because
they: map $[0,1]$ to $[0,1]$, are polynomial in $x,\epsilon$, are
strictly increasing in $x$ for fixed $\epsilon\neq0$, and are non-decreasing
in $\epsilon$. It is a proper admissible system because $h^{(0,1)}(x;\epsilon)=g(x)>0$
for $x\in(0,1]$. Since $h(x;\epsilon)=\epsilon g(x)$, we also see
that $\epsilon(x)=x/g(x)$. For consistency with Definition~\ref{def:Xf_ex_Q},
we use $\ex x$ to denote the restriction of $\epsilon(x)$ to the
domain $\Xc_{f}=\left\{ x\in(0,1]\mid\epsilon(x)\in[0,1]\right\} $.

Using \eqref{eq:uni_scalar_pot}, we find that
\[
\Us(x;\epsilon)=xg(x)-\int_{0}^{x}g(z)\d z-\epsilon\frac{1}{2}g(x)^{2}.
\]
 Likewise, the fixed-point potential is given by
\begin{align}
Q(x) & =xg(x)-\int_{0}^{x}g(z)\d z-\frac{1}{2}\frac{x}{g(x)}g(x)^{2}\nonumber \\
 & =\frac{1}{2}xg(x)-\int_{0}^{x}g(z)\d z,\label{eq:gldpc_Q_simple}\\
 & =\frac{1}{2}\left(\frac{2t}{n}-x\right)I_{x}(t,n-t)-\frac{x^{t}(1-x)^{n-t}}{nB(t,n-t)},\nonumber 
\end{align}
where the last expression is stated without derivation and can be
verified using well-known properties of $I_{x}(a,b)$ \cite[Sec.~8.17]{Olver-2010}.
Since $I_{0}(a,b)=0$ and $I_{1}(a,b)=1$, we find that $Q(0)=0$
and $Q(1)=-(1-2t/n)/2$. For the BEC case, the EBP EXIT function is
$h^{EBP}(x)=x^{2}$ and the ``trial entropy'' is given by 
\begin{align*}
P(x) & =\int_{0}^{x}h^{\mathrm{EBP}}(g(z))\epsilon'(z)\d z\\
 & =\int_{0}^{x}g(z)^{2}\epsilon'(z)\d z\\
 & =-xg(x)+2I_{x}(t,n-t)\\
 & =-2Q(x).
\end{align*}
Therefore, we have $P(0)=0$ and $P(1)=1-2t/n$.

It also turns out that $P(x)$ has a unique root $x^{*}$ in $(0,1]$
(e.g., see Lemma~\ref{lem:gldpc_unique_root}). For LDPC codes, this
typically means that $\epsilon(x^{*})$ is a (tight) upper bound on
the MAP threshold. But, the bounded-distance decoder associated with
$g(x)$ is suboptimal for BCH codes. Therefore, we find that $P(1)>r_{\mathrm{BEC}}$
and, hence, $\epsilon(x^{*})$ is not an upper bound on the MAP threshold
of the system. Instead, we will find that $\ec=\epsilon(x^{*})$ is
the threshold of the SC system defined in \cite{Jian-isit12}.
\begin{lem}
\label{lem: P'x&P''x} For a fixed $t\geq2$, $P'(x)<0$ for all $x\in(0,\frac{t-1}{n-2})$,
and $P''(x)\geq0$ for all $x\in[\frac{t-1}{n-2},1)$.\end{lem}
\begin{IEEEproof}
See Appendix \ref{sec:app_app_proof}.\end{IEEEproof}
\begin{lem}
\label{lem:gldpc_unique_root} For any $2\leq t\leq\left\lfloor \frac{n-1}{2}\right\rfloor $,
$P(x)$ has an unique root $x^{*}$ in $(0,1]$ and $\ec=\epsilon(x^{*})<1$.\end{lem}
\begin{IEEEproof}
First, we observe that $P(0)=0$. Next, we see that $P(x)<0$ for
all $x\in(0,\frac{t-1}{n-2})$ because Lemma~\ref{lem: P'x&P''x}
shows that $P'(x)<0$ in this range. Since $P(1)>0$ for $t$ in the
stated range, one can see that $P(x)$ has at least one root in $[\frac{t-1}{n-2},1)$.
Finally, we see that $P(x)$ is convex on $[\frac{t-1}{n-2},1)$ because
Lemma~\ref{lem: P'x&P''x} shows that $P''(x)\geq0$ for all $x\in[\frac{t-1}{n-2},1)$.
Therefore, $P(x)$ has exactly one root in $[\frac{t-1}{n-2},1)$.
Since $Q(1)=-(1-2t/n)/2<0$ for $t\leq\left\lfloor \frac{n-1}{2}\right\rfloor $,
we can apply Lemma~\ref{lem:alt_thresh}(ii) to see that $\ec<\ex 1=1/g(1)=1$.
Also, $h'(0;1)=0$ implies that the system is unconditionally stable
and, hence, Lemma~\ref{lem:e(x)_properties} shows that $\Xc_{f}$
is closed. Therefore, we can use Lemma~\ref{lem:alt_thresh}(iii)
to see that
\[
\ec=\min\left\{ \ex x\mid x\in\Xc_{f},P(x)=0\right\} =\epsilon(x^{*})
\]
because $P(x)=-2Q(x)$ and $x^{*}\in\Xc_{f}$.\end{IEEEproof}
\begin{cor}
If $\epsilon<\ec$, then there is a $w_{0}<\infty$ such that the
DE recursion for the SC GLDPC in ensemble defined in \cite{Jian-isit12}
converges to the zero vector for all $w>w_{0}$.\end{cor}
\begin{IEEEproof}
Since $\epsilon<\ec<\est=1$, we can apply Lemma~\ref{lem:threshold_lemma}(iv)
to see that there is a $w_{0}<\infty$ such that, for all $w<w_{0}$,
the SC DE recursion will converge to the zero vector.
\end{IEEEproof}

\subsection{Application to ISI Channels with Erasure Noise}

\label{sec:gec-ldpc} In~\cite{Pfister-jsac08}, a family of intersymbol-interference
(ISI) channels with erasure noise is investigated as an analytically
tractable model of joint iterative decoding of LDPC codes and channels
with memory. Let $\phi(x;\epsilon)$ be the function that maps the
\emph{a priori} erasure rate $x$ from the code and the channel erasure
rate $\epsilon$ to the erasure rate of extrinsic messages from the
channel detector to the bit nodes%
\footnote{The $\phi$ function is denoted by $f$ in \cite{Pfister-jsac08}. %
}. Then, the resulting DE update equation for the erasure rate, $x^{(\ell)}$,
of bit-to-check messages can be written in the form of \eqref{eq:scalar_recursion},
where $f(x;\epsilon)=\phi(L(x);\epsilon)\lambda(x)$ and $g(x;\epsilon)=1-\rho(1-x)$~\cite{Pfister-jsac08}.
If $\phi(x;\epsilon)$ is continuously differentiable on $\Xc\times\Ec$,
then these equations define an admissible system with $\emax=\xmax=\ymax=1$
and 
\begin{align*}
\Us(x;\epsilon) & =x\left(1-\rho(1-x)\right)-\left(x-\frac{1-R(1-x)}{R'(1)}\right)\\
 & \quad-\frac{1}{L'(1)}\Phi(L(1-\rho(1-x));\epsilon),
\end{align*}
where $\Phi(x;\epsilon)=\int_{0}^{x}\phi(z;\epsilon)\text{d}z$. If
the DE equations define a proper admissible system, then we can also
compute the fixed-point potential $Q(x)$ using Lemma~\ref{lem:Q_ebp_integral}.
The main benefit that the calculation is the same (up to a scale factor)
as the calculation of the ``trial entropy'' $P(x)$ in \cite{Nguyen-arxiv11,Nguyen-2012}.
Assuming that $h(x;\epsilon)$ is defined and strictly increasing
in $\epsilon$ for $\epsilon\in[0,\infty]$, we use $\epsilon(x)$
to denote the unique $\epsilon\in[0,\infty]$ associated with a fixed
point $x\in[0,1]$. For consistency with Definition~\ref{def:Xf_ex_Q},
we use $\ex x$ to denote the restriction of $\epsilon(x)$ to the
domain $\Xc_{f}=\left\{ x\in(0,1]\mid\epsilon(x)\in[0,1]\right\} $.
Adjusting notation from \cite{Nguyen-arxiv11,Nguyen-2012}, we find
that $P(x)=-L'(1)Q(x)$ for $x\in\Xc_{f}$, where $Q(x)=\Us(x;\ex x)$.
Using this, the natural generalization of the Maxwell threshold is
given by

\begin{align*}
\emaxwell\! & \triangleq\!\min\left\{ \est,\min\left\{ \epsilon(x)\,|\, P(x)\!=\!0,x\!\in\!(0,1]\right\} \right\} \\
 & =\min\left\{ \epsilon(x)\,|\, P(x)\!=\!0,x\!\in\![0,1]\right\} ,
\end{align*}
where $\epsilon(0)\triangleq\est$. If $P(x)$ has a unique root $x^{*}\in(0,1]$
and the threshold is not determined by stability, then the EXIT area
theorem shows that $\emap\leq\emaxwell=\epsilon(x^{*})$ because $P(1)=1-L'(1)/R'(1)$
equals the code rate~\cite{Nguyen-arxiv11,Nguyen-2012}. For the
dicode erasure channel (DEC), the counting argument in \cite{Measson-it08}
has also been extended to prove the tightness of this upper bound~\cite{Nguyen-arxiv11,Nguyen-2012}. 

The analysis of irregular LDPC codes on ISI channels with erasure
noise was extended to SC ensembles in \cite{Kudekar-isit11-DEC,Nguyen-arxiv11,Nguyen-icc12,Nguyen-2012}
and the resulting DE equations depend on bit ordering during transmission.
If the bits in each spatial group are clustered together in the channel
input (rather than interleaved across the entire input), then the
SC DE equations are given by~\eqref{eq:coupled_scalar_recursion}
and we can apply our analysis to the SC system. If the channel is
the DEC, precoded DEC, or a class-2 partial-response channel with
erasure noise (PR2EC), then there are closed-form expressions for
$\phi(x;\epsilon)$ and it is easy to verify that the DE equations
define proper admissible systems with strict stability thresholds~\cite{Pfister-jsac08,Sekido-sita11,Nguyen-arxiv11,Nguyen-icc12}. 
\begin{lem}
\label{lem:maxwell_gec} Consider the ensemble LDPC$(\lambda,\rho)$
on a channel defined by $\phi(x;\epsilon)$ and assume the code rate
$r=1-L'(1)/R'(1)$ satisfies $r>0$. If $\phi(x;\epsilon)$ is continuously
differentiable on \textup{$\Xc\times\Ec$,} $\frac{\d}{\d\epsilon}\phi(x;\epsilon)>0$
for $(x,\epsilon)\in\Xm\times\Ec$, $\phi'(0;\epsilon)$ is strictly
increasing in $\Ec$, and $\phi(1;1)=1$, then the potential threshold
$\ec$ given by \eqref{eq:potential_threshold_0} equals the Maxwell
threshold $\emaxwell$. \end{lem}
\begin{IEEEproof}
Under the given conditions, it is easy to verify that the DE equations
define a proper admissible system. From $\phi(1;1)=1$, we see that
$h(1;1)=1$ and $\ex 1=1$. Using $Q(1)=-P(1)/L'(1)=-r/L'(1)<0$,
we can apply Lemma~\ref{lem:alt_thresh}(ii) to see that $\ec<\ex 1=1$.
Since $\phi'(0;\epsilon)$ is strictly increasing in $\Ec$, the expansion
$h(x;\epsilon)=\phi'(0,\epsilon)\lambda'(0)\rho'(1)x+o(x)$ shows
that the system has a strict stability threshold. Therefore, we can
apply Lemma~\ref{lem:alt_thresh}(iii) to see that
\begin{alignat*}{1}
\ec & =\min\left\{ \ex x\,|\, Q(x)=0,x\in\overline{\Xc_{f}}\right\} \\
 & =\min\left\{ \epsilon(x)\,|\, P(x)=0,x\in[0,1]\right\} =\emaxwell
\end{alignat*}
because $P(x)=-L'(1)Q(x)$ for $x\in\Xc_{f}$ and $\epsilon(x)>1$
for $x\in(0,1]\backslash\Xc_{f}$.\end{IEEEproof}
\begin{cor}
If the conditions of Lemma~\ref{lem:maxwell_gec} hold and $\epsilon<\emaxwell$,
then there is a $w_{0}<\infty$ such that the SC DE recursion converges
to the zero vector for all $w>w_{0}$.\end{cor}
\begin{IEEEproof}
Since $\emaxwell=\ec\leq\est$ by Lemma~\ref{lem:maxwell_gec} and
$\epsilon<\emaxwell$, we can apply Lemma~\ref{lem:threshold_lemma}(iv)
to see that there is a $w_{0}<\infty$ such that, for all $w<w_{0}$,
the SC DE recursion will converge to the zero vector.
\end{IEEEproof}

\subsection{Application to Compressed Sensing}

Spatial coupling can also be used to improve the performance of compressed
sensing~\cite{Kudekar-aller10,Krzakala-physrevx,Donoho-it13,Krzakala-jsm12}.
Under some conditions, one can analyze belief-propagation reconstruction
for compressed sensing using a Gaussian approximation~\cite{Guo-aller09,Donoho-pnas09,Bayati-it11,Donoho-it13,Krzakala-physrevx,Krzakala-jsm12}.
This leads to a simple scalar recursion for the mean-square error
(MSE) in the estimate of each signal component. In particular, we
consider the reconstruction of a length-$n$ signal vector, whose
entries are i.i.d. copies of a random variable $X$, from $\delta n$
linear measurements in the limit as $n\to\infty$. We assume that
$\mathbb{E}[X^{2}]<\infty$ and that each measurement is corrupted
by independent Gaussian noise with variance $\sigma^{2}$. 

Some results from \cite{Guo-it05} are required to understand the
next few equations. Let $Y=\sqrt{\snr}X+Z$ be a scaled observation
of $X$ with standard Gaussian noise $Z$. If $\mathbb{E}[X^{2}]=1$,
then the variable $\snr$ corresponds to the standard notion of signal-to-noise
ratio for $Y$. For general $X$, $\snr$ can be seen simply as a
scaling factor. Regardless, the scalar estimate, $\hat{X}=\mathbb{E}\left[X|Y\right]$,
minimizes the MSE, $\mathbb{E}[(X-\hat{X})^{2}]$, and the resulting
MSE is denoted $\mmse(\snr)$. This function is bounded because $0\leq\mmse(\snr)\leq\mmse(0)\leq\mathbb{E}[X^{2}]<\infty$.
It also satisfies the differential relationship, with the mutual information,
given by
\begin{equation}
\frac{1}{2}\mmse(\snr)=\frac{\d}{\d\snr}I\left(X;\sqrt{\snr}X+Z\right).\label{eq:mmse_mutual_deriv}
\end{equation}
Using this, the recursion given by~\cite[(33)]{Donoho-it13} for
the MSE of SC compressed sensing is essentially%
\footnote{The details of the boundary conditions and coupling coefficients differ
in negligible ways. %
} equal to 
\begin{equation}
x_{i}^{(\ell+1)}=\sum_{j=1}^{N}A_{j,i}\,\mmse\left(\sum_{k=1}^{M}A_{j,k}\,\frac{1}{\sigma^{2}+\frac{1}{\delta}x_{k}^{(\ell)}}\right).\label{eq:cs_coupled_recursion}
\end{equation}

Now, we will construct an uncoupled recursion that results in the
same coupled recursion. Since the $\mmse$ function is non-increasing
and $(\sigma^{2}+\frac{1}{\delta}x)^{-1}\in(0,1/\sigma^{2}]$ for
$x\in[0,\infty)$, we introduce a slight twist and define 
\[
f(y)=\mmse\left(\frac{1}{\sigma^{2}}-y\right).
\]
This implies that $f(y)$ is non-decreasing in $y$ and satisfies
$f(y)\in[0,\mmse(0)]$ for $y\in[0,1/\sigma^{2}]$. Likewise, to undo
this small twist, we define 
\[
g(x)=\frac{1}{\sigma^{2}}-\frac{1}{\sigma^{2}+\frac{1}{\delta}x},
\]
where $g(x)\in(0,1/\sigma^{2}]$ for $x\in[0,\infty]$. The given
bounds on $f$ and $g$ also allow us to choose $\xmax=\mmse(0)$
and $\ymax=g(\xmax)<1/\sigma^{2}$. From this, we see that the uncoupled
recursion is given by
\begin{equation}
f(g(x))=\mmse\left(\frac{1}{\sigma^{2}+\frac{1}{\delta}x}\right).\label{eq:cs_overall_scalar}
\end{equation}

For Theorems~\ref{thm:main_scalar_ub} and~\ref{thm:main_scalar_lb},
we must also verify that $f$ and $g$ satisfy some technical conditions.
First, it is easy to verify that $g(x)$ is strictly increasing in
$x$ and $C^{2}$ on $\Xc$. One can also show that $f(y)$ is $C^{1}$
on $\Yc$ (e.g., see \cite{Guo-it11}) and, hence, the recursion satisfies
the necessary conditions of the theorems for $\sigma^{2}\in(0,\infty)$.
Since $\sum_{k=1}^{M}A_{j,k}=1$, one can also verify that these definitions
make the coupled recursion in \eqref{eq:coupled_scalar_recursion}
identical to \eqref{eq:cs_coupled_recursion}. Lastly, we note that
the boundary conditions assumed by \eqref{eq:cs_coupled_recursion}
are correct if we have perfect estimates from signal nodes beyond
the boundaries. This is easily achieved by choosing the signal values
to be zero outside the range of the SC system.

Now, we can use \eqref{eq:uni_scalar_pot} to construct the potential
function. For $f$, we note that
\[
F(y)=2I\left(X;\sqrt{1/\sigma^{2}}\, X\!+\! Z\right)\!-\!2I\left(X;\sqrt{1/\sigma^{2}-y}\, X\!+\! Z\right)
\]
 satisfies $F(0)=0$ and it is easy to verify that $F'(y)=f(y)$ follows
from \eqref{eq:mmse_mutual_deriv}. For $g$, we observe that
\[
G(x)=\frac{1}{\sigma^{2}}x-\delta\ln\left(1+\frac{x}{\delta\sigma^{2}}\right)
\]
satisfies $G(0)=0$ and it is easy to check that $G'(x)=g(x)$. Combining
these, we see that
\begin{align}
\Us & (x)=-\frac{x}{\sigma^{2}+\frac{1}{\delta}x}+\delta\ln\left(1+\frac{x}{\delta\sigma^{2}}\right)\label{eq:cs_potential}\\
 & \!-\!2I\left(X;\sqrt{\frac{1}{\sigma^{2}}}X\!+\! Z\right)\!+\!2I\left(X;\sqrt{\frac{1}{\sigma^{2}\!+\! x/\delta}}X\!+\! Z\right).\nonumber 
\end{align}

\begin{cor}
Consider the MSE achieved by SC compressed sensing with BP reconstruction
(e.g., see \cite{Krzakala-physrevx,Donoho-it13,Krzakala-jsm12}).
For any $\delta>0$, there is a $w_{0}<\infty$ such that this MSE
is upper bounded by $\overline{x}^{*}+\delta$ for all $w>w_{0}$,
where 
\[
\overline{x}^{*}=\max\left(\argmin_{x\in\Xc}\Us(x)\right)
\]
and $\Us$ is given by \eqref{eq:cs_potential}.\end{cor}
\begin{IEEEproof}
This follows directly from the above definitions and Theorem~\ref{thm:main_scalar_ub}.\end{IEEEproof}
\begin{rem}
After we developed this example, we discovered that Kudekar et al.
also added a very similar example to the recent update of their paper~\cite{Kudekar-unpub13}. 
\end{rem}

\section{Conclusions and Future Work}

In this paper, we study a class of coupled scalar recursions and provide
a tight characterization of their behavior based on an underlying
uncoupled recursion. This result enables one to easily compute the
asymptotic decoding thresholds for a variety of iterative decoding
systems. We also demonstrate a precise connection between the threshold,
$\ec$, below which the coupled system converges to the zero vector
and the Maxwell threshold, $\emaxwell$, which is the conjectured
MAP threshold for LDPC ensembles.

The proof techniques used in this paper can also be extended, with
some complications, to coupled recursions on vectors (e.g., see \cite{Yedla-itw12})
and log-likelihood-ratio densities (e.g., see \cite{Kumar-aller12,Kumar-itsub13}).
Since the cited works only consider threshold saturation, we are currently
working to prove Maxwell saturation for these more general systems.

\appendices

\section{Proofs from Section \ref{sec:MaxwellSat}}

\label{sec:app_maxsat_proof}

\subsection{Proof of Proposition \ref{prop:finitew}}

For (i), we note that a bounded subset of real numbers has a limit
point iff it contains infinitely many points. Therefore, $\mathcal{F}\cap[\overline{x}^{*},\xmax]$
cannot have a limit point if $\mathcal{F}$ is a finite set. From
Theorem~\ref{thm:main_scalar_ub}, we see that $w_{0}<\infty$ whenever
$\mathcal{F}\cap[\overline{x}^{*},\xmax]$ does not have a limit point.
For (ii), recall that the composition of real analytic functions is
real analytic and that a real analytic function is identically zero
if its set of zeros has a limit point~\cite{Krantz-2002}. Since
$\mathcal{F}$ is the zero set of the real analytic function $x-f(g(x))$
and $x-f(g(x))\neq0$ for some $x\in\Xc$, it follows that $\mathcal{F}$
cannot have a limit point and must be finite. For (iii), the condition
implies that $\mathcal{F}\cap[\overline{x}^{*},\overline{x}^{*}+\gamma]=\{\overline{x}^{*}\}$
and therefore $\overline{x}^{*}$ is an isolated point in $\mathcal{F}\cap[\overline{x}^{*},\xmax]$.
Hence, $\overline{x}^{*}$ is not a limit point of $\mathcal{F}$.
For (iv), the Taylor expansion of $f(g(x))$ about $x=\overline{x}^{*}$
shows that there exists a $\gamma>0$ such that $f(g(x))<x$ for all
$x\in(\overline{x}^{*},\overline{x}^{*}+\gamma]$ and, thus, we can
apply (iii).

\subsection{Proof of Proposition \ref{prop:half_iter_shift}}

For (i), we note that $g(x)$ is invertible and write
\begin{align*}
y\in g(\mathcal{F}) & \Leftrightarrow g^{-1}(y)\in\mathcal{F}\\
 & \Leftrightarrow f(g(g^{-1}(y)))=g^{-1}(y)\\
 & \Leftrightarrow f(y)=g^{-1}(y)\\
 & \Leftrightarrow g(f(y))=y\\
 & \Leftrightarrow y\in\mathcal{F}'.
\end{align*}
A very similar argument shows $x\in f(\mathcal{F}')\Leftrightarrow x\in\mathcal{F}$
because $f(y)$ is invertible. For (ii), we use $x=f(g(x))$ to simplify
$\Vs(g(x))$ and this gives 
\begin{align*}
\Vs(g(x)) & =g(x)f(g(x))-F(g(x))-G(f(g(x)))\\
 & =xg(x)-G(x)-F(g(x)).
\end{align*}
Similarly, $y=g(f(y))$ implies that $\Us(f(y))$ simplifies to $\Vs(y)$.
For (iii), we note that all minimizers of $\Us(x)$ must satisfy $x=f(g(x))$
(see Lemma~\ref{lem:scalar_potential_minima_fp}) and, hence, (ii)
implies $\Us(x)=\Vs(g(x))$. Likewise, all minimizers of $\Vs(x)$
must satisfy $x=g(f(x))$ and $\Vs(x)=\Us(f(x))$. Therefore, one
gets a contradiction if $\Us(x)$ and $\Vs(x)$ have different minimum
values. Next, we let $m=\min_{x\in\Xc}\Us(x)=\min_{y\in\Yc}\Vs(y)$
be that minimum value and write
\begin{align*}
y\in g(\mathcal{M}) & \Leftrightarrow g^{-1}(y)\in\mathcal{M},\,\Us(g^{-1}(y))=m\\
 & \Leftrightarrow f(g(g^{-1}(y)))=g^{-1}(y),\,\Vs(y)=m\\
 & \Leftrightarrow f(y)=g^{-1}(y),\,\Vs(y)=m\\
 & \Leftrightarrow g(f(y))=y,\,\Vs(y)=m\\
 & \Leftrightarrow y\in\mathcal{M}'.
\end{align*}
A very similar argument shows $x\in f(\mathcal{M}')\Leftrightarrow x\in\mathcal{M}$.

\subsection{Proof of Lemma \ref{lem:translated_recursion}}

First, we verify that the translated scalar system is well defined.
In particular, we define $\sYc=[0,\ymax-g(\xf)]$ and observe that
$\sfs:\sYc\to\sXc$ because $\sfs(0)=f(0+g(\xf))-\xf=f(g(\xf))-\xf=0$
and $\sfs(\ymax-g(\xf))=f(\ymax-g(\xf)+g(\xf))-\xf=f(\ymax)-\xf\leq\xmax-\xf$.
Likewise, we observe that $\sgs:\sXc\to\sYc$ because $\sgs(0)=g(0+\xf)-g(\xf)=0$
and $\sgs(\xmax-\xf)=g(\xmax-\xf+\xf)-g(\xf)=g(\xmax)-g(\xf)\leq\ymax-g(\xf)$.
Moreover, it is easy to verify that $\sfs$ and $\sgs$ inherit their
monotonicity and  differentiability directly from $f$ and $g$.

For translated coupled system, we derive
\begin{align*}
 & \shc(\x)\triangleq\A^{\t}\sfc\left(\A\sgc(\x)\right)\\
 & =\A^{\t}\left(\f\left(\A\left(\g(\x\!+\!\xf\bm{1}_{M})\!-\! g(\xf)\bm{1}_{M}\right)\!+\! g(\xf)\bm{1}_{N}\right)\!-\!\xf\bm{1}_{N}\right)\\
 & \stackrel{(a)}{=}\A^{\t}\left(\f\left(\A\g(\x+\xf\bm{1}_{M})\right)-\xf\bm{1}_{N}\right)\\
 & =\A^{\t}\f\left(\A\g(\x+\xf\bm{1}_{M})\right)-\xf\bm{1}_{M}+\xf\left(\bm{1}_{M}-\A^{\t}\bm{1}_{N}\right)\\
 & =\left(\h(\x+x_{0}\bm{1}_{M})-\xf\bm{1}_{M}\right)+\xf\left(\bm{1}_{M}-\A^{\t}\bm{1}_{N}\right),
\end{align*}
where $(a)$ uses $\A\bm{1}_{M}=\bm{1}_{N}$. If $\xf>0$, then this
is strictly different than simply translating $\h(\x)$ to $\h(\x+\xf\bm{1}_{M})-x_{0}\bm{1}_{M}$.
In particular, the term $\xf(\bm{1}_{M}-\A^{\t}\bm{1}_{N})$ is exactly
the adjustment required to move the implied boundary value from 0
to $\xf$.

For the potentials, it is easy to verify that $\sFs'(y)=\sfs(y)$
and $\sGs'(x)=\sgs(x)$. For $\sUs(x)$, we write
\begin{align*}
\sUs(x) & =x\sgs(x)-\sGs(x)-\sFs(\sgs(x))\\
 & =x\left(g(x+\xf)-g(\xf)\right)-G(x+\xf)+xg(\xf)\\
 & \quad\quad+G(\xf)-F(g(x+\xf)-g(\xf)+g(\xf))\\
 & \quad\quad+\left(g(x+\xf)-g(\xf)\right)\xf+F(g(\xf))\\
 & =(x+\xf)g(x+\xf)-G(x+\xf)-F(g(x+\xf))\\
 & \quad\quad-\left(\xf g(\xf)-G(\xf)-F(g(\xf))\right)\\
 & =\Us(x+\xf)-\Us(\xf).
\end{align*}

\section{Proofs from Section \ref{sub:proof_ub_theorem}}

\label{sec:app_ub_proof}

\subsection{Proof of Lemma \ref{lem:scalar_potential_minima_fp}}

First, we note that $F(x)$ is convex because $f(x)$ is non-decreasing
and $G(x)$ is strictly convex because $g(x)$ is strictly increasing.
For any $x_{0}\in\Xc$, this implies that $-F(g(x))\leq-F(g(x_{0}))-f(g(x_{0}))(g(x)-g(x_{0}))$
and $-G(x)\leq-G(x_{0})-g(x_{0})(x-x_{0})$ with equality iff $x=x_{0}$.
Using these to upper bound $\Us(x)$ gives
\begin{align*}
\Us(x) & =xg(x)-G(x)-F(g(x))\\
 & \leq xg(x)-G(x_{0})-g(x_{0})(x-x_{0})\\
 & \quad-F(g(x_{0}))-f(g(x_{0}))(g(x)-g(x_{0}))\\
 & =\Us(x_{0})+\left(x-f(g(x_{0}))\right)\left(g(x)-g(x_{0})\right),
\end{align*}
with equality iff $x=x_{0}$. Now, choosing $x=f(g(x_{0}))$ shows
that $\Us\left(f(g(x_{0}))\right)\leq\Us(x_{0})$ with equality iff
$f(g(x_{0}))=x_{0}$. Hence, one step of the recursion from $x_{0}$
must strictly decrease the potential if $x_{0}$ is not a fixed point.

Next, we prove that, if $x_{0}$ is a local minimum of $\Us(x)$ on
$\Xc$, then $x_{0}$ is a fixed point. We do this by showing the
contrapositive: if $x_{0}$ is a not a fixed point, then $x_{0}$
is not a local minimum of $\Us(x)$ on $\Xc$. If $x_{0}$ is not
a fixed point, then $x_{0}-f(g(x_{0}))\neq0$ and there are two cases.
If $x_{0}-f(g(x_{0}))<0$, then $x_{0}<f(g(x_{0}))\leq\xmax$ and
the continuity of $x-f(g(x))$ implies that there is a $\delta>0$
and $\eta>0$ such that $x-f(g(x))<-\eta$ for all $x\in[x_{0},x_{0}+\delta]$.
This implies that, for any $x\in(x_{0},x_{0}+\delta]$, we have
\begin{align}
\Us(x)-\Us(x_{0}) & =\int_{x_{0}}^{x}(x-f(g(x)))g'(x)\d x\label{eq:U_inc_stable_fp}\\
 & <-\eta\int_{x_{0}}^{x}g'(x)\d x\nonumber \\
 & <-\eta\left(g(x)-g(x_{0})\right)<0,\nonumber 
\end{align}
where the last step holds because $g$ is strictly increasing. This
shows that $x_{0}$ is not a local minimum. If $x_{0}-f(g(x_{0}))>0$,
then a very similar argument shows that, for some $\delta>0$ and
any $x\in[x_{0}-\delta,x_{0})$, we have $\Us(x)<\Us(x_{0})$. Thus,
$x_{0}$ is not a local minimum and the proof is complete.

\subsection{Proof of Lemma \ref{lem:coupled_scalar_dec_sym}}

For $\ell=0$, property (i) follows from
\begin{align*}
\x{}^{(1)} & =\A^{\t}\f(\A\g(\x_{\max}))\preceq\A^{\t}\f(\A\bm{1}_{M}\cdot\ymax)\\
= & \;\A^{\t}\f(\bm{1}_{N}\cdot\ymax)\preceq\A^{\t}\bm{1}_{N}\cdot\xmax\preceq\x_{\max},
\end{align*}
where $\A\bm{1}_{M}=\bm{1}_{N}$ and $\A^{\t}\bm{1}_{N}\preceq\bm{1}_{M}$.
The inductive step for $\ell>1$ follows from the fact that $\h(\x)=\A^{\t}\f(\A\g(\x))$
is isotone. 

Let $\T_{M}$ denote the matrix, defined by $\left[\T_{M}\x\right]_{i}=x_{M-i+1}$,
that reverses the order of elements in a vector. This matrix can be
represented by $\left[\T_{M}\right]_{i,j}=\delta_{j,M-i+1}$ because
\[
\sum_{j=1}^{M}\left[\T_{M}\right]_{i,j}x_{j}=\sum_{j=1}^{M}\delta_{j,M-i+1}x_{j}=x_{M-i+1}.
\]
Since $j=M-i+1$ implies $i=M-j+1$, it follows that $\left[\T_{M}\right]_{i,j}=\delta_{i,M-j+1}$
and $\T_{M}^{\t}=\T_{M}$. Also, $\A$ is symmetric under simultaneous
row-column reversal (i.e., $A_{j,k}=A_{N-j+1,M-k+1}$) and this implies
that $\A\T_{M}=\T_{N}\A$. 

Since $\x^{(0)}=\T_{M}\x^{(0)}$ by definition and property (ii) is
equivalent to $\x^{(\ell)}=\T_{M}\x^{(\ell)}$, this property follows
by induction using

\begin{align*}
\x{}^{(\ell+1)} & =\A^{\t}\f(\A\g(\T_{M}\x^{(\ell)}))=\A^{\t}\f(\A\T_{M}\g(\x^{(\ell)}))\\
 & =\A^{\t}\f(\T_{N}\A\g(\x^{(\ell)}))=\A^{\t}\T_{N}\f(\A\g(\x^{(\ell)}))\\
 & =\T_{M}^{\t}\A^{\t}\f(\A\g(\x^{(\ell)}))=\T_{M}\x^{(\ell+1)}.
\end{align*}

For property (iii), we note that each mapping $\f,\g,\A,\A^{\t}$
is closed on the set of symmetric unimodal vectors. For $\f,\g$,
this holds because $f,g$ are monotone and operate on each element
of the vector. For $\A,\A^{\t}$, symmetry follows from $\A\T_{M}=\T_{N}\A$
above. For unimodality, we have
\begin{align*}
\left[\A\x\right]_{i+1}-\left[\A\x\right]_{i} & =\frac{1}{w}\sum_{j=0}^{w-1}x_{i+1+j}-\frac{1}{w}\sum_{j=0}^{w-1}x_{i+j}\\
 & =\frac{1}{w}\left(x_{i+w}-x_{i}\right)\geq0
\end{align*}
if $i\leq M-(i+w)+1$ (due to symmetry) and this is equivalent to
$i\leq N-i$. Therefore, if the length-$M$ vector $\x$ is symmetric
and unimodal, then the length-$N$ vector $\A\x$ is symmetric and
unimodal. The proof for $\A^{\t}$ is very similar and, hence, omitted.

\subsection{Proof of Lemma \ref{lem:mod_coupled_scalar_dec_inc}}

First, we note that $\x\preceq\z$ implies $\q(\x)\preceq\q(\z)$
and $\q\left(\h\left(\x\right)\right)\preceq\q\left(\h\left(\z\right)\right)$.
One can verify this by observing that, for each $j\in\ir 1M$, $\left[\q(\x)\right]_{j}$
is a non-decreasing function of each $x_{i}$. Now, we consider property
(i). Since $\hat{\x}^{(1)}=\q(\x^{(1)})$ and $\x^{(1)}\preceq\x^{(0)}$,
we have 
\[
\hat{\x}^{(1)}=\q(\h(\x^{(0)}))=\q(\x^{(1)})\preceq\q(\x^{(0)})=\x^{(0)}=\hat{\x}^{(0)}.
\]
Assuming $\hat{\x}^{(\ell)}\preceq\hat{\x}^{(\ell-1)}$, we proceed
by induction and observe that
\[
\hat{\x}^{(\ell+1)}=\q\left(\h\left(\hat{\x}^{(\ell)}\right)\right)\preceq\q\left(\h\left(\hat{\x}^{(\ell-1)}\right)\right)=\hat{\x}^{(\ell)}.
\]
Next, we consider property (ii). Since $\hat{\x}^{(0)}=\x^{(0)}$,
induction on $\hat{\x}^{(\ell)}\succeq\x^{(\ell)}$ shows that
\[
\hat{\x}^{(\ell+1)}=\q\left(\h\left(\hat{\x}^{(\ell)}\right)\right)\succeq\q\left(\h\left(\x^{(\ell)}\right)\right)=\x^{(\ell+1)}.
\]
To show property (iii), we note that $\hat{\x}^{(0)}$ is non-decreasing
and proceed by induction. Informally, the vector update is symmetric
about $\ti$ but the argument vector is asymmetric and flat after
$\ti$. Due to monotonicity, it is not too hard to see that $\z=\A^{\t}\f\left(\A\g(\x)\right)$
will increase to a maximum and then decrease due to the 0-boundary.
The key observation is that the maximum will occur at $i\geq\ti$
so that $\q(\z)$ is non-decreasing. Mathematically, we observe that,
for  a non-decreasing $\x$ (i.e., $\left[\x\right]_{i+1}\geq\left[\x\right]_{i}$
for $i\in\ir 1{M-1}$), it follows that $\g(\x)$, and $\A\g(\x)$,
$\y=\f(\A\g(\x))$ are non-decreasing. For $\z=\A^{\t}\y$, the subtlety
is treating the 0-boundary properly and we find
\begin{align*}
\left[\A^{\t}\y\right]_{i}-\left[\A^{\t}\y\right]_{i-1} & =\frac{1}{w}\sum_{j=0}^{w-1}y_{i-j}-\frac{1}{w}\sum_{j=0}^{w-1}y_{i-1-j}\\
 & =\frac{1}{w}(y_{i}-y_{i-w}).
\end{align*}
Therefore, we need to show that $y_{i}\geq y_{i-w}$ for $i\in\ir 1{\ti}$.
Since $\y$ is a length-$N$ non-decreasing vector, this holds unless
$y_{i}=0$ due to the boundary because $i\notin\ir 1N$. The key observation
is that $y_{i}=0$ if $i>N$ and this can cause a problem if $N<i\leq\ti$.
Since $\ti=\left\lceil (N+w-1)/2\right\rceil $, this requires $w\geq2i-N+1$.
But, $i-w\leq N-i-1\leq0$ if $i\leq N+1$ and, hence, there is no
problem because $y_{i}=0$ due to the boundary only if we also have
$y_{i-w}=0$ due to the boundary.

\subsection{Proof of Lemma \ref{lem:coupled_scalar_hessian_bound}}

Let $u_{m,i}(\x)=\left[\Uc''(\x)\right]_{m,i}$ and recall that $\Uc'(\x)=\g'(\x)\left(\x-\A^{\t}\f(\A\g(\x)\right)$.
This implies that

\begin{align*}
 & u_{m,i}(\x)=\frac{\mathrm{d}}{\mathrm{d}x_{m}}\left[\g'(\x)\left(\x-\A^{\t}\f(\A\g(\x)\right)\right]_{i}\\
 & =\frac{\mathrm{d}}{\mathrm{d}x_{m}}\left[\left(x_{i}-\sum_{j=1}^{N}A_{j,i}\, f\left(\sum_{k=1}^{M}A_{j,k}\, g(x_{k})\right)\right)g'(x_{i})\right]\\
 & =\delta_{i,m}g''(x_{i})\left(x_{i}-\sum_{j=1}^{N}A_{j,i}\, f\left(\sum_{k=1}^{M}A_{j,k}\, g(x_{k})\right)\right)\\
 & +g'(x_{i})\!\left(\delta_{i,m}\!-\!\sum_{j=1}^{N}A_{j,i}\, f'\left(\sum_{k=1}^{M}A_{j,k}\, g(x_{k})\right)\! A_{j,m}g'(x_{m})\right)\!.
\end{align*}
Hence, we have the upper bound
\begin{align*}
\left|u_{m,i}(\x)\right| & \leq\delta_{i,m}\left\Vert g''\right\Vert _{\infty}\xmax\\
 & \quad+\left\Vert g'\right\Vert _{\infty}\left(\delta_{i,m}+\left\Vert f'\right\Vert _{\infty}\left\Vert g'\right\Vert _{\infty}\sum_{j=1}^{N}A_{j,i}A_{j,m}\right).
\end{align*}
Since $\left\Vert \Uc''(\x)\right\Vert _{1}$ equals the maximum absolute
column sum,
\begin{align*}
 & \left\Vert \Uc''(\x)\right\Vert _{1}=\max_{i\in\ir 1M}\sum_{m=1}^{M}\left|u_{m,i}(\x)\right|\\
 & \leq\max_{i\in\ir 1M}\Bigg(\sum_{m=1}^{M}\delta_{i,m}\left\Vert g''\right\Vert _{\infty}\xmax\\
 & \quad+\sum_{m=1}^{M}\Bigg(\left\Vert g'\right\Vert _{\infty}\delta_{i,m}+\left\Vert f'\right\Vert _{\infty}\left\Vert g'\right\Vert _{\infty}^{2}\sum_{j=1}^{N}A_{j,i}A_{j,m}\Bigg)\Bigg)\\
 & \leq\left\Vert g''\right\Vert _{\infty}\xmax+\left\Vert g'\right\Vert _{\infty}+\left\Vert f'\right\Vert _{\infty}\left\Vert g'\right\Vert _{\infty}^{2}.
\end{align*}
 Since the Hessian $\Uc''(\x)$ is symmetric, it follows that $\left\Vert \Uc''(\x)\right\Vert _{\infty}=\left\Vert \Uc''(\x)\right\Vert _{1}$.
Hence, the bound on $\left\Vert \Uc''(\x)\right\Vert _{2}$ follows
from the standard inequality 
\[
\left\Vert \Uc''(\x)\right\Vert _{2}\leq\sqrt{\left\Vert \Uc''(\x)\right\Vert _{1}\left\Vert \Uc''(\x)\right\Vert _{\infty}}.
\]

\section{Proofs from Section \ref{sec:dependence}}

\label{sec:app_dep_proof}

\subsection{Proof of Lemma \ref{lem:lipshcitz_FGUP}}

Since $\frac{\d}{\d\epsilon}\Us(x;\epsilon)$ exists and is continuous
on $\Xc\times\Ec$, there exists a $\beta<\infty$ such that $|\Us(x;t_{0})-\Us(x;t_{1})|\leq\beta|t_{0}-t_{1}|$
(i.e., $\Us$ is uniformly Lipschitz continuous in $\epsilon$). For
arbitrary $t_{0},t_{1}\in\Ec$, we can assume without loss of generality
that $\Psi(t_{0})\geq\Psi(t_{1})$. Therefore, the Lipschitz continuity
of $\Psi(\epsilon)$ follows from 
\begin{align*}
\left|\Psi(t_{0})-\Psi(t_{1})\right| & =\left|\min_{x_{0}\in\Xc}\Us(x_{0};t_{0})-\min_{x_{1}\in\Xc}\Us(x_{1};t_{1})\right|\\
 & =\Us(x_{0}^{*};t_{0})-\Us(x_{1}^{*};t_{1})\\
 & \leq\Us(x_{1}^{*};t_{0})-\Us(x_{1}^{*};t_{1})\\
 & \leq\beta\left|t_{0}-t_{1}\right|,
\end{align*}
where the respective minima are achieved (at points $x_{0}^{*},x_{1}^{*}$)
because $\Us(x;\epsilon)$ is continuous in $x$ and $\Xc$ is compact.
Next, we show that $F^{(0,1)}(x;\epsilon)$ is non-negative and non-decreasing
in $x$. This holds because $f^{(0,1)}(x;\epsilon)$ is continuous
and non-negative on $\Xc\times\Ec$ and therefore, we can write 
\begin{align*}
F^{(0,1)}(x;\epsilon) & \triangleq\frac{\d}{\d\epsilon}\int_{0}^{x}\int_{0}^{\epsilon}f^{(0,1)}(s,t)\d t\d s\\
 & =\int_{0}^{x}f^{(0,1)}(s;\epsilon)\d s\geq0.
\end{align*}
The argument for $G^{(0,1)}(x;\epsilon)$ is identical.

\subsection{Proof of Lemma \ref{lem:e(x)_derivative}}

The definition of $\Xc_{f}$ shows that there exists an $\epsilon\in\Ec$
such that $h(x;\epsilon)-x=0$ for each $x\in\Xc_{f}$. This $\epsilon$
is unique and denoted by $\ex x$ because $h(x;\epsilon)$ is strictly
increasing in $\epsilon$ for $x\in\Xc_{f}\subset\Xm$ (e.g., $h^{(0,1)}(x;\epsilon)>0$).
Now, consider any $x_{0}\in\Xc_{f}$ satisfying $\ex{x_{0}}\in(0,\emax)$.
Since $h(x;\epsilon)$ is strictly increasing in $\epsilon$ for $x\in\Xc_{f}$,
it follows that $h(x_{0};0)-x_{0}<h(x_{0};\ex{x_{0}})-x_{0}=0$ and,
likewise, $h(x_{0};\emax)-x_{0}>0$. Using the continuity of $h$,
we see that there must be a $\delta>0$ such that $h(x;0)\leq x\leq h(x;\emax)$
for all $x\in I=[x_{0}-\delta,x_{0}+\delta]$. Therefore, the definition
of $\Xc_{f}$ shows that $I\subseteq\Xc_{f}$ and, hence, $x_{0}$
lies in the interior of $\Xc_{f}$. 

For $x\in\Xc_{f}$, the positive derivative condition also allows
us to apply the implicit function theorem for continuously differentiable
functions to $h(x;\epsilon)-x=0$. The result is that $\ex x$ is
continuously differentiable on $\Xc_{f}$ with derivative
\begin{align}
\epx x & =\frac{1-h^{(1,0)}(x;\ex x)}{h^{(0,1)}(x;\ex x)}.\label{eq:epx}
\end{align}

\subsection{Proof of Lemma \ref{lem:Q_ebp_integral}}

Under the stated assumptions, Lemma~\ref{lem:e(x)_derivative} shows
that $\ex x$ is a continuously differentiable function satisfying
the fixed-point equation $h(x;\epsilon(x))=x$ for all $[x_{1},x_{2}]\in\Xc_{f}$.
Since $\Us(x;\ex x)$ is differentiable, we can write
\begin{align*}
Q(x_{2}) & -Q(x_{1})=\Us(x_{2};\ex{x_{2}})-\Us(x_{1};\ex{x_{1}})\\
 & =\int_{x_{1}}^{x_{2}}\left(\frac{\d}{\d x}\Us(x;\ex x)\right)\d x\\
 & =\int_{x_{1}}^{x_{2}}\left(\Us^{(1,0)}(x;\ex x)+\Us^{(0,1)}(x;\epsilon)\epx x\right)\d x\\
 & \stackrel{(a)}{=}\int_{x_{1}}^{x_{2}}\Us^{(0,1)}(x;\ex x)\epx x\d x\\
 & \stackrel{(b)}{=}-\!\int_{x_{1}}^{x_{2}}\!\left(G^{(0,1)}\left(x;\ex x\right)\!+\! F^{(0,1)}\left(x;\ex x\right)\right)\!\epx x\d x,
\end{align*}
where $(a)$ follows from $\Us^{(1,0)}(x;\ex x)=0$ because $h(x;\ex x)=x$
and $(b)$ follows from using \eqref{eq:dUde_simple} to expand $\Us^{(0,1)}(x;\ex x)$
under the condition $h(x;\ex x)=x$.

\subsection{Proof of Lemma \ref{lem:e(x)_properties}}

 To see (i), we observe that $A=\left\{ (x,\epsilon)\in\Xc\times\Ec\mid h(x;\epsilon)=x\right\} $
is compact because $\Xc\times\Ec$ is compact and $h(x;\epsilon)-x$
is jointly continuous. Since the $x$-projection $\pi\colon\Xc\times\Ec\to\Xc$
defined by $(x,\epsilon)\mapsto x$ is continuous, we also find that
$\pi(A)=\left\{ x\in\Xc\mid\exists\epsilon\in\Ec,h(x;\epsilon)=x\right\} $
is compact. Therefore, the set 
\[
\Xc_{f}=\left\{ x\in\Xc\mid\exists\epsilon\in\Ec,h(x;\epsilon)=x\right\} \backslash\{0\}
\]
 is either closed or missing a single limit point at $x=0$.

Consider (ii). Consider the sequences $\underline{\epsilon}_{n}=(1-\frac{1}{n})\est$
and $\overline{\epsilon}_{n}=\est+\frac{\emax-\est}{n}$ for $n\in\mathbb{N}$.
For each $n\in\mathbb{N}$, the condition $0\leq\underline{\epsilon}_{n}<\est$
(or $\underline{\epsilon}_{n}=\est=0$ if $\est=0$) implies that
there is a $\delta_{n}>0$ such that $h(x;\underline{\epsilon}_{n})<x$
for all $x\in(0,\delta_{n}]$. Likewise, the condition $\est<\overline{\epsilon}_{n}\leq\emax$
implies that, for any $\gamma_{n}>0$, there is an $x_{n}\in(0,\gamma_{n})$
such that $h(x_{n};\overline{\epsilon}_{n})\geq x_{n}$. Starting
from any $x_{1}<\delta_{1}$, we can generate a decreasing sequence
$x_{n}\in(0,x_{n-1})$ by choosing $x_{n}$ in the second step based
on the parameter $\gamma_{n}=\min\{x_{n-1}/2,\delta_{n}\}$ given
by the first step. By construction, the $x_{n}$ sequence satisfies
$h(x_{n};\underline{\epsilon}_{n})<x_{n}\leq h(x_{n};\overline{\epsilon}_{n})$
and, hence, the continuity of $h$ shows that there is a sequence
$\epsilon_{n}\in[\underline{\epsilon}_{n},\overline{\epsilon}_{n}]$
such that $h(x_{n};\epsilon_{n})=x_{n}$. Finally, we observe that
$\epsilon_{n}\to\est$ and $x_{n}\to0$.

Consider (iii). If a system is unconditionally stable, then there
is a $\delta>0$ such that $h(x;\emax)<x$ for all $x\in(0,\delta]$.
From this, we see that $\est=\emax$ and $\ec\leq\est$ is satisfied
trivially. This also implies that $x\in(0,\delta]$ cannot support
a fixed point and $\Xc_{f}\cap[0,\delta]=\emptyset$. Along with (i),
we see that $\Xc_{f}$ is closed.

Consider (iv). If a system has a strict stability threshold $\est\in[0,\emax)$,
then, for $\epsilon\in(\est,\emax]$, there is a $\delta>0$ such
that $h(x;\epsilon)>x$ for all $x\in(0,\delta]$. Since $g(x;\epsilon)$
is strictly increasing in $x$, it follows that $g(\delta;\epsilon)-g(0;\epsilon)=\int_{0}^{\delta}g'(z;\epsilon)\d z>0$.
Using this and $x-h(x;\epsilon)<0$ for all $x\in(0,\delta]$, we
find that 
\[
\Us(\delta;\epsilon)-\Us(0;\epsilon)=\int_{0}^{\delta}(x-h(x;\epsilon))g'(x;\epsilon)\d x<0.
\]
This implies that $\epsilon\geq\ec$ because $0$ is not an $x$-minimizer
of $\Us(x;\epsilon)$ and, hence, $\overline{x}^{*}(\epsilon)>0$.
As $\epsilon\in(\est,\emax]$ was arbitrary, we conclude that $\est\geq\ec$.
For the second part, consider any sequence $x_{n}\in\Xc_{f}$ such
that $x_{n}\to0$. Such as sequence exists by (ii). If $\liminf_{n}\ex{x_{n}}<\est$,
then there is a subsequence $x_{n_{k}}\in\Xc_{f}$ and an $\eta>0$
such that $x_{n_{k}}\to0$ and $h(x_{n_{k}};\est-\eta)\geq x_{n_{k}}$.
But, this contradicts the definition of the stability threshold. A
similar argument shows that $\limsup_{n}\ex{x_{n}}>\est$ contradicts
the strict stability threshold condition. Therefore, we find that
$\lim_{n}\ex{x_{n}}=\est$.

\subsection{Proof of Lemma \ref{lem:alt_thresh}}

 Consider (i). For $\epsilon<\ec$, Lemma~\ref{lem:threshold_lemma}(iii)
shows that $\overline{x}^{*}(\epsilon)=0$ and this implies that $0$
is a fixed point of the uncoupled recursion. For $\epsilon<\ec$,
Proposition~\ref{prop:fp0_U0} shows that $\Us(0;\epsilon)=0$ and,
hence, $\Psi(\epsilon)=\Us(\overline{x}^{*}(\epsilon),\epsilon)=0$.
Moreover, $\Psi(\ec)=0$ by the continuity of $\Psi$. Finally, $\Psi(\epsilon)<0$
implies $\epsilon>\ec$ and the definition of $\ec$ implies that
$\overline{x}^{*}(\epsilon)>0$.

Consider (ii). For a proper admissible system, Theorem~\ref{thm:scalar_min_potential_integral}
shows that $\Psi(\epsilon)$ is strictly decreasing for $\epsilon\in[\ec,\emax]$.
On the other hand, (i) shows that $\Psi(\epsilon)=0$ for $\epsilon\leq\ec$.
From this, it also easy to verify that 
\begin{align}
\ec & =\sup\left\{ \epsilon\in\Ec\mid\Psi(\epsilon)\geq0\right\} \nonumber \\
 & =\sup\left\{ \epsilon\in\Ec\mid\Psi(\epsilon)=0\right\} \nonumber \\
 & =\min\left\{ \inf\left\{ \epsilon\in\Ec\mid\Psi(\epsilon)<0\right\} ,\emax\right\} ,\label{eq:ec_psi_lt0}
\end{align}
where the last expression requires the outer minimum because $\inf\left\{ \epsilon\in\Ec\mid\Psi(\epsilon)<0\right\} =\infty$
if $\Psi(\epsilon)=0$ for all $\epsilon\in\Ec$. Since $\Psi(\epsilon)={\textstyle \min_{x\in\Xc}}\Us(x;\epsilon)$,
the first of these proves the first stated result. For the second
statement, if $Q(x)<0$ for $x\in\Xc_{f}$, then $\Psi(\ex x)=\min_{x'\in\Xc}\Us(x';\ex x)\leq Q(x)$
implies that $\Psi(\ex x)\leq Q(x)<0$ and, hence, $\ex x>\ec$. 

Consider (iii). Let $A=\left\{ \epsilon\in\Ec\mid\Psi(\epsilon)<0\right\} $
and observe that~\eqref{eq:ec_psi_lt0} implies $\ec=\min\{\inf A,\emax\}$.
From this, we see that, if $\ec<\emax$, then $A\neq\emptyset$ and
$\ec=\inf A$. Now, we will show that $B=\left\{ \epsilon\in\Ec\mid\exists x\in\Xc_{f},\ex x=\epsilon,Q(x)<0\right\} $
satisfies $A=B$. To see this, we note that (i) shows $\Psi(\epsilon)<0$
(i.e., $\epsilon\in A$) implies $\overline{x}^{*}(\epsilon)>0$ and,
hence, $\overline{x}^{*}(\epsilon)\in\Xc_{f}$. Since $\ex{\overline{x}^{*}(\epsilon)}=\epsilon$
implies $Q(\overline{x}^{*}(\epsilon))=\Psi(\epsilon)<0$, we find
that $\epsilon\in B$. For the other direction, we consider the following.
If there is an $\epsilon\in\Ec$ such that $\ex x=\epsilon$ for some
$x\in\Xc_{f}$ with $Q(x)<0$ (i.e., $\epsilon\in B$), then 
\begin{equation}
\Psi(\epsilon)=\min_{x'\in\Xc}\Us(x';\ex x)\leq Q(x)<0.\label{eq:Psi_lt_Q}
\end{equation}
Hence, $\epsilon\in A$. Therefore, $A=B$ and $\inf A=\inf B=\ec$. 

Since $A\neq\emptyset$, there is a sequence $\epsilon_{n}\in A$
such that $\lim_{n}\epsilon_{n}=\inf A=\ec$. This implies that $\Psi(\epsilon_{n})\to0$
because $\Psi$ is continuous and $\Psi(\ec)=0$. Likewise, consider
any sequence $x_{n}\in\Xc_{f}$ where $Q(x_{n})<0$ and $\lim\ex{x_{n}}=\inf B=\ec$.
Using the same idea as \eqref{eq:Psi_lt_Q}, we see that $\Psi(\ex{x_{n}})\leq Q(x_{n})<0$.
Since $\epsilon_{n}=\ex{x_{n}}\in A$ and $\epsilon_{n}\to\ec$, it
follows that $\Psi(\epsilon_{n})\to0$. But, this also implies $\lim_{n}Q(x_{n})=0$
because $\limsup_{n}Q(x_{n})\leq0$ by construction and $\liminf_{n}Q(x_{n})\geq\liminf_{n}\Psi(\ex{x_{n}})=0$.
By considering subsequences where $\overline{x}=\lim_{n}x_{n}$ exists,
we see that $\ex{\overline{x}}=\ec$ and $Q(\overline{x})=0$ for
any limit point of the sequence $x_{n}$. From this, we see that 

\begin{equation}
\min\left\{ \ex x\mid x\in\overline{\Xc_{f}},Q(x)=0\right\} \leq\ec.\label{eq:maxwell_lt_ec}
\end{equation}

The possibility of strict inequality is due only to the fact that
an isolated point $x\in\overline{\Xc_{f}}$ with $Q(x)=0$ might decrease
the LHS because there would exist no sequence $x_{n}\in\Xc_{f}$ with
$Q(x_{n})<0$ that approaches $x$. But, any $x\in\Xc_{f}$ with $Q(x)=0$
must also satisfy $x\neq0$ and $\Psi(\ex x)\leq Q(x)=0$. Hence,
if $\Psi(\ex x)=0$, then $\overline{x}^{*}(\ex x)\geq x>0$ and $\ex x\geq\ec$.
Likewise, if $\Psi(\ex x)<0$ then the monotonicity of $\Psi$ implies
that $\ex x>\ec$. Therefore, we have equality in \eqref{eq:maxwell_lt_ec}
if $\Xc_{f}$ is a closed set because $x_{n}\to x\in\Xc_{f}$.

Lastly, we consider the case where $0$ is a limit point of $\Xc_{f}$.
In this case, the system has a strict stability threshold $\est\in[0,\emax]$
by assumption. If $\est\in[0,\emax)$, then Lemma~\ref{lem:e(x)_properties}(iv)
shows that $\est\geq\ec$ and that the limit $\ex 0\triangleq\lim_{n}\ex{x_{n}}=\est$
exists for all sequences $x_{n}\in\Xc_{f}$ such that $x_{n}\to0$.
From this, we see that $\ex 0\geq\ec$ and, hence, we must have equality
in \eqref{eq:maxwell_lt_ec}. If $\est=\emax$ and $0$ is a limit
point of $\Xc_{f}$, then the limit $\ex 0\triangleq\lim_{n}\ex{x_{n}}=\est$
also exists because $\ex{x_{n}}\leq\emax$ holds trivially and $\liminf_{n}\ex{x_{n}}<\est$
leads to the same contradiction as in the proof of Lemma~\ref{lem:e(x)_properties}(iv).
In this case, $\ex 0=\emax\geq\ec$ holds trivially.

\section{Proofs from Section \ref{sec:applications}}

\label{sec:app_app_proof}

\subsection{Proof of Lemma \ref{lem: P'x&P''x}}

From $P(x)=-2Q(x)$ and \eqref{eq:gldpc_Q_simple}, we can see that
\begin{align*}
P'(x) & =g(x)-xg'(x)\\
 & =\frac{1}{B(t,n-t)}\Big(\int_{0}^{x}z^{t-1}(1-z)^{n-t-1}\d z\\
 & \quad\quad-x^{t}(1-x)^{n-t-1}\Big).
\end{align*}
Since the derivative $\frac{\d}{\d x}x^{t}(1-x)^{n-t-1}$ is given
by 
\begin{align*}
 & =tx^{t-1}(1-x)^{n-t-1}-(n-t-1)x^{t}(1-x)^{n-t-2}\\
 & =(t(1-x)-(n-t-1)x)x^{t-1}(1-x)^{n-t-2}\\
 & =(t-(n-1)x)x^{t-1}(1-x)^{n-t-2},
\end{align*}
and the expression $x^{t}(1-x)^{n-t-1}$ can be written as 
\[
\int_{0}^{x}\left(t-(n-1)z\right)z^{t-1}(1-z)^{n-t-2}\d z,
\]
we find that $P'(x)$ can be expressed as 
\begin{align}
P'(x) & =\frac{1}{B(t,n-t)}\int_{0}^{x}\bigg(z^{t-1}(1-z)^{n-t-1}\nonumber \\
 & \quad\quad-\left(t-(n-1)z\right)z^{t-1}(1-z)^{n-t-2}\bigg)\d z\nonumber \\
 & =\frac{1}{B(t,n-t)}\int_{0}^{x}\bigg((1-z)\nonumber \\
 & \quad\quad-\left(t-(n-1)z\right)z^{t-1}(1-z)^{n-t-2}\bigg)\d z\nonumber \\
 & =\frac{1}{B(t,n-t)}\int_{0}^{x}\big(1-t\label{eq: P'x}\\
 & \quad\quad+(n-2)z\big)z^{t-1}(1-z)^{n-t-2}\d z.\nonumber 
\end{align}
Using $z^{t-1}(1-z)^{n-t-2}>0$, it can be shown $P'(x)<0$ for all
$0<x<\frac{t-1}{n-2}$. By applying fundamental theorem of calculus
to \eqref{eq: P'x}, we find that
\begin{align*}
P''(x) & =\frac{1}{B(t,n-t)}\left(1-t+(n-2)x\right)x^{t-1}(1-x)^{n-t-2},
\end{align*}
and, hence, $P''(x)\geq0$ for all $\frac{t-1}{n-2}\leq x<1$.

\section*{Ackowledgements}

The authors would like to thank the Associate Editor, David Burshtein,
and the two anonymous reviewers for their time and valuable suggestions.
They would also like to thank Santhosh Kumar for his careful reading
of an initial draft of this paper.



\begin{IEEEbiographynophoto}{Arvind Yedla} (S'08--M'12) received his Ph.D.\ in electrical engineering from Texas A\&M University, College Station in 2012 and joined the Mobile Solutions Lab, Samsung in 2012, where he is currently a Senior Engineer. His current research interests include information theory, channel coding, and iterative information processing with applications in wireless communications. \end{IEEEbiographynophoto}

\begin{IEEEbiographynophoto}{Yung-Yih Jian} received his Ph.D. degree in electrical engineering from Texas A\&M University in 2013 and joined Qualcomm, Inc. in Santa Clara. Prior to joining Texas A\&M University, he spent four years in Industrial Technology Research Institute (ITRI), Chutung, Taiwan. His reserach interests include information theory, error correcting codes, signal processing with  applications in wireless communications.
\end{IEEEbiographynophoto}

\begin{IEEEbiographynophoto}{Phong S. Nguyen}   received the B.Eng. degree in electronics and telecommunications from Hanoi University of Technology, Hanoi, Vietnam, in 2004, and the Ph.D. degree in electrical and computer engineering from Texas A\&M University, College Station, in 2012.
He joined Marvell Semiconductor, Inc., Santa Clara, CA, in 2012 where he is currently a Staff Engineer in the Data Storage Signal Processing group. His research interests include information theory, signal processing and channel coding with applications in data storage and wireless communications.
\end{IEEEbiographynophoto}

\begin{IEEEbiographynophoto}{Henry D. Pfister} (S'99--M'03--SM'09) received his Ph.D. in electrical engineering from UCSD in 2003 and is currently an associate professor in the electrical and computer engineering department of Duke University.  Prior to that, he was a professor at Texas A\&M University (2006-2014), a post-doc at EPFL (2005-2006), and a senior engineer at Qualcomm Corporate R\&D in San Diego (2003-2004).
He received the NSF Career Award in 2008, the Texas A\&M ECE Department Outstanding Professor Award in 2010, and was a coauthor of the 2007 IEEE COMSOC best paper in Signal Processing and Coding for Data Storage.  He is currently an associate editor in coding theory for the IEEE Transactions on Information Theory.
\end{IEEEbiographynophoto}
\end{document}